\documentclass{aa}
\usepackage{txfonts}

\usepackage{graphbox}   
\usepackage{cellspace}  
\usepackage{multirow}  
\usepackage{color}
\usepackage{url}

\DeclareRobustCommand{\ion}[2]{\textup{#1\,\textsc{\lowercase{#2}}}}

\begin{document}

\title{The Lyman Alpha Reference Sample} \subtitle{XV. Relating Ionised Gas
  Kinematics with Lyman-$\alpha$ observables}

\author{
  E.~C.~Herenz\inst{\ref{IUCAA}}
  \and
  A.~Schaible\inst{\ref{H1},\ref{H2}}
  \and
  P.~Laursen\inst{\ref{DAWN},\ref{NB}}
  \and
  A.~Runnholm\inst{\ref{SU}}
  \and
  J.~Melinder\inst{\ref{SU}}
  \and
  A.~Le Reste\inst{\ref{MP}}
  \and
  M.~J.~Hayes\inst{\ref{SU}}
  \and
  G.~\"{O}stlin\inst{\ref{SU}}
  \and
  J.~Cannon\inst{\ref{SP}}
  \and
  G.~Micheva\inst{\ref{AIP}}
  \and
  M.~Roth\inst{\ref{AIP}}
  \and
  K.~Saha\inst{\ref{IUCAA}}
}

\institute{ Inter-University Centre for Astronomy and Astrophysics (IUCAA), Pune
  University Campus, Pune 411 007,
  India  \\ e-mail: \url{edmund.herenz@iucaa.in} \label{IUCAA}
  \and
  Universit\"{a}t Heidelberg,
  Interdisziplin\"{a}res Zentrum f\"{u}r Wissenschaftliches Rechnen,
  Im Neuenheimer Feld 205, D-69120 Heidelberg, Germany
  \label{H1}
  \and
  Universit\"{a}t Heidelberg,
  Zentrum f\"{u}r Astronomie, Institut f\"{u}r Theoretische Astrophysik,
  Albert-Ueberle-Straße 2,
  D-69120 Heidelberg, Germany
  \label{H2}
  \and
  Cosmic Dawn Center (DAWN), Copenhagen, Denmark \label{DAWN}
  \and
  Niels Bohr Institute, University of Copenhagen, Copenhagen,
  Denmark \label{NB}
  \and
  Department of Astronomy, Stockholm
  University, AlbaNova University Centre, SE-106 91, Stockholm,
  Sweden \label{SU}
  \and
  Minnesota Institute for Astrophysics, University of Minnesota, 116 Church
  Street SE, Minneapolis, MN 55455, USA \label{MP}
  \and
  Department of Physics \& Astronomy, Macalester College, 1600 Grand
  Avenue, Saint Paul, MN\,55105, USA \label{SP}
  \and
  Leibniz Institute for Astrophysics (AIP),
  An der Sternware 16, 14482 Potsdam, Germany \label{AIP}
}

\abstract{Gas kinematics affect the radiative transfer and escape of hydrogen Lyman-$\alpha$ (Ly$\alpha$) emission from galaxies.  We investigate this interplay empirically by relating the ionised gas kinematics of 42 galaxies in the extended Ly$\alpha$ Reference Sample (eLARS) with their Ly$\alpha$ escape fractions, $f_\mathrm{esc}^\mathrm{Ly\alpha}$, Ly$\alpha$ equivalent widths, $\mathrm{EW}_\mathrm{Ly\alpha}$, and Ly$\alpha$ luminosities, $L_\mathrm{Ly\alpha}$.  To this aim we use PMAS integral-field spectroscopic observations of the Balmer-$\alpha$ line.  Our sample contains 18 rotating disks, 13 perturbed rotators, and 13 galaxies with more complex kinematics.  The distributions of $f_\mathrm{esc}^\mathrm{Ly\alpha}$, $\mathrm{EW}_\mathrm{Ly\alpha}$, and $L_\mathrm{Ly\alpha}$ do not differ significantly between these kinematical classes, but the largest Ly$\alpha$ observables are found amongst the kinematically complex systems.  We find no trends of $f_\mathrm{esc}^\mathrm{Ly\alpha}$ and $\mathrm{EW}_\mathrm{Ly\alpha}$ with kinematic or photometric inclinations.  We calculate shearing velocities, $v_\mathrm{shear}$, and intrinsic velocity dispersions, $\sigma_0^\mathrm{obs}$ (empirically corrected for beam-smearing effects), as global kinematical measures for each galaxy.  The sample is characterised by highly turbulent motions ($30$\,km\,s$^{-1} \lesssim \sigma_0^\mathrm{obs} \lesssim 80$\,km\,s$^{-1}$) and more than half of the sample shows dispersion dominated kinematics. We uncover clear trends between Ly$\alpha$ observables and global kinematical statistics: $\mathrm{EW}_\mathrm{Ly\alpha}$ and $L_\mathrm{Ly\alpha}$ correlate with $\sigma_0^\mathrm{obs}$, while $f_\mathrm{esc}^\mathrm{Ly\alpha}$ anti-correlates with $v_\mathrm{shear}$ and $v_\mathrm{shear} / \sigma_0^\mathrm{obs}$.  Moreover, we find, that galaxies with $\mathrm{EW}_\mathrm{Ly\alpha} \geq 20$\,\AA{} are characterised by higher $\sigma_0$ and lower $v_\mathrm{shear} / \sigma_0^\mathrm{obs}$ than galaxies below this threshold.  We discuss statistically the importance of $v_\mathrm{shear}$, $\sigma_0^\mathrm{obs}$, and $v_\mathrm{shear}/\sigma_0^\mathrm{obs}$ for regulating the Ly$\alpha$ observables in comparison to other galaxy parameters.  It emerges that $\sigma_0^\mathrm{obs}$ is the dominating parameter for regulating $\mathrm{EW}_\mathrm{Ly\alpha}$ and that is as important as nebular extinction, gas covering fraction, and ionising photon production efficiency in regulating $f_\mathrm{esc}^\mathrm{Ly\alpha}$.  A simple scenario where the starburst age is simultaneously regulating turbulence, $\mathrm{EW}_\mathrm{Ly\alpha}$, and $f_\mathrm{esc}^\mathrm{esc}$ does not find support by our observations.  However, we show that the small scale distribution of dust appears to be influenced by turbulence in some galaxies.  In support of our observational result we discuss how turbulence is theoretically expected to play a significant role in modulating $f_\mathrm{esc}^\mathrm{Ly\alpha}$.}


\maketitle


\defcitealias{Herenz2016}{H16}
\defcitealias{Runnholm2020}{R20}
\defcitealias{Melinder2023}{M23}

\section{Introduction}
\label{sec:intro}

Studying the early stages of galaxy formation is fundamental for understanding their evolution and diversity in general.  This requires observing the most distant galaxies.  Although the field has recently been revolutionised with the launch of the \emph{James Webb} Space Telescope (JWST), galaxies in these early phases still appear as very faint and close-to-unresolved sources.  While JWST has provided rest-frame optical emission line detections out to redshifts $z\sim7 - 11$ \citep[e.g.,][]{Bunker2023,ArrabalHaro2023,Heintz2023,Saxena2023,Tang2023,Boyett2024}, there is one hydrogen line that continues to push the frontier of cosmology: the ultra-violet Lyman-$\alpha$ (Ly$\alpha$; $\lambda_\mathrm{Ly\alpha} = 1215.68$\,\AA{}) resonant line \citep[see reviews by][]{Hayes2015,Dijkstra2019,Ouchi2020}.  Originating mainly from the recombining gas in the vicinity of hot O- and B-stars Ly$\alpha$ is associated in particular with young galaxies that are in the process of rapidly converting gas into stars, as famously envisioned by \citet{Partridge1967}.

An observational census of Ly$\alpha$-emitting galaxies (LAEs) has been established for galaxies with Ly$\alpha$ luminosities $L_\mathrm{Ly\alpha} \gtrsim 10^{41.5}$\,erg\,s$^{-1}$ throughout most of the history of the Universe by means of the LAE luminosity function \citep[e.g.,][]{Ouchi2008,Blanc2011,Sobral2017,Wold2017,Herenz2019,Liu23,Thai2023} up into the Epoch of Reionisation \citep[e.g.,][]{Ouchi2010,Matthee2015,Santos2016,Konno2018,Ning2022,Wold2022}.  These censuses ascertain that LAEs are an abundant population of high-redshift ($z \gtrsim 2$) galaxies that can serve as a tracer for addressing fundamental questions of galaxy formation and cosmology.  For example, the HETDEX survey will measure redshifts of more than a million LAEs from $z \sim 1.8 $ to $z \sim 2.5$ to investigate the nature of dark energy \citep{Gebhardt2021,MentuchCoooper2023}.  Moreover, LAEs also serve as an important probe to constrain the timeline and the topology of the intergalactic medium during the Epoch of Reionisation (e.g., reviews by \citealt{Dijkstra2014} and \citealt{Choudhury2022}; see also recent results by \citealt{Nakane2023}, \citealt{Witstok2024}, and \citealt{Witten2024}). Finally, large area cosmological line intensity mapping experiments, that attempt to measure baryonic acoustic oscillations as a function of the Universe's age, will also rely on the Ly$\alpha$ signal from the unresolved population of LAEs \citep[e.g.,][]{Bernal2022,Lujan-Niemeyer2023}.

Given the high Ly$\alpha$ absorption cross-section of neutral hydrogen, the resulting Ly$\alpha$ optical depths in typical interstellar media (ISM) are extremely high ($\tau_\mathrm{Ly\alpha} > 10^6$ for neutral columns $N_\mathrm{HI} > 10^{20}$\,cm$^{-2}$).  Hence the Ly$\alpha$ line is resonant.  The resulting complex radiative transfer within the ISM redistributes the initial Ly$\alpha$ radiation field spatially and in frequency space.  Moreover, the radiative transfer increases a Ly$\alpha$ photons' path length and makes it prone to be absorbed by dust.  Theoretical works demonstrate the resulting intricate relationships between Ly$\alpha$ observables (e.g., Ly$\alpha$ spectral morphology, Ly$\alpha$ surface brightness distribution, and total Ly$\alpha$ line intensity) and the density distribution, dust distribution, as well as the kinematics of the interstellar scattering medium \citep[see, e.g., recent papers by][and references therein]{Remolina-Guti2019,Smith2019,Lao2020,Song2020,Gronke2021,Li2022,Blaizot2023,Munirov2023}.

A key quantity in the context of LAEs is the Ly$\alpha$ escape fraction, $f_\mathrm{esc}^\mathrm{Ly\alpha}$, defined as the ratio between the observed Ly$\alpha$ luminosity to the intrinsic Ly$\alpha$ luminosity.  An understanding of the galactic characteristics that regulate $f_\mathrm{esc}^\mathrm{Ly\alpha}$ is required to answer the question ``\emph{What makes a star-forming galaxy a LAE?}''.  Many observational \citep[e.g.,][]{Giavalisco1996,Keel2005,Finkelstein2009,Hayes2011,Malhotra2012,Atek2014,Hathi2016,Hagen2016,Oyarzun2016,Paulino-Afonso2018,Matthee2021,McCarron2022,ChavezOrtiz2023,Napolitano2023,Hayes2023b,Lin2024} and modelling \citep[e.g.,][]{Schaerer2008,Verhamme2008,Forero-Romero2011,Dijkstra2012,Dayal2012,Garel2015,Garel2016,Gurung-Lopez2019} efforts were conducted in the spirit of answering this question.  It is now established that the physical and morphological characteristics of LAEs are significantly different compared to star-forming galaxies where $f_\mathrm{esc}^\mathrm{Ly\alpha} \sim 0$.  As reviewed by \cite{Ouchi2020}, high-$z$ LAEs are commonly compact ($r_e \sim 1$\,kpc) star-forming galaxies ($\mathrm{SFR} \sim 1 \dots 10$\,M$_\odot$yr$^{-1}$) with low-dust content ($E(B-V) \lesssim 0.2$) that harbour young stellar populations ($\mathrm{age} \sim 10$\,Myr) of low stellar mass ($\sim 10^8 \dots 10^9$\,M$_\odot$).  Moreover, the ISM of those galaxies is often metal-poor, with metallicities ranging from $Z \sim 0.1$\,Z$_\odot$ to $Z \sim 0.3$\,Z$_\odot$.  This differentiates LAEs from non-LAEs and/or continuum selected high-$z$ galaxies, that are typically more evolved and more massive systems, but there is certainly overlap between both populations.

The defining criterion for LAEs in surveys is often based on a threshold in the equivalent width of the Ly$\alpha$ line, $\mathrm{EW}_\mathrm{Ly\alpha}$.  Interestingly, $\mathrm{EW}_\mathrm{Ly\alpha}$ is linearly correlated with $f_\mathrm{esc}^\mathrm{Ly\alpha}$ \citep{Sobral2019,Melinder2023}.  \cite{Sobral2019} explain the measured slope of this relation in a simple scenario where, on average, galaxies with higher $\mathrm{EW}_\mathrm{Ly\alpha}$ and $f_\mathrm{esc}^\mathrm{Ly\alpha}$ are dominated by stellar populations with a harder ionising spectrum, as characterised by the ionising photon production efficiency $\xi_\mathrm{ion}$, and lower $E(B-V)$.  This idea is consistent with results that show how the overall dust-poor high-$\mathrm{EW}_\mathrm{Ly\alpha}$ LAEs are indeed characterised by higher $\xi_\mathrm{ion}$ \citep[e.g.,][]{Trainor2016,Nakajima2016,Nakajima2018,Maseda2020,Hayes2023b,Kramarenko2023}.  Certainly, $\xi_\mathrm{ion}$ and $E(B-V)$ are galaxy characteristics that we expect to directly influence the radiative transfer.  That these characteristics occur predominantly in compact star-forming systems of low-metallicity and low-mass informs us more about the processes of star-formation and, more generally, galaxy formation physics in such systems.  However, the causal connection between Ly$\alpha$ radiative transport physics and stellar mass or star-formation rate appears to involve some other process.  In this respect we recall that galaxy and ISM kinematics are causally connected with star-formation and stellar mass \citep[see, e.g., review by][]{Glazebrook2013}.  Thus we may expect also causal relationships between galaxy kinematics and $f_\mathrm{esc}^\mathrm{Ly\alpha}$ or $\mathrm{EW}_\mathrm{Ly\alpha}$.

Significant effort has been devoted to understand the effect of kinematics and gas distribution on the Ly$\alpha$ line profile morphology \citep[e.g.,][and references therein]{Blaizot2023}.  The kinematics of the scattering medium are also frequently studied in absorption, especially using low-ionisation state absorption lines of Si$^+$ and C$^+$ ions in the UV \citep[][]{Kunth1998,Wofford2013,Henry2015,Rivera-Thorsen2015,Hayes2023,Hayes2023b}.  These line profile analyses and absorption line studies reveal how low neutral \ion{H}{I} covering fractions and outflow kinematics appear to promote Ly$\alpha$ escape.  In principle, the neutral phase can also be probed by the 21\,cm line.  However, the spatial resolution of the radio beam is usually significantly lower then the spatial scales that can be probed by imaging and spectroscopy \citep{Cannon2004,Pardy2014}.  Nevertheless, such analyses reveal that mergers offset large amounts of \ion{H}{I} via tidal interactions and it appears likely that the thereby reduced neutral column towards the star-forming regions promotes Ly$\alpha$ escape \citep{Purkayastha2022,LeReste2024}.  A promising way forward are highly spatially resolved aperture synthesis observations in combination with other observational probes of the emitting and the scattering medium.  Especially the recent work by \cite{LeReste2022} informs us about the role of the interplay between \ion{H}{II} and \ion{H}{I} gas on inter-stellar medium scales in facilitating Ly$\alpha$ escape.

To date few studies aimed at understanding the role of the gas kinematics in shaping the global Ly$\alpha$ characteristics $f_\mathrm{esc}^\mathrm{Ly\alpha}$, $\mathrm{EW}_\mathrm{Ly\alpha}$, and Ly$\alpha$ luminosity ($L_\mathrm{Ly\alpha}$). To this aim we need to resort to galaxy averaged kinematical statistics.  Importantly, this averaging should be done on spatial scales most relevant for the radiative transfer problem.  Such an analysis was attempted by \cite{Herenz2016}, hereafter \citetalias{Herenz2016}, who used integral-field spectroscopic observations of the H$\alpha$ line of the Lyman-$\alpha$ reference sample \citep[LARS;][]{Ostlin2014}.  This study found that, galaxies whose ionised gas kinematics are dominated by unordered (turbulent) motions can appear as Ly$\alpha$ emitters, whereas galaxies that are dominated by large scale coherent motions never appear to show significant Ly$\alpha$ escape.  Evidence for turbulence driven Ly$\alpha$ escape was also presented by \cite{Puschnig2020} from a study of the dust and the molecular gas content of LARS.  Moreover, a connection between global galaxy kinematics and Ly$\alpha$ escape was reported very recently by \cite{Foran2023} for the first time in high-$z$ galaxies.  These authors analysed the global kinematical characteristics of $z\sim2$ and $z\sim3$ galaxies.  For these galaxies also consistent $\mathrm{EW}_\mathrm{Ly\alpha}$ measurements were available.  Objects that showed no significant Ly$\alpha$ emission or even Ly$\alpha$ in absorption (i.e. low or negative $\mathrm{EW}_\mathrm{Ly\alpha}$) were found to be a mix of dispersion and rotation dominated systems, whereas galaxies with significant Ly$\alpha$ emission were always found to be dispersion dominated.

The goal of the present study is to revisit the relationship between galaxy kinematics and Ly$\alpha$ kinematics observationally, with the main aim of removing the shortcomings of the rather small sample used in the kinematical study of \citetalias{Herenz2016}.  To this aim we will be analysing IFS observations of the H$\alpha$ line for all 42 galaxies of the LARS + extended LARS programmes \citep[eLARS;][herafter M23]{Melinder2023}.  The enlarged dataset now allows to discuss the importance of galaxy kinematics for Ly$\alpha$ observables with respect to other galaxy characteristics in a quantitative manner.

We structure this paper as follows: In Sect.~\ref{sec:obsred} we describe the sample and data used and, especially, the newly obtained IFS observations (the data reduction is outlined in Appendix~\ref{sec:pmas-data-reduction}).  In Sect.~\ref{sec:analysis-results} we present the kinematic analysis and our relations between galaxy kinematics and Ly$\alpha$ observables (the creation of the kinematic maps and the calculation of the intrinsic dispersion are described in more detail in Appendix~\ref{sec:deta-descr-kinem} and Appendix~\ref{sec:meas-intr-disp}, respectively).  We then discuss in Sect.~\ref{sec:disc} the importance of kinematic parameters with respect to the \emph{``soup of quantities''} that have been found correlate with Ly$\alpha$ properties of galaxies.  There we also interpret our results.  We conclude in Section~\ref{sec:conc}.

\section{Data}
\label{sec:obsred}

\subsection{The (Extended) Lyman-$\alpha$ Reference Sample}
\label{sec:extended-lyman-alpha}

The sample analysed here consists of 42 star-forming galaxies selected from the SDSS spectroscopic database.  The main selection criteria are cuts in the H$\alpha$ equivalent width, $\mathrm{EW}_\mathrm{H\alpha}$ (as a proxy for stellar-age), and FUV luminosity (at $\lambda \approx 1500$\,\AA{} from GALEX), $L_\mathrm{UV}$ (as a proxy for star-formation rate).  Fourteen of the 42 galaxies had to satisfy a cut in $\mathrm{EW}_\mathrm{H\alpha} > 100$\,\AA{}.  These 14 galaxies (LARS1 -- LARS14) comprised the initial Ly$\alpha$ Reference Sample \citep[LARS;][]{Ostlin2014,Hayes2014}.  As detailed in \cite{Ostlin2014}, additional considerations for the sample selection were that the galaxies cover a range in FUV luminosities typical of star-forming galaxies in the early universe, and the availability of relevant Hubble Space Telescope (HST) archival data to design the HST imaging and spectroscopic campaign as economically as possible.

The initial sample was extended with 28 star-forming galaxies (eLARS 1 -- 28) that had to satisfy a cut of $\mathrm{EW}_\mathrm{H\alpha}> 40$\,\AA{}, and again the sample was picked to include a dynamical range in $L_\mathrm{UV}$ \citepalias{Melinder2023}.  Only 7 out of those 28 galaxies show $\mathrm{EW}_\mathrm{H\alpha}> 100$\,\AA{}.  The final sample, LARS+eLARS, now spans a range in $\log_{10} (L_\mathrm{FUV}\,[\mathrm{L}_\odot])$ from 9 to 11 and in $\mathrm{EW}_\mathrm{H\alpha}$ from 41\,\AA{} to 578\,\AA{}.  Stellar mass and H$\alpha$ star-formation rate distributions are presented in Sect.~\ref{sec:corr-betw-galaxy}.  All galaxies show emission line ratios consistent with being powered by a young stellar population according to the extreme starburst classification demarcation line in the [\ion{O}{III}]$\lambda5007$]/H$\beta$ vs. [\ion{N}{II}]$\lambda6548$/H$\alpha$ diagram provided by \cite{Kewley2001}.  However, five galaxies could be classified as harbouring an AGN according to the criterion of \cite{Kauffmann2003}.  The HST imaging observations of all 42 galaxies were used to synthesise images in Ly$\alpha$, H$\alpha$, and H$\beta$.  We use for the present analysis global measurements derived from these images (see Sect.~\ref{sec:glob-meas-from} below).

\subsection{PMAS Observations of eLARS}
\label{sec:pmas-observ-elars}

\begin{table}
  \caption{Log of Calar-Alto 3.5m PMAS eLARS observations.}
  \centering
  \begin{tabular}{lcrccl}
    \hline \hline \noalign{\smallskip}
    eLARS  & Night   &  $t_\mathrm{obs}$ & $R$ & DIMM             & Clouds \\
    ID     &      &  [s]              & {}  & [\arcsec{}] & {} \\ 
    \hline \noalign{\smallskip}
    1      &  16/03/31  &  1800          & 3890    & 2.0  & thick \\
    2      &  16/04/02  &  1800          & 4348    & 2.2  & clear \\
    3A     &  16/04/01  &  1800          & 2975    & 2.2  & clear \\
    3B     &  16/04/01  &  1800          & 2931    & 2.2  & clear \\
    4      &  16/04/02  &   900          & 2727    & 2.3  & clear \\
    5A     &  16/04/01  &  1800          & 3580    & 2.3  & clear \\
    5B     &  16/04/01  &  1800          & 3395    & 2.3  & clear \\
    6      &  17/03/26  &  2$\times$1800 & 6262    & 2.0  & thin   \\
    7      &  16/03/30  &  3$\times$900  & 3572    & 0.8  & clear \\
    8      &  16/04/02  &  1800          & 3997    & 1.5  & thick \\
    9      &  16/04/03  &  1800          & 3831    & 1.3  & clear  \\
    10     &  16/04/01  &  1800          & 3092    & 2.3  & thin  \\
    11     &  17/03/22  &  2$\times$1800 & 6174    & 2.2  & clear  \\
    12     &  17/03/25  &  2$\times$1800 & 5400    & 1.7  & clear  \\
    13     &  16/03/30  &  3$\times$900  & 3591    & 1.3  & clear \\
    14     &  16/03/30  &  2$\times$1800 & 3680    & 1.2  & clear  \\
    15     &  17/03/25  &  2$\times$1800 & 5732    & 1.4  & clear \\
    16     &  16/04/03  &  2$\times$1800 & 4398    & 1.4  & thin \\
    17     &  16/04/03  &  2$\times$900  & 5494    & 1.4  & thick \\
    18     &  16/04/03  &  1800          & 4361    & 1.2  & thin \\
    19     &  16/03/31  &  2$\times$1800 & 3843    & 1.3  & thin \\
    20     &  17/03/26  &  2$\times$1800 & 6469    & 1.9  & thin \\
    21     &  17/03/24  &  1800          & 4475    & 2.2  & clear \\
    22     &  16/04/01  &  1800          & 2677    & 1.7  & clear \\
    23     &  16/04/02  &  1800          & 2626    & 1.7  & clear \\
    24     &  16/04/02  &  1800          & 2951    & 2.1  & clear \\
    25     &  17/03/26  &  3$\times$1800 & 6375    & 2.1  & thin \\
    26A    &  17/03/25  &  2$\times$1800 & 5019    & 1.7  & clear \\
    26B    &  17/03/25  &  2$\times$1800 & 5301    & 1.8  & clear \\
    27     &  17/03/25  &  2$\times$1800 & 5528    & 1.4  & clear \\
    28     &  17/03/22  &  2$\times$1800 & 5960    & 1.1  & clear \\ \noalign{\smallskip}
    \hline \hline
  \end{tabular}
  \label{tab:log}
\end{table}

We observed all 28 eLARS galaxies with the Potsdam Multi Aperture Spectrophotometer \citep[PMAS; ][]{Roth2005} at the Calar Alto 3.5\,m telescope.  An overview of the campaign is provided in Table~\ref{tab:log}.  We used PMAS in its Lens Array configuration with double magnification -- in this configuration the 256 spectral pixels (spaxels) of 1\arcsec{}$\times$1\arcsec{} size sample contiguously a 16\arcsec{}$\times$16\arcsec{} field of view.  Within this field of view the bulk of the high surface-brightness H$\alpha$ emission of the galaxies in the sample could be covered with a single pointing.  Exceptions are eLARS 3, 5, and 26, which required two pointings (pointings are differentiated by the suffixes A and B in Table~\ref{tab:log}).  We used the $R1200$ backward-blazed grating as in \citetalias{Herenz2016}, but different to \citetalias{Herenz2016}, here the grating angle could be held fixed so to disperse the first spectral order around H$\alpha$ close to the centre of the detector for all galaxies.  According to the PMAS grating table\footnote{Retrieved online from \url{http://www.caha.es/pmas/PMAS_COOKBOOK/TABLES/pmas_gratings.html}.}  the expected nominal resolving power for the adopted position (\texttt{GROTPOS = 109}) is $R \sim 5000$.  In order to sample the spectroscopic line spread function adequately at this resolving power \citep[see][]{Robertson2017} we did not bin the read out of the PMAS 4k$\times$4k CCD along the dispersion direction (\texttt{XBIN = 1}).  This set up was chosen to obtain data that is optimal for a kinematic study of the ionised gas.

The observations took place during the course of two visitor mode observing runs in spring 2016 (run350) and spring 2017 (run362).  We observed all our targets at air-masses $\lesssim 1.3$.  The observing conditions were not always optimal, with mediocre seeing and occasional thick cloud cover.  Nevertheless, by visually monitoring the incoming read outs of the PMAS 4k$\times$4k CCD after each exposure, we ensured that there is sufficient signal in the H$\alpha$ lines that could be used for our analysis.  Observations that did not contain enough signal due to cloud cover were repeated.  If the conditions permitted, we took an additional 400s blank sky exposure close to the target.  We ensured that such blank sky exposures were always taken when the redshifted H$\alpha$ line was in spectral proximity to a bright telluric line.  Sky flat exposures were taken at the beginning or end of each night.  On-sky exposures were always flanked by a continuum lamp exposure and an arc-lamp exposure (HgNe).

Due to strong ambient temperature fluctuations the spectrograph was not always optimally focused. Thus, especially during the 2016 run, sometimes lower values of $R$ are measured (see Table~\ref{tab:log} -- the resolving power determination and correction is described in Appendix~\ref{sec:deta-descr-kinem}).  Fortunately, the galaxy's H$\alpha$ profiles are always resolved significantly, also in the slightly sub-optimal $R\sim3000$ datasets.

We reduced the observational raw data with the general reduction pipeline for fibre-fed integral-field spectrographs \texttt{p3d} \citep{Sandin2010,Sandin2012}.  Our reduction strategy is detailed in Sect.~3 of \citetalias{Herenz2016}, and for completeness we provide a summary in Appendix~\ref{sec:pmas-data-reduction}.  The fully reduced PMAS cuboids for eLARS are made availble via the Strasbourg astronomical Data Center  as supplemental data to this publication\footnote{\textcolor{blue}{URL / reference here, when the datacubes are uploaded.}}.

\subsection{The original LARS PMAS data}
\label{sec:original-lars-pmas}

In this study we combine our new observational results with the results from our study of H$\alpha$ kinematics in the original LARS sample \citepalias{Herenz2016}.  The LARS datacubes are available via the CDS\footnote{\url{https://doi.org/10.26093/cds/vizier.35870078} \citep{2015yCat..35870078H}}.  In the following we reanalyse these LARS datacubes with the methods used for the eLARS observations presented in Sect.~\ref{sec:line-sight-velocity} and described in detail in Appendix~\ref{sec:deta-descr-kinem}.  While there are slight quantitative differences in the recovered velocity fields and velocity dispersion maps (see Sect.~\ref{sec:comp-glob-kinem}), their visual appearance is not significantly different from what was presented in \citetalias{Herenz2016}.  Thus, we will streamline the presentation here by not presenting those maps again; the interested reader is referred to Figs.~5--7 in \citetalias{Herenz2016}.

\subsection{Global Measurements from HST Imaging}
\label{sec:glob-meas-from}

For relating the ionised gas kinematics with the galaxies physical parameters (stellar mass $M_\star$, star-formation rate $\mathrm{SFR}_\mathrm{H\alpha}$, starburst age) and their Ly$\alpha$ observables (equivalent width $\mathrm{EW}_\mathrm{Ly\alpha}$, escape fraction $f_\mathrm{esc}^\mathrm{Ly\alpha}$, and observed Ly$\alpha$ luminosity, $L_\mathrm{Ly\alpha}$) we rely on the integrated measurements provided in \citetalias{Melinder2023}.  As described there in detail, the required global quantities were measured within circular apertures on Voronoi-binned maps that were created from UV and optical PSF-matched HST imaging data that was processed with an updated version LaXs pipeline \citep{Hayes2009}.  The radii of these apertures, denoted $r_\mathrm{tot}$ in \citetalias{Melinder2023}, are defined on the Ly$\alpha$ maps, as Ly$\alpha$ is often more extended then H$\alpha$ or the FUV continuum.  The $\mathrm{EW}_\mathrm{Ly\alpha}$ is defined as the ratio between the aperture integrated Ly$\alpha$ luminosity and the UV continuum luminosity at $\lambda_\mathrm{Ly\alpha}$, and $f_\mathrm{esc}^\mathrm{Ly\alpha}$ is defined as
\begin{equation}
  \label{eq:fesc}
  f_\mathrm{esc}^\mathrm{Ly\alpha} = \frac{L_\mathrm{Ly\alpha} / L_\mathrm{H\alpha}}{(L_\mathrm{Ly\alpha} / L_\mathrm{H\alpha})_\mathrm{int}  \cdot 10^{0.4 \cdot k(\lambda_\mathrm{H\alpha}) \cdot E(B-V)_n}} \;\text{,}
\end{equation}
where $k(\lambda_\mathrm{H\alpha})$ is the extinction coefficient at the wavelength of H$\alpha$, and $E(B-V)$ is the aperture integrated colour excess that attenuates nebular emission and is derived from the Balmer-decrement,
\begin{equation}
  \label{eq:1}
  E(B-V) = \frac{2.5}{k(\lambda_{\mathrm{H\beta}}) -
    k(\lambda_\mathrm{H\alpha})} \cdot \log_{10}
  \frac{L_\mathrm{H\alpha} /  L_\mathrm{H\beta}}{(L_\mathrm{H\alpha} / L_\mathrm{H\beta})_\mathrm{int}} \;\text{,}
\end{equation}
using the integrated luminosities derived from the reconstructed H$\alpha$ and
H$\beta$ images.  The intrinsic ratios of Ly$\alpha$ / H$\alpha$ and H$\alpha$ /
H$\beta$ luminosities are given by the canonical Case-B recombination values
$(L_\mathrm{Ly\alpha} / L_\mathrm{H\alpha})_\mathrm{int} = 8.7$ (see, e.g.,
Fig.~4.2 in \citealt{Hayes2015}) and
$(L_\mathrm{H\alpha} / L_\mathrm{H\beta})_\mathrm{int} = 2.86$, respectively.
Moreover, \citetalias{Melinder2023} use the \cite{Cardelli1989} extinction curve
for $k(\lambda)$ \footnote{At the relevant wavelengths in this work we have $k(\lambda_\mathrm{H\alpha}) = 2.535$, $k(\lambda_\mathrm{H\beta}) = 3.793$, and $k(\lambda_\mathrm{Ly\alpha}) = 10.862$.}
in Eq.~\eqref{eq:fesc} and Eq.~\eqref{eq:1}.  Nebular
extinction is thus assumed to be caused by a homogeneous dust screen in front of
the line emitting gas.

We note that no significant Ly$\alpha$ emission could be detected for six galaxies in the sample (LARS 6, LARS 10, LARS13, ELARS 12, ELARS 14, and ELARS 16).  Hence, these galaxies only have upper limits in their Ly$\alpha$ observables.  Furthermore, we note that \citetalias{Melinder2023} performed a refined and more accurate re-processing of the HST imaging data also for the 14 galaxies constituting the original LARS sample.  Hence, the global Ly$\alpha$ luminosities, equivalent widths, and escape fractions are different and supersede the values presented in \cite{Hayes2014} and \cite{Runnholm2020}.  Adopting the a cut of $\mathrm{EW}_\mathrm{Ly\alpha} \geq 20$\,\AA{} to define a galaxy as LAE, the sample contains 12 LAEs.

\section{Analysis}
\label{sec:analysis-results}

\subsection{Line-of-sight velocity and velocity dispersion maps}
\label{sec:line-sight-velocity}

\begin{table*}
  \caption{Overview of the global kinematical parameters derived from the PMAS observations of the LARS+eLARS galaxies.}
  \centering
  {\renewcommand{\arraystretch}{1.3}
  \begin{tabular}{lccrccc}
\hline \hline \noalign{\smallskip}
ID & kin. & $\cos(i)$ & $v_\mathrm{shear}$ & $\sigma_0^\mathrm{obs}$ & $v_\mathrm{shear} / \sigma_0^\mathrm{obs}$ & $\sigma_\mathrm{tot}$ \\
{} & class & {} &  \multicolumn{1}{c}{[km\,s$^{-1}$]} & [km\,s$^{-1}$] & {} & [km\,s$^{-1}$]\\ \hline \noalign{\smallskip}
L01 & PR &  \dots  & $ 52.4^{+1.6}_{-2.8} $ & $ 47.2 \pm 0.6 $ &$ 1.11^{+0.04}_{-0.06} $  &  60.5  \\
L02 & CK &  \dots  & $ 15.4^{+8.0}_{-2.9} $ & $ 38.2 \pm 0.7 $ &$ 0.40^{+0.21}_{-0.08} $  &  46.0  \\
L03 & PR &  \dots  & $ 130.2^{+5.4}_{-6.7} $ & $ 77.3 \pm 2.9 $ &$ 1.68^{+0.10}_{-0.10} $  &  130.9  \\
L04 & CK &  \dots  & $ 68.2^{+2.9}_{-5.7} $ & $ 43.0 \pm 0.3 $ &$ 1.58^{+0.07}_{-0.13} $  &  55.1  \\
L05 & PR &  \dots  & $ 30.6^{+5.8}_{-1.2} $ & $ 51.2 \pm 0.5 $ &$ 0.60^{+0.11}_{-0.02} $  &  54.7  \\
L06 & CK &  \dots  & $ 50.6^{+1.9}_{-1.0} $ & $ 26.2 \pm 0.5 $ &$ 1.94^{+0.08}_{-0.05} $  &  58.9  \\
L07 & PR &  \dots  & $ 26.0^{+4.9}_{-2.7} $ & $ 60.6 \pm 0.8 $ &$ 0.43^{+0.08}_{-0.04} $  &  67.4  \\
L08 & RD &  0.63 & $ 148.9^{+4.4}_{-3.2} $ & $ 41.2 \pm 0.4 $ &$ 3.61^{+0.11}_{-0.09} $  &  111.8  \\
L09$^\dagger$ & CK &  \dots  & $ 159.6^{+3.0}_{-8.5} $ & $ 56.1 \pm 0.4 $ &$ 2.84^{+0.06}_{-0.15} $  &  112.7  \\
L10 & PR &  \dots  & $ 32.4^{+5.5}_{-6.0} $ & $ 36.5 \pm 0.6 $ &$ 0.89^{+0.15}_{-0.16} $  &  49.2  \\
L11 & RD &  0.00 & $ 149.1^{+11.2}_{-6.1} $ & $ 60.9 \pm 1.4 $ &$ 2.45^{+0.19}_{-0.12} $  &  114.8  \\
L12 & CK &  \dots  & $ 81.3^{+18.7}_{-3.3} $ & $ 64.8 \pm 1.1 $ &$ 1.26^{+0.29}_{-0.06} $  &  81.3  \\
L13$^\dagger$ & CK &  \dots  & $ 168.5^{+18.7}_{-3.3} $ & $ 61.0 \pm 0.9 $ &$ 2.77^{+0.31}_{-0.07} $  &  97.3  \\
L14 & CK &  \dots  & $ 38.8^{+2.3}_{-0.1} $ & $ 59.0 \pm 1.2 $ &$ 0.66^{+0.04}_{-0.02} $  &  72.9  \\
E01 & CK &  \dots  & $ 43.0^{+2.1}_{-2.0} $ & $ 55.6 \pm 0.3 $ &$ 0.77^{+0.04}_{-0.04} $  &  58.7  \\
E02 & PR &  \dots  & $ 17.7^{+2.1}_{-0.3} $ & $ 31.2 \pm 0.4 $ &$ 0.57^{+0.07}_{-0.01} $  &  45.2  \\
E03 & PR &  \dots  & $ 181.8^{+1.1}_{-2.8} $ & $ 40.7 \pm 0.2 $ &$ 4.46^{+0.04}_{-0.07} $  &  129.0  \\
E04 & RD &  0.91 & $ 60.7^{+2.4}_{-1.7} $ & $ 43.7 \pm 0.4 $ &$ 1.39^{+0.06}_{-0.04} $  &  66.7  \\
E05 & RD &  \dots  & $ 145.9^{+1.9}_{-1.3} $ & $ 37.7 \pm 0.3 $ &$ 3.88^{+0.06}_{-0.05} $  &  121.2  \\
E06 & RD &  0.74 & $ 61.3^{+1.2}_{-1.7} $ & $ 28.7 \pm 0.2 $ &$ 2.14^{+0.05}_{-0.06} $  &  59.5  \\
E07$^\dagger$ & CK &  \dots  & $ 93.1^{+1.6}_{-19.3} $ & $ 30.6 \pm 0.5 $ &$ 3.03^{+0.07}_{-0.63} $  &  71.7  \\
E08 & RD &  0.83 & $ 88.0^{+1.6}_{-3.3} $ & $ 25.4 \pm 0.3 $ &$ 3.46^{+0.08}_{-0.14} $  &  64.3  \\
E09 & PR &  \dots  & $ 31.9^{+6.2}_{-4.4} $ & $ 32.8 \pm 0.7 $ &$ 0.97^{+0.19}_{-0.14} $  &  47.4  \\
E10 & RD &  0.39 & $ 121.6^{+2.2}_{-2.2} $ & $ 43.7 \pm 0.4 $ &$ 2.78^{+0.06}_{-0.06} $  &  94.4  \\
E11 & RD &  0.85 & $ 77.4^{+1.3}_{-2.1} $ & $ 46.7 \pm 0.4 $ &$ 1.66^{+0.03}_{-0.05} $  &  65.8  \\
E12 & RD &  0.41 & $ 126.4^{+1.7}_{-1.5} $ & $ 35.4 \pm 0.3 $ &$ 3.57^{+0.06}_{-0.05} $  &  93.5  \\
E13 & CK &  \dots  & $ 21.7^{+3.3}_{-2.2} $ & $ 61.7 \pm 1.3 $ &$ 0.35^{+0.05}_{-0.04} $  &  55.7  \\
E14 & RD &  0.91 & $ 55.2^{+1.5}_{-2.5} $ & $ 31.9 \pm 0.7 $ &$ 1.73^{+0.06}_{-0.09} $  &  54.1  \\
E15 & PR &  \dots  & $ 50.4^{+4.5}_{-3.2} $ & $ 31.1 \pm 0.4 $ &$ 1.62^{+0.15}_{-0.11} $  &  47.4  \\
E16 & RD &  0.39 & $ 76.9^{+4.3}_{-1.6} $ & $ 23.9 \pm 0.4 $ &$ 3.22^{+0.19}_{-0.09} $  &  59.7  \\
E17 & RD &  0.64 & $ 85.3^{+1.8}_{-5.0} $ & $ 24.7 \pm 0.6 $ &$ 3.43^{+0.12}_{-0.22} $  &  72.8  \\
E18 & RD &  0.08 & $ 74.9^{+2.2}_{-3.5} $ & $ 26.4 \pm 0.6 $ &$ 2.84^{+0.11}_{-0.14} $  &  63.9  \\
E19 & RD &  0.78 & $ 52.0^{+6.4}_{-1.6} $ & $ 28.2 \pm 0.6 $ &$ 1.85^{+0.23}_{-0.07} $  &  50.4  \\
E20 & PR &  \dots  & $ 37.1^{+1.2}_{-1.8} $ & $ 36.9 \pm 0.5 $ &$ 1.00^{+0.04}_{-0.05} $  &  45.4  \\
E21 & RD &  0.37 & $ 50.6^{+7.1}_{-3.6} $ & $ 26.6 \pm 0.8 $ &$ 1.90^{+0.27}_{-0.15} $  &  57.7  \\
E22 & CK &  \dots  & $ 38.6^{+18.3}_{-3.0} $ & $ 39.5 \pm 1.3 $ &$ 0.98^{+0.47}_{-0.08} $  &  55.5  \\
E23 & PR &  \dots  & $ 31.6^{+6.2}_{-2.7} $ & $ 23.9 \pm 0.7 $ &$ 1.32^{+0.26}_{-0.12} $  &  49.6  \\
E24$^\dagger$ & RD &  \dots  & $ 95.0^{+8.2}_{-2.8} $ & $ 78.0 \pm 2.9 $ &$ 1.23^{+0.12}_{-0.06} $  &  94.8  \\
E25 & PR &  \dots  & $ 48.9^{+3.1}_{-5.7} $ & $ 34.1 \pm 0.5 $ &$ 1.43^{+0.09}_{-0.17} $  &  45.0  \\
E26 & RD &  \dots  & $ 89.0^{+0.8}_{-2.3} $ & $ 30.9 \pm 0.7 $ &$ 2.86^{+0.08}_{-0.10} $  &  65.6  \\
E27 & RD &  0.95 & $ 48.9^{+0.7}_{-1.8} $ & $ 26.6 \pm 0.6 $ &$ 1.83^{+0.05}_{-0.08} $  &  58.9  \\
E28 & PR &  \dots  & $ 14.7^{+0.9}_{-2.4} $ & $ 25.8 \pm 0.5 $ &$ 0.57^{+0.04}_{-0.09} $  &  33.5  \\
\hline \hline
\end{tabular}}
  \label{tab:kcgz}
  \tablefoot{Kinematical classification (RD = rotating disk, PR = perturbed rotator, CK = complex kinematics; see Sect.~\ref{sec:meth-class}), kinematic inclinations from GalPak$^\mathrm{3D}$ modelling of RDs ($\cos(i)$; Appendix \ref{sec:galpak3d-modelling}), and kinematical parameters ($v_\mathrm{shear}$, $\sigma_0^\mathrm{obs}$, and $v_\mathrm{shear}/\sigma_0^\mathrm{obs}$; see Sect.~\ref{sec:comp-glob-kinem}). Galaxies marked by a $\dagger$ in the ID column show double component H$\alpha$ profiles in some spaxels.}
\end{table*}

\begin{figure*}
  \centering
  \includegraphics[width=0.487\textwidth]{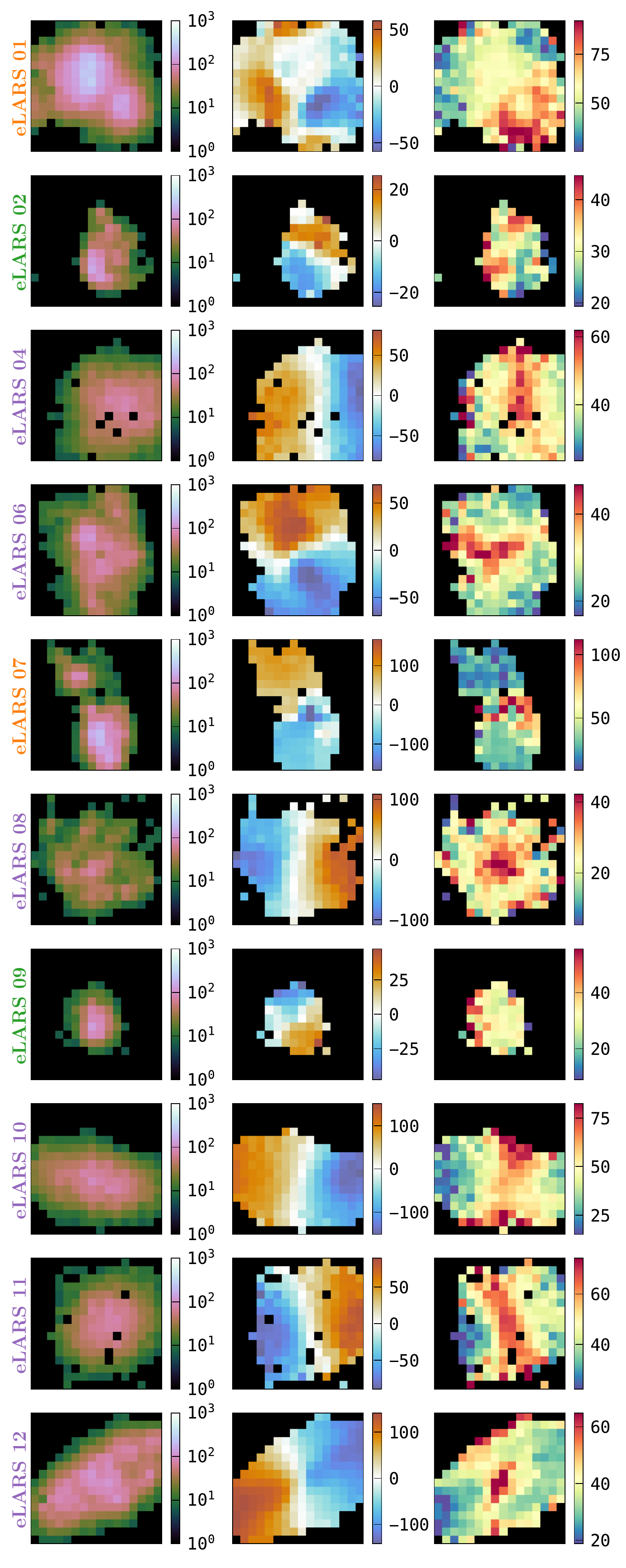}
  \includegraphics[width=0.48\textwidth]{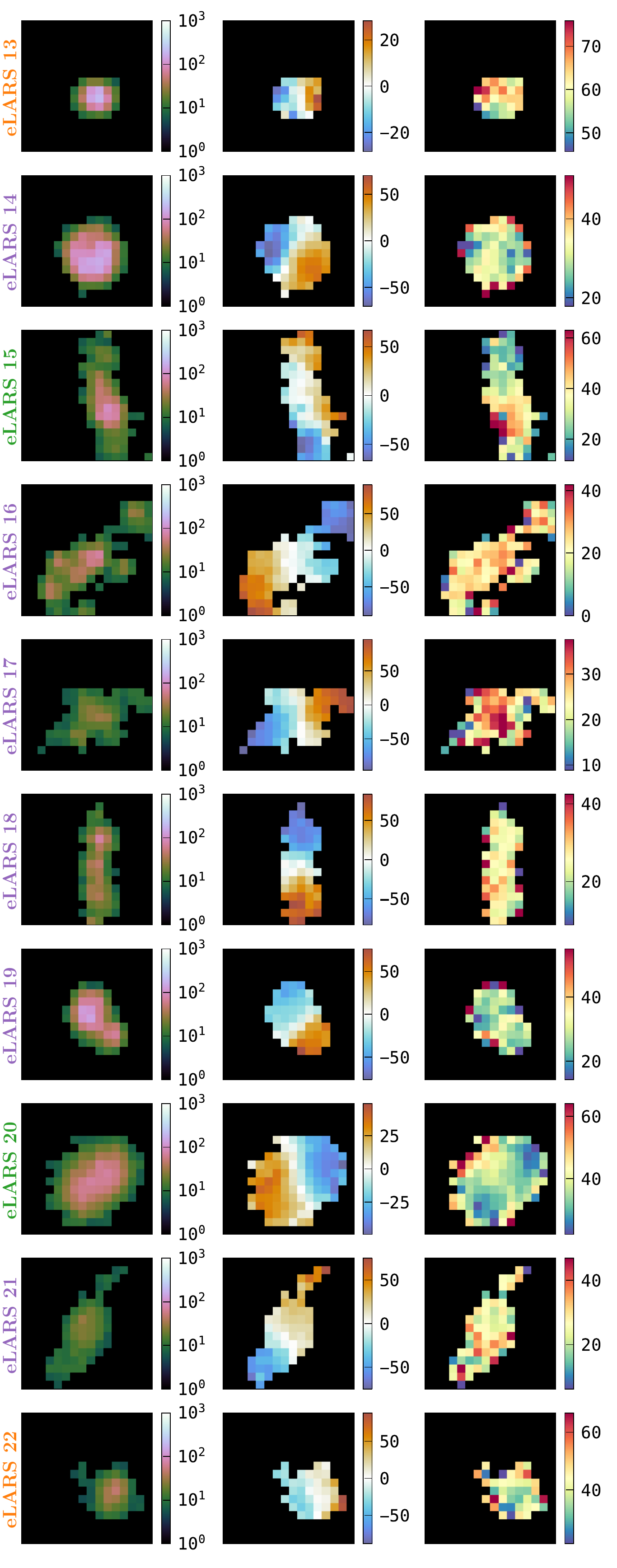}
  \caption{H$\alpha$ kinematics of eLARS from PMAS (16\arcsec{}$\times$16\arcsec{} field of view).  For each galaxy we show H$\alpha$ $S/N$ (\textit{left panels}; colour coded in log-scale from 1 to 1000), $v_\mathrm{los}$ [km\,s$^{-1}$] (\textit{centre panels}; approaching to receding velocities are colour coded linearly from blue to red with the centre at $v_\mathrm{los} = 0$ in white and scaled symmetrically to the absolute maximum), and $\sigma_v$ [km\,s$^{-1}$] (\textit{right panels}; colour coded linearly from the 2th-percentile, median, and 98-th percentile of the observed $\sigma_v$ distribution per galaxy).  The name of each galaxy is coloured according to the visual kinematic classification (Sect.~\ref{sec:visu-class-veloc}): rotating disks in purple, perturbed rotators in green, and galaxies with complex kinematics in orange.  Galaxies with two PMAS pointings are assembled at the bottom of this two-page figure.}
  \label{fig:1}
\end{figure*}
\begin{figure*}
  \centering
  \includegraphics[width=0.887\textwidth]{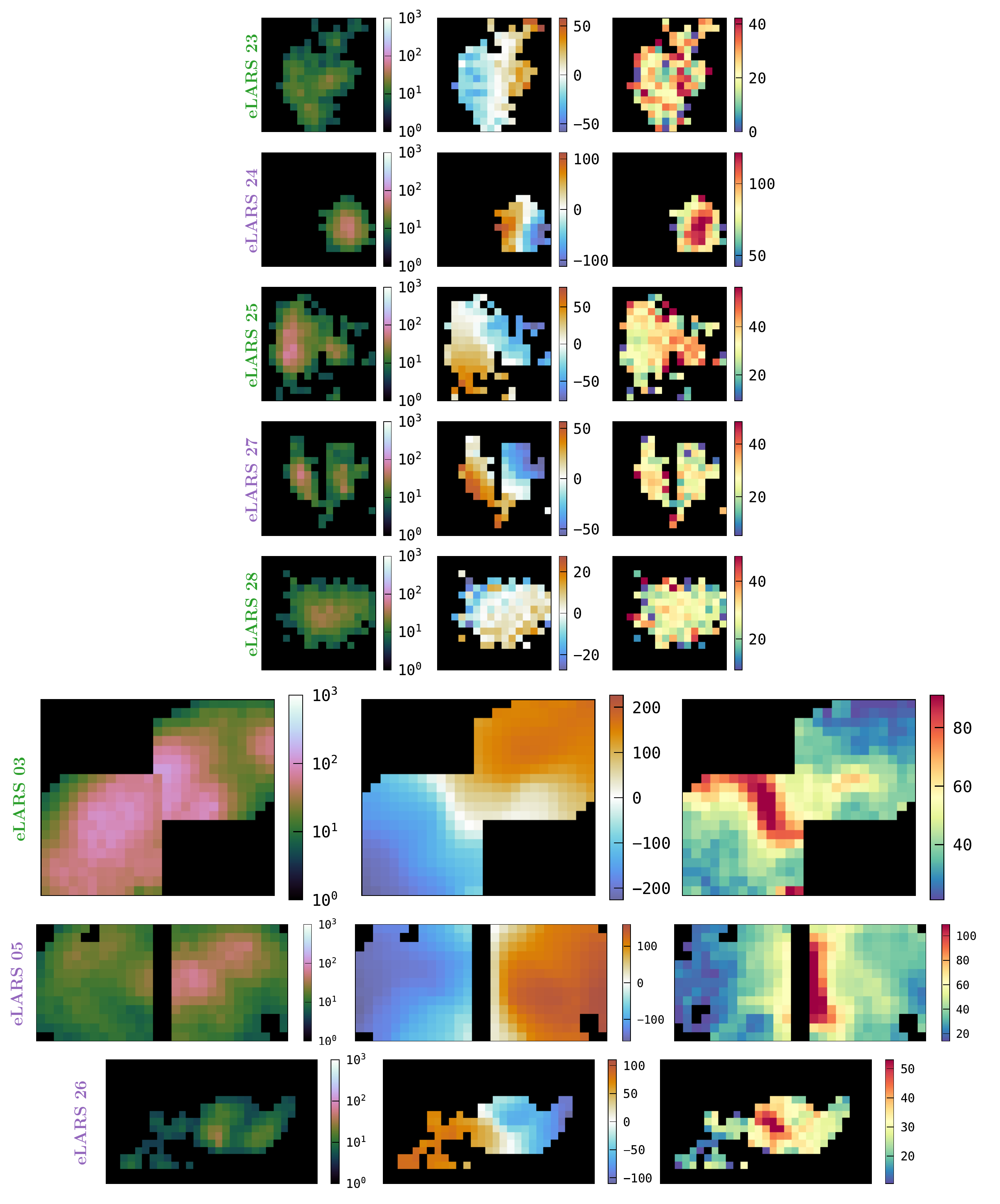}
  \setcounter{figure}{0}  
  \caption{\textit{continued.}}
\end{figure*}

We project the kinematical properties traced by H$\alpha$ in the IFS cuboids into 2D maps of line-of-sight velocity $v_\mathrm{los}$ and velocity dispersion $\sigma_v$.  Our procedures for creating the $S/N$, $v_\mathrm{los}$, and $\sigma_v$ maps closely follow the well established convention for this task that involves fitting single-component Gaussian profiles to the continuum subtracted data \citep[e.g.,][]{Alonso-Herrero2009}.  We describe the creation of the maps in more detail in Appendix~\ref{sec:deta-descr-kinem}.

We here note that only for the galaxies eLARS 7 and eLARS 24 numerous spaxels are not well described by a single Gaussian component.  The spectral profile in those spaxels show strong bimodality indicating the projection of two distinct kinematic components along the line of sight\footnote{The morphology of eLARS 7 indicates that it is a merging system (see Figure~4 in \citetalias{Melinder2023}).  Here the double component profiles occur in the region where the northern and the southern member of the interacting pair overlap.  The situation is different in eLARS 24, where the morphology appears more disk-like (see Figure~5 in \citetalias{Melinder2023}).  Here the double component profiles appear most pronounced towards south-western part, perhaps indicating an infalling star-forming satellite galaxy.  Furthermore, according to the catalogue of \cite{Tempel2017}, eLARS 24 resides in a group environment where frequent mergers would be expected, and an \ion{H}{I} envelope that encompasses the nearest neighbour 30\arcsec{} towards the south west confirms the merger hypothesis  (Le Reste et al., in prep.).}.  Thus the width and position of the Gaussian fits becomes unreliable parameters.  A similar situation is also encountered in LARS 9 and LARS 13 from the \citetalias{Herenz2016} sample.  All four galaxies with double component profiles are excluded in the following statistical analyses involving velocity dispersion measurements.

We show the $v_\mathrm{los}$ and $\sigma_v$ maps alongside their signal-to-noise maps ($S/N$) in H$\alpha$ for the eLARS galaxies in Figure~\ref{fig:1}.  An atlas that allows for a visual comparison between the HST imaging data with the kinematical maps is presented in Appendix~A.1 of \cite{Schaible2023}.  To the best of our knowledge, kinematic maps of the ionised phase of the galaxies analysed in this study have not been presented elsewhere, with the exception being eLARS 1, for which H$\alpha$ kinematics from scanning Fabry-Perot observations were analysed in \cite{Sardaneta2020}.

\subsection{Visual classification of the velocity fields}
\label{sec:visu-class-veloc}

\subsubsection{Classification Scheme and Results}
\label{sec:meth-class}

We visually characterise the $v_\mathrm{los}$ maps shown in Figure~\ref{fig:1} according to three qualitative kinematical categories \citep[see review by][]{Glazebrook2013}: ``rotating disk'' (RD), ``perturbed rotator'' (PR), and ``complex kinematics'' (CK).

For RD galaxies the bulk of the H$\alpha$ emitting gas needs to be dominated by orbital motion.  The resulting velocity field shows a smooth gradient which steepens in the centre and this gradient is aligned parallel to the major axis of the optical image.  Projection of a galaxy's differential rotation lead to a classical symmetric ``spider-leg'' morphology of observed iso-velocity contours (not shown here; see, e.g., Figure~8 in \citealt{Gnerucci2011} or Figure~6 in \citealt{Epinat2010} for examples).  Importantly, the receding and approaching velocities mirror each other.  Moreover, for RD galaxies the velocity dispersion map often show elevated $\sigma_v$ values near the kinematical centre (the reasons for this are provided in Appendix~\ref{sec:meas-intr-disp}).  The velocity fields of PR galaxies are still dominated by orbital motions, but they show significant deviations from a perfect disk.  Lastly, galaxies characterised by CK exhibit no clear gradient indicative of simple orbital motions.  For CK galaxies the gradient also appears less steep in comparison to RDs or PRs.

This visual classification is certainly subjective as there is a not a well defined boundary between the classes.  Hence, the classification was done by six authors individually (ECH, AS, PL, MH, ALR, and G\"{O}) and then finalised in a consolidation session.  Moreover, as described in detail in Appendix~\ref{sec:galpak3d-modelling} the classification of the RDs is confirmed by successfully modelling the RD systems by simple disk models with GalPak$^\mathrm{3D}$ \citep{Bouche2015} and by the symmetric appearance of the line-of-sight velocity profiles along the kinematical major axis (Appendix~\ref{sec:empirical-methods}; Figure~\ref{fig:yu21}).  The classifications for the combined LARS and eLARS sample are listed in Table~\ref{tab:kcgz}.  Moreover, we colour the label of each galaxy in Figure~\ref{fig:1} according to its kinematical class.

According to our visual classification, the eLARS sample contains 4 CK galaxies (14\% of eLARS), 8 PR galaxies (29\% of eLARS), and 16 RD galaxies (57\% of eLARS).  This is different compared to the original LARS sample, where 7 galaxies have complex kinematics (50\% of LARS), 5 are a PR (36\% of LARS), and only 2 are a RD (14\% of LARS).  This difference is a consequence of the sample selection.  Whereas LARS galaxies had to satisfy an H$\alpha$ equivalent width cut of $EW_\mathrm{H\alpha} \geq 100$\AA{}, this requirement was lowered to $EW_\mathrm{H\alpha} \geq 40$\AA{} for eLARS to remove the bias towards irregular and merging systems \citepalias{Melinder2023}, which are characterised by CK.  The combined sample of all 42 eLARS and LARS galaxies consists of 18 (43\%) RD galaxies, 13 (31\%) PR galaxies, while 11 (26\%) galaxies of the sample exhibit CK.

This almost three-way split amongst the kinematical categories bears resemblance to the split found in early IFU studies of star-forming galaxies at similar masses in the high-$z$ ($z \gtrsim 1$) universe \citep[][and references therein]{Glazebrook2013}.  More recent surveys, that use stricter criteria to define rotating disks, report disk fractions in excess of 50~\% \citep{Wisnioski2015,Wisnioski2019}, but that could be an overestimate due to mergers acting as disk impostors \citep{Simons2019}.

\subsubsection{Comparing Ly$\alpha$ observables between kinematical classes}
\label{sec:comp-lyalpha-observ}

\begin{figure*}
  \includegraphics[width=\textwidth]{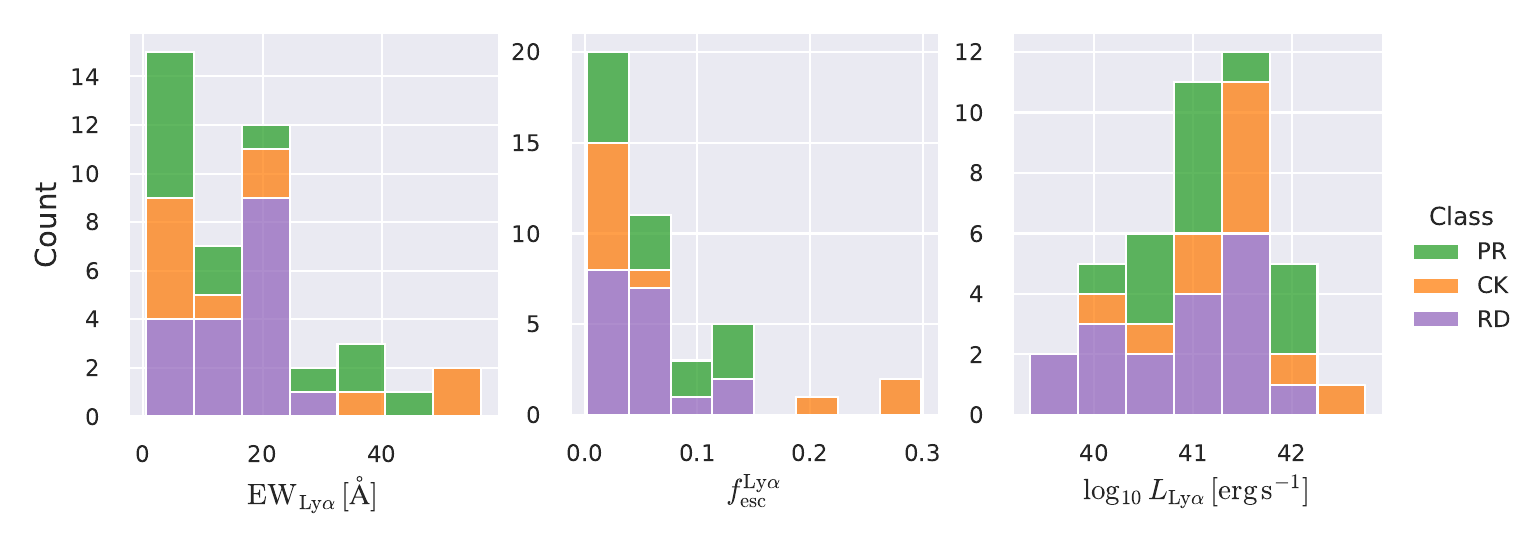} \vspace{-0.7cm}
  \caption{Stacked histograms, colour-coded by kinematical class (rotating disks: purple; perturbed rotators: green; complex kinematics: orange), of the Ly$\alpha$ observables $\mathrm{EW}_\mathrm{Ly\alpha}$ (\emph{left panel}), $f_\mathrm{esc}^\mathrm{Ly\alpha}$ (\emph{middle panel}), and $L_\mathrm{Ly\alpha}$ (\emph{right panel}).}
  \label{fig:hists}
\end{figure*}

We may imagine that interstellar medium conditions in galaxies with complex velocity fields are favourable for Ly$\alpha$ escape.  For example, irregular and merging systems are often characterised by high specific SFRs, which lead to spatially concentrated injection of momentum and energy from stellar feedback processes.  The resulting winds and outflows then may lead to a more effective clearing of escape channels for Ly$\alpha$ (and also Lyman continuum) radiation.  The 42 galaxies of the combined LARS and eLARS sample allow for a statistical exploration of such a scenario.

We begin by showing stacked histograms of the Ly$\alpha$ observables colour-coded by their kinematical class in our sample in Figure~\ref{fig:hists}.   The maximum $\mathrm{EW}_\mathrm{Ly\alpha}$ (56.7\,\AA; LARS 2), maximum $f_\mathrm{esc}^\mathrm{Ly\alpha}$ (0.3; LARS 2), and maximum $L_\mathrm{Ly\alpha}$ ($5.6\times10^{42}$\,erg\,s$^{-1}$; LARS 14) values are found indeed among the CK objects.  Moreover, also the averages of those observables are larger in the CK class ($\langle \mathrm{EW}_\mathrm{Ly\alpha} \rangle = 23.15$\,\AA: $\langle f_\mathrm{esc}^\mathrm{Ly\alpha} \rangle = 0.102$; $\langle L_\mathrm{Ly\alpha} \rangle = 1.02 \times 10^{42}$\,erg\,s$^{-1}$) than the averages found for the PR ($\langle \mathrm{EW}_\mathrm{Ly\alpha} \rangle = 17.47$\,\AA; $\langle f_\mathrm{esc}^\mathrm{Ly\alpha} \rangle = 0.067$; $\langle L_\mathrm{Ly\alpha} \rangle = 2.75 \times 10^{41}$\,erg\,s$^{-1}$) and for the RD ($\langle \mathrm{EW}_\mathrm{Ly\alpha} \rangle = 16.36$\,\AA; $\langle f_\mathrm{esc}^\mathrm{Ly\alpha} \rangle = 0.057$; $\langle L_\mathrm{Ly\alpha} \rangle = 3.10 \times 10^{41}$\,erg\,s$^{-1}$) galaxies, respectively.  However, also non-detections of Ly$\alpha$ emission are found amongst objects in all three classes.

In order to quantify the difference of the observed distributions of Ly$\alpha$ observables amongst the different classes we use the two-sample Kolmogorov-Smirnov (KS) test.  Here the six galaxies with upper limits in their Ly$\alpha$ observables are excluded from calculating the KS-test statistic.  As mentioned in Sect.~\ref{sec:meth-class}, the boundary between our three classes is not well defined, and some PRs could also be classified either as CKs or RDs.  For the KS-test we thus regroup our sample into two groups, $\mathcal{A}$ and $\mathcal{B}$.  First, we combine the RDs and PRs into group $\mathcal{A}$ (27 galaxies) and the CKs into group $\mathcal{B}$ (9 galaxies), then we associate only the RDs with group $\mathcal{A}$ (15 galaxies) and combine the PRs with the CKs in group $\mathcal{B}$ (21 galaxies).  For both parings and for all three Ly$\alpha$ observables the KS-test is consistent with the null-hypothesis that the samples in the kinematic class are drawn from the same parent population (p-values in the range $p_\mathrm{KS} \sim0.2\dots0.3$).  This exercise shows that there is significant overlap in the Ly$\alpha$ observables amongst the kinematical classes.
The qualitative nature of a galaxy's line-of-sight ionised gas velocity field in our sample does not reflect whether it is as a strong Ly$\alpha$ emitter or not.

\subsection{Global kinematical measures: $v_\mathrm{shear}$,
  $\sigma_0^\mathrm{obs}$, and $\sigma_\mathrm{tot}$}
\label{sec:comp-glob-kinem}

\begin{figure*}
  \includegraphics[width=\textwidth]{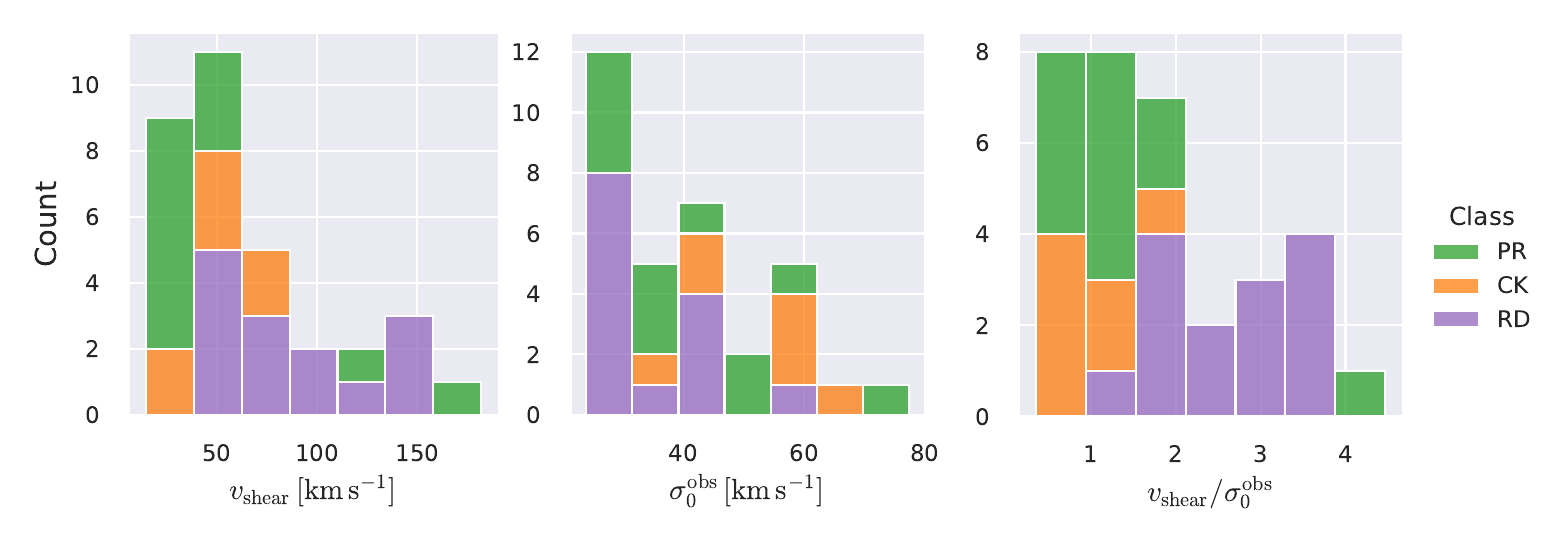} \vspace{-0.7cm}
  \caption{Stacked histograms of the global kinematical measures $v_\mathrm{shear}$ (\emph{left panel}) , $\sigma_0^\mathrm{obs}$ (\emph{middle panel}), and $v_\mathrm{shear} / \sigma_0^\mathrm{obs}$ (\emph{right panel}), colour coded by kinematical class as in Figure~\ref{fig:hists}.}
  \label{fig:hists_kin}
\end{figure*}

We now turn to a more quantitative analysis of the observed line-of-sight velocity- and velocity dispersion maps.  Therefore we derive two main summarising measures from those maps: the apparent shearing velocity, $v_\mathrm{shear}$, and the intrinsic velocity dispersion, $\sigma_0$ \citep[see review by][]{Glazebrook2013}.

For the calculation of $v_\mathrm{shear}$ we account for extreme outliers from the distribution of observed velocities.  To this aim we calculate $v_\mathrm{shear} = \frac{1}{2} ( v_\mathrm{max} - v_\mathrm{min} )$ for each galaxy, with $v_\mathrm{max}$ and $v_\mathrm{min}$ being the values of the 95th percentile and the 5th percentile of the $v_\mathrm{los}$ values, respectively.  The upper- and lower errors on $v_\mathrm{los}$ are estimated from the difference of the 95th and 5th percentiles, respectively, to the mean of the values outside of each percentile.

For $\sigma_0$ we adopt the notion of a galaxy's average line-of-sight velocity dispersion.  This simplistic definition implicitly assumes that the $\sigma_0$ is isotropic.  For measuring this quantity from the observed dispersion fields we need to account for inflated values in individual spaxels due to the convolution of the galaxies' intrinsic H$\alpha$ spatial- and spectral- morphology with the atmospheres point spread function and the coarse sampling by 1\arcsec{}$\times$1\arcsec{} spaxels.  In \citetalias{Herenz2016} we made no such correction, but the large fraction of RD galaxies in the combined LARS+eLARS sample demands us to be more cautious in this respect.  Various methods that attempt to correct for this ``beam smearing'' or, more correctly, ``point-spread function smearing'' effect have been presented in the literature \citep[e.g.,][]{Davies2011,Varidel2016}.  We compare several such methods against results from morphokinematical modelling of our observations with the GalPaK$^\mathrm{3D}$ software \citep{Bouche2015} in Appendix~\ref{sec:meas-intr-disp}.  This modelling accounts for the spectral and spatial convolution of the observed H$\alpha$ kinematics and it adopts also the notion that $\sigma_0$ is isotropic.  However, GalPak$^\mathrm{3D}$ can only be applied to rotating systems.  Given the large fraction of more complex systems in our sample we thus require an empirical method for estimating $\sigma_0$ directly from the data.  In Appendix~\ref{sec:meas-intr-disp} we find that the model-based velocity dispersions can be reliably recovered by averaging the $\sigma_v$ maps, provided that certain spaxels are masked according to a local gradient-measure in the $v_\mathrm{los}$ map.  The uncertainty on $\sigma_{0}^\mathrm{obs}$ is calculated by propagating the $\Delta \sigma_0$'s of the individual spaxels (Sect.~\ref{sec:line-sight-velocity}) in quadrature.

The so determined $v_\mathrm{shear}$ and $\sigma_0^\mathrm{obs}$ values are listed in Table~\ref{tab:kcgz}.  We comment that there is a small difference between the $v_\mathrm{shear}$ values reported for the LARS galaxies in \citetalias{Herenz2016} and the values listed here.  This difference is because \citetalias{Herenz2016} applied a Voronoi-binning technique to the datacube, whereas here we use each spaxel as is.  Moreover, we here used the intervals outside the 95th and 5th percentiles individually to estimate the uncertainty, whereas \citetalias{Herenz2016} symmetrised the error-bars.  That being so, the here reported updated $v_\mathrm{shear}$ measurements and error estimates agree within the error-bars of \citetalias{Herenz2016}.  In contrast, the here tabulated $\sigma_0^\mathrm{obs}$ values were computed more thoroughly as explained above (see also Appendix~\ref{sec:meas-intr-disp}) and are thus slightly different.

Furthermore, we list in Table~\ref{tab:kcgz} the integrated velocity dispersion $\sigma_\mathrm{tot}$.  We obtain $\sigma_\mathrm{tot}$ by calculating the square root of the second central moment of the H$\alpha$ profile that is obtained by summing all spaxels with $S/N_\mathrm{H\alpha} \geq 6$.  This measure provides a useful comparison with observations that do not resolve the kinematics on the spatial scales probed here, e.g. \ion{H}{I} observations where the beam-size is close to the FoV of the PMAS observations or measurements of unresolved galaxies at high $z$.  This column in Table~\ref{tab:kcgz} is provided for reference and we do not use this quantity in the analyses presented below.

We find maximum shearing velocities in eLARS 3 ($v_\mathrm{shear} = 181.8^{+1.1}_{-2.8}$\,km\,s$^{-1}$) and LARS 13 ($168.5^{+18.7}_{-3.3}$\,km\,s$^{-1}$), with the former being a rotating disk and the latter being a merger that exhibits complex kinematics.  Minimum shearing velocities are observed in eLARS 28 ($v_\mathrm{shear} = 14.7^{+0.9}_{-2.4}$\,km\,s$^{-1}$) and LARS 2 ($v_\mathrm{shear} = 15.4^{+8.0}_{-2.9}$\,km\,s$^{-1}$), a perturbed rotator and a complex system, respectively.  In Figure~\ref{fig:hists_kin} (left panel) we show a stacked histogram of $v_\mathrm{shear}$ colour-coded by kinematical class.  It can be appreciated, that the distribution of the RDs differs markedly in comparison to the PRs and CKs, with the latter two classes showing overall lower $v_\mathrm{shear}$ values.  Despite the eLARS extension enlarging the fraction of disks compared to LARS, only one new $v_\mathrm{shear}$ extremum is introduced into the combined sample (eLARS 03).  Thus, we now sample the $v_\mathrm{shear}$ range at hand more densely, as reflected also by the mean and the median of $v_\mathrm{shear}$ (57.96\,km\,s$^{-1}$ and 73.13\,km\,s$^{-1}$, respectively) being similar or identical to the respective values of the original LARS sample (52.4\,km\,s$^{-1}$ and 73.13\,km\,s$^{-1}$).  Given that the selection of our sample is limited to young star-bursts, the shearing velocities in our sample are unsurprisingly lower than what is observed for more massive disk galaxies, even when not corrected for inclination.

For the velocity dispersions the maxima are observed in eLARS 24 ($\sigma_0^\mathrm{obs} = 78.0 \pm 2.9$\,km\,s$^{-1}$) and LARS 3 ($\sigma_0^\mathrm{obs} = 77.3 \pm 2.9$\,km\,s$^{-1}$) and the minima are seen in eLARS 16 ($\sigma_0^\mathrm{obs} = 23.9 \pm 0.9$\,km\,s$^{-1}$) and eLARS 23 ($\sigma_0^\mathrm{obs} = 23.9 \pm 0.7$\,km\,s$^{-1}$).  We show a stacked histogram of $\sigma_0^\mathrm{obs}$, colour coded by kinematical class, in Figure~\ref{fig:hists_kin} (middle panel).  There appears not much difference regarding the distributions of $\sigma_0^\mathrm{obs}$ in the different kinematical classes, except that most of the lowest velocity dispersions are found amongst the RDs and PRs.  The eLARS extension introduces a larger fraction of lower velocity dispersion ($\sigma_0^\mathrm{obs} \sim 35$\,km\,s$^{-1}$) galaxies into the sample; this is apparent from Table~\ref{tab:kcgz}, and can also be seen by comparing mean and median $\sigma_0^\mathrm{obs}$ of the full sample ($41.1$\,km\,s$^{-1}$ and $37.3$\,km\,s$^{-1}$) to the original sample ($55.6$\,km\,s$^{-1}$ and $51.9$\,km\,s$^{-1}$).  The on average lower $\sigma_0^\mathrm{obs}$ values are again a consequence of the lower $\mathrm{EW}_\mathrm{H\alpha}$ cut used to define the extended sample (cf. Sect.~\ref{sec:meth-class}).  This is due to the fact that $\mathrm{EW}_\mathrm{H\alpha}$ traces specific SFR (sSFR $=$ SFR per galaxy mass), which tightly correlates with SFR ($\tau =0.51$, $p_\tau \sim 10^{-6}$ for LARS+eLARS; see also analysis by \citealt{Law2022}), and since the latter, as discussed already in the introduction, tightly correlates with SFR.  This is analysed further Sect.~\ref{sec:corr-betw-galaxy} below.

We also use the ratio $v_\mathrm{shear}/\sigma_0^\mathrm{obs}$ that combines the above measurements to a summarising statistic of a galaxy's kinematical state.  These ratios are also listed in Table~\ref{tab:kcgz}.  The asymmetric error-bars on $v_\mathrm{shear}$ were propagated into $v_\mathrm{shear}/\sigma_0^\mathrm{obs}$ after moving into log-space and then using method presented in the appendix of \cite{Laursen2019}.  Our sample is characterised by low $v_\mathrm{shear}/\sigma_0^\mathrm{obs}$ values, ranging from 0.35 to 4.46 (median: 1.6, average: 1.8).  We show the corresponding stacked-histogram with colour coding by kinematical class in the right panel of Figure~\ref{fig:hists_kin}.  The distributions of $v_\mathrm{shear}/\sigma_0^\mathrm{obs}$ differ markedly when comparing the RDs with the CKs and PRs, with the later clearly dominating the low- $v_\mathrm{shear}/\sigma_0^\mathrm{obs}$ range.  Still, the overall low ratios in the whole sample are below that of typical disk galaxies ($v_\mathrm{rot}/\sigma_0 \gtrsim 8$, where $v_\mathrm{rot}$ is the maximum rotational velocity; e.g., \citealt{Epinat2008}) and are more akin to low-mass star-forming galaxies at low- and high-redshifts.  Galaxies with $v_\mathrm{rot}/\sigma_0 < \sqrt{3.36} \approx 1.86$ can be classified as dispersion dominated galaxies, as here the turbulent motions exceed the value that would correspond to equal contributions of random motions and rotation to the dynamical support of a turbulent disc \citep{FoersterSchreiber2020,Bik2022}.  According to this criterion 24 of the 42 galaxies in LARS+eLARS are dispersion dominated; most of these are PRs or CKs.

We remark again, that $v_\mathrm{shear}$ is not inclination corrected, as for the non-rotating complex systems such a correction is often ill-defined.  Thus, strictly speaking, $v_\mathrm{shear}/\sigma_0^\mathrm{obs}$ characterises the projected kinematical state.  However, we argue that this projected kinematical state, rather than the inclination corrected ratio, is the relevant parameter for relating kinematics with Ly$\alpha$ observables.  Especially for disk galaxies the Ly$\alpha$ escape fraction is theoretically expected to be strongly inclination dependent, with edge-on systems having absorbed most of the Ly$\alpha$ photons along the line of sight whereas a face-on system may show higher escape fractions \citep{Laursen2007,Laursen2009a,Behrens2014,Verhamme2012,Smith2022b}.  At comparable masses, this could lead to an anti-correlation between $v_\mathrm{shear}/\sigma_0^\mathrm{obs}$ and $f_\mathrm{esc}^\mathrm{Ly\alpha}$ for disk galaxies, since face-on disks should have preferentially lower $v_\mathrm{shear}/\sigma_0^\mathrm{obs}$.  A deprojected ratio could not trace such types of relations that are the subject of this study.  We analyse the inclination dependence on the Ly$\alpha$ observables further in Sect.~\ref{sec:incl-depend-lyalpha}.


\subsection{Inclination dependence on Ly$\alpha$ escape}
\label{sec:incl-depend-lyalpha}

\begin{figure}
  \centering
  \includegraphics[width=0.4\textwidth]{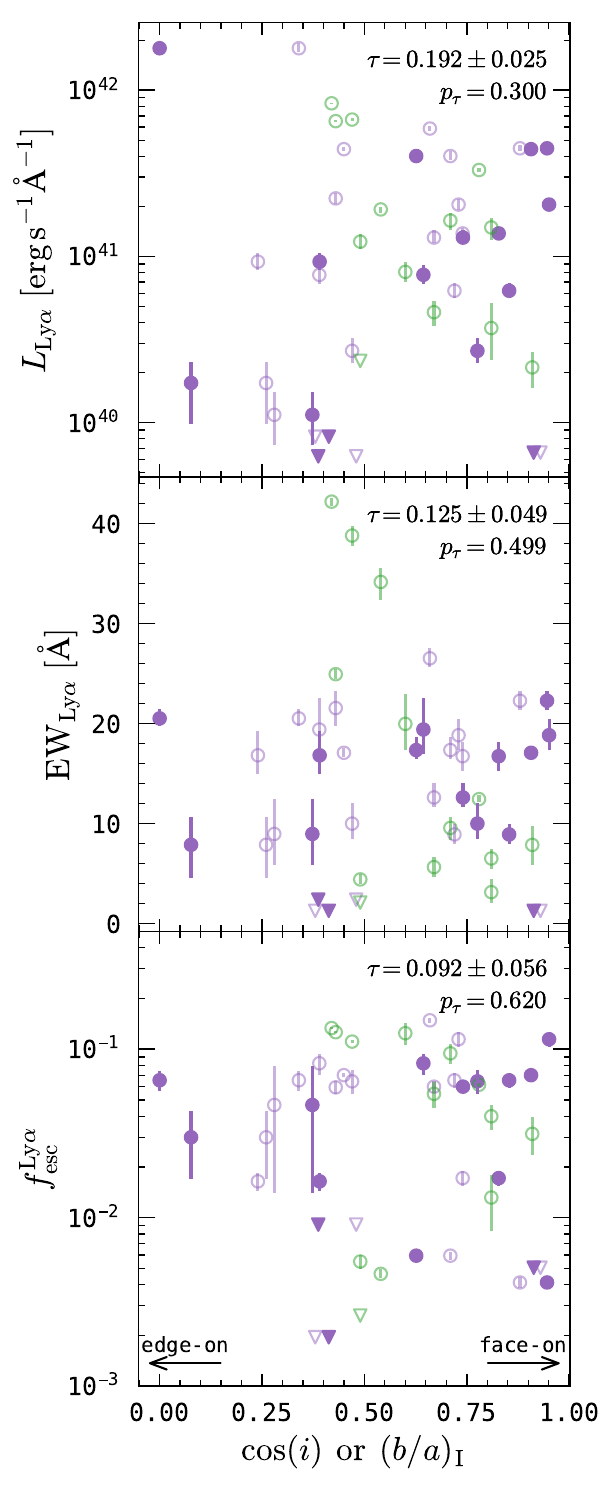}
  \vspace{-1em}
  \caption{Ly$\alpha$ observables (\emph{top panel}: $L_\mathrm{Ly\alpha}$; \emph{middle panel}: $\mathrm{EW}_\mathrm{Ly\alpha}$; \emph{bottom panel}: $f_\mathrm{esc}^\mathrm{Ly\alpha}$ ) vs. galaxy inclination.  Filled purple symbols are based on the kinematical inclination, $i$, from the disk modelling of the RDs with GalPak$^\mathrm{3D}$.  Unfilled symbols are based on photometric inclinations (purple: RD; green: PR) according to $\cos(i) \approx (b/a)_\mathrm{I}$ where $ (b/a)_\mathrm{I}$ is the $I$-band axis ratio from \cite{Rasekh2022}.  Thus, each RD is represented by a filled and an open symbol in these plots.  Circles with error-bars indicate measurements of the Ly$\alpha$ observables, whereas downward pointing triangles indicate upper limits.  Correlation coefficients (Kendall's $\tau$) and corresponding $p$-values for the kinematical inclinations are provided in the top-right corner of each panel.  }
  \label{fig:cosilya}
\end{figure}

To date the most detailed Ly$\alpha$ radiative transfer simulations were carried out for massive disk galaxies \citep{Behrens2014,Verhamme2012,Smith2022b}.  These single galaxy simulations found a strong inclination dependence of $\mathrm{EW}_\mathrm{Ly\alpha}$ and $f_\mathrm{esc}^\mathrm{Ly\alpha}$ \citep[see also][]{Laursen2007}.  The highest $\mathrm{EW}_\mathrm{Ly\alpha}$ and $f_\mathrm{esc}^\mathrm{Ly\alpha}$ are found when the simulated galaxies are seen face-on, whereas Ly$\alpha$ is significantly extinguished when the galaxies are viewed edge-on.  A plausible conjecture from those simulation results is that high-$z$ samples selected on Ly$\alpha$ emission might be biased towards face-on systems.  However, observational results on high-$z$ LAEs, which use the photometric major- to minor axis ratio, $b/a$, or the flattening, $1 - b/a$, as proxy for the inclination (since $\cos(i) = b/a$ for infinitely thin disks), are not supportive of this idea \citep{Gronwall2011,Shibuya2014a,Paulino-Afonso2018}.

In the present contribution we are equipped with kinematical inclinations, $i$, from the GalPak$^\mathrm{3D}$ disk modelling of the 15 RDs that was performed to verify the empirical masking technique for the calculation of $\sigma_0^\mathrm{obs}$ (Appendix~\ref{sec:galpak3d-modelling}).  We list $\cos(i)$ in Table~\ref{tab:kcgz}.  We compare these measurements to the photometric ratio determined in the I-band by \cite{Rasekh2022}, as this quantity minimises the mean difference between photometric axis ratios and kinematic inclinations ($\langle \cos(i) - (b/a) \rangle = 0.07$).  Of course, no perfect match is expected, given that the simple approximation $\cos(i) \approx b/a$ does not account for the finite thickness of the disk.  Thus the photometric inclination is biased low for systems that are observed close to edge-on; LARS 11 and eLARS 18 are such cases in our sample.  Moreover, \cite{Rasekh2022} calculate the axis ratio from the image moments.  However, these measures are sensitive to sub-structures (i.e., clumps and spiral arms) that are not homogeneously distributed throughout the disk.  The moment-based axis ratios can then measure the galaxy as a more elongated structure and thus overestimate the actual inclination; eLARS 4, 17, 19, and 27 are such cases in our sample.  Kinematical modelling provides a more accurate measure of the disk inclination.

The inclination dependence of the Ly$\alpha$ observables in our sample is plotted in Figure~\ref{fig:cosilya}.  It can be seen that there is no preference for larger Ly$\alpha$ observables towards more face-on systems in the LARS+eLARS sample.  This is also confirmed by a correlation analysis\footnote{Throughout we measure correlations with the generalised Kendall-$\tau$ correlation coefficient according to \cite{Isobe1986} as this formalism allows for treating upper limits sensibly in the analysis.  Moreover, following \cite{Curran2014} we use a Monte-Carlo procedure for the calculation of the error on $\tau$.  More details are given in Appendix~\ref{sec:robustn-corr}..} using Kendall's $\tau$ \citep[e.g.,][]{Puka2011}, with $p$-values significantly in excess of 0.05.  Moreover, no trend is observed when we consider the photometric axis ratios as a proxy for the inclination.  Here it needs to be kept in mind that the concept of inclination can not be applied to CK systems, hence CK systems are omitted from the plot.  Our data obtained for the LARS+eLARS sample does not support the idea that LAE samples are biased towards face-on disks.

\subsection{Relations between galaxy parameters, $M_\star$ and
  $\mathrm{SFR}_\mathrm{H\alpha}$, and global kinematical statistics}
\label{sec:corr-betw-galaxy}

\begin{figure*}
  \sidecaption
  \includegraphics[width=12cm]{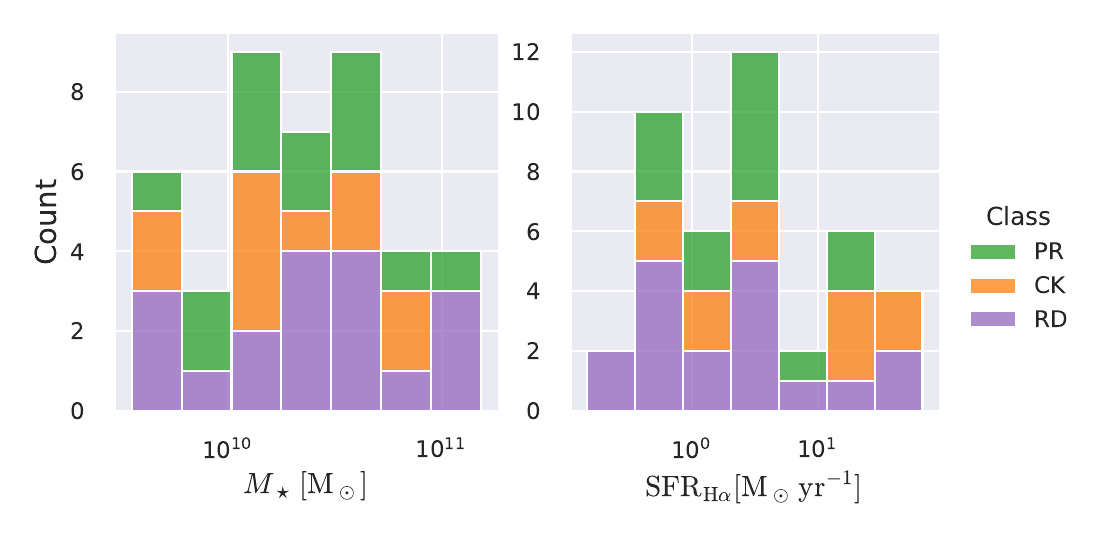} 
  \caption{Stacked histograms, colour-coded by kinematical class (as in Figure~\ref{fig:hists}) of $M_\star$ (\emph{left panel}) and $\mathrm{SFR}_\mathrm{H\alpha}$ (\emph{right panel}) from \citetalias{Melinder2023}.}
  \label{fig:hist_m_sfr}
\end{figure*}

\begin{figure*}
  \centering
  \includegraphics[width=0.4\textwidth]{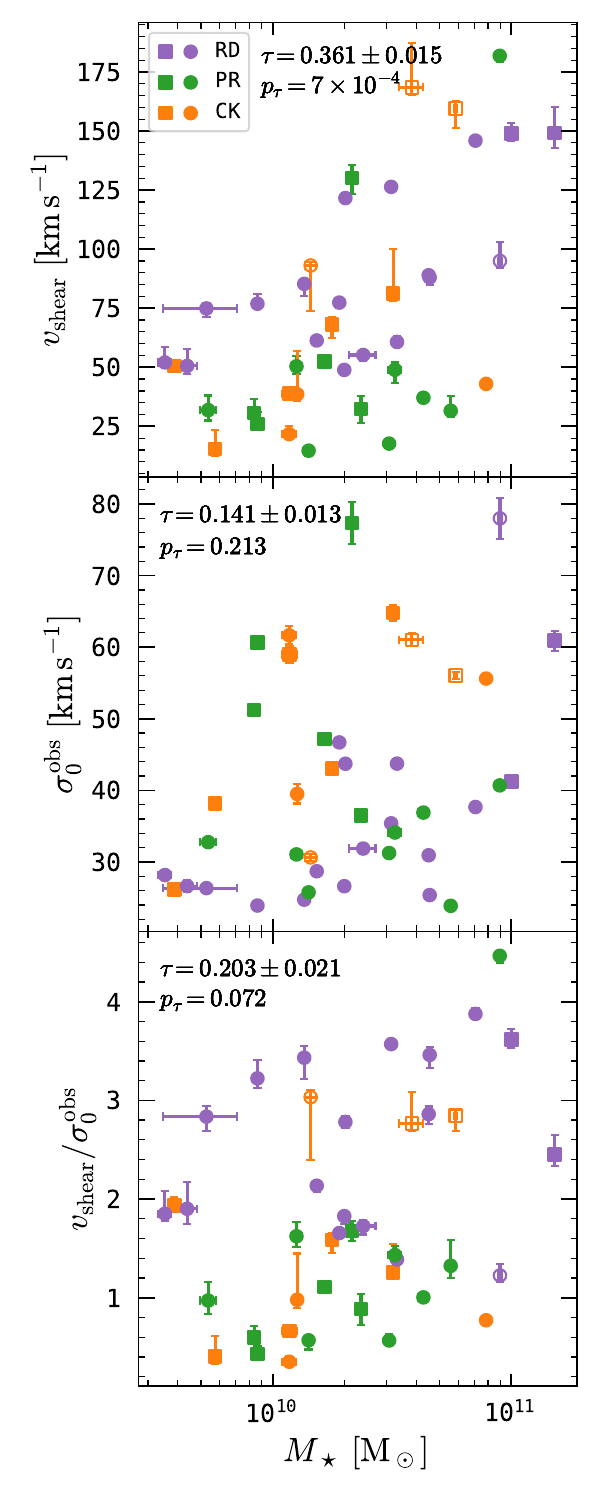} \quad
  \includegraphics[width=0.4\textwidth]{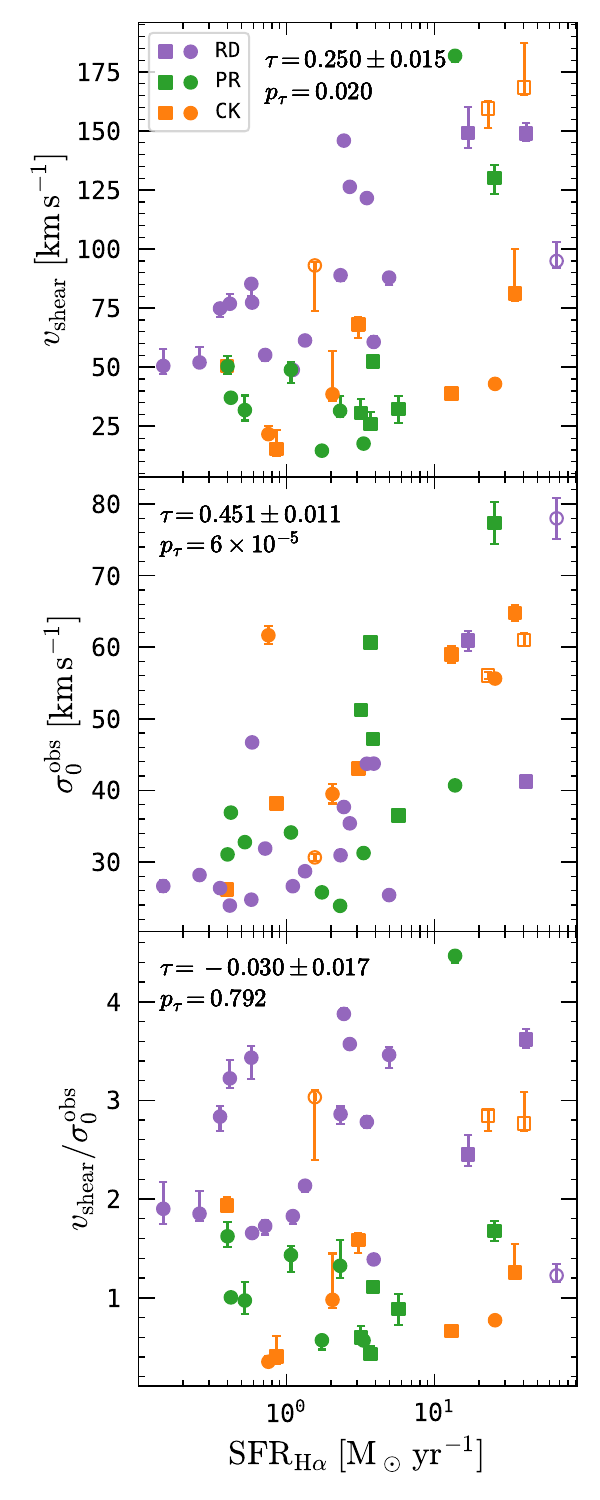}
  \vspace{-1em}
  \caption{Global kinematical measures vs. stellar mass (\emph{left panels}) and star-formation rate (from H$\alpha$; \emph{right panels}) for the LARS (squares) and eLARS galaxies (circles).  The left-hand top-, middle-, and bottom panel show $v_\mathrm{shear}$ vs. $M_\star$, $\sigma$ vs. $M_\star$, and $v_\mathrm{shear}/\sigma_0^\mathrm{obs}$ vs. $M_\star$, respectively.  The right-hand top-, middle-, and bottom panel, show $v_\mathrm{shear}$ vs. $\mathrm{SFR}_\mathrm{H\alpha}$, $\sigma_0^\mathrm{obs}$ vs.  $\mathrm{SFR}_\mathrm{H\alpha}$, and $v_\mathrm{shear} / \sigma_0^\mathrm{obs}$ vs. $\mathrm{SFR}_\mathrm{H\alpha}$.  Rotating disks, perturbed rotators, and systems with complex kineamtics are shown with purple, green-, and orange symbols, respectively.  Systems with double component H$\alpha$ profiles in some spaxels are shown with open symbols; these galaxies are not included in the calculation of Kendal's $\tau$ and $p_\tau$ for relations involving $\sigma_0^\mathrm{obs}$ and $v_\mathrm{shear}/\sigma_\mathrm{obs}$.}
  \label{fig:kin_vs_msfr}
\end{figure*}

It is established that LAEs at high-$z$ are preferentially low-mass ($M_\star \lesssim 10^{10}$\,M$_\odot$) galaxies with star-formation rates on and slightly above the star-forming main sequence \citep[$\mathrm{SFR} \sim 10$\,M$_\odot$\,yr$^{-1}$; e.g.,][]{Rhoads2014,Oyarzun2017,Kusakabe2018,Ouchi2019,Pucha2022,ChavezOrtiz2023}.  Kinematically, the SFR is known to tightly correlate with the intrinsic velocity dispersion \citep[e.g.,][]{Green2010,Green2014,Law2022}.   The rotation velocity, $v_\mathrm{max}$, directly traces mass.  For rotating disks $v_\mathrm{shear} \leq v_\mathrm{max} \cdot \sin(i)$ holds, where $i$ is the inclination and $v_\mathrm{max}$ is the maximum value of the rotation curve; equality holds when the sampled velocity field reaches into the flat part of the rotation curve.  Thus, before establishing and interpreting any relation between ionised gas kinematics and Ly$\alpha$ observables, we first need to investigate whether such correlations between physical galaxy parameters and galaxy kinematics exist in LARS+eLARS.

We begin by showing in Figure~\ref{fig:hist_m_sfr} histograms of the distribution of $M_\star$ and $\mathrm{SFR}_\mathrm{H\alpha}$.  These global quantities are measured from SED fitting to the HST images of the galaxies and the relevant details can be found in \citetalias{Melinder2023}.  The sample spans a $M_\star$ ($\mathrm{SFR}_\mathrm{H\alpha}$) range from $3.5\times10^9$\,M$_\odot$ (0.15\,M$_\odot$yr$^{-1}$) to $1.52\times10^{11}$\,M$_\odot$ (67\,M$_\odot$yr$^{-1}$), with the mean and median being $3.2\times10^{10}$\,M$_\odot$ (8.6\,M$_\odot$yr$^{-1}$) and $2\times10^{10}$\,M$_\odot$ (2.4\,M$_\odot$yr$^{-1}$), respectively.  The masses sample rather uniformly a decade above and half a decade below the characteristic stellar mass, $M_\star^*$, of the stellar mass function of $z\sim 3$ LAEs \citep[$M_\star^* \approx 4 \times 10^{10}$\,M$_\odot$;][]{Santos2021}.  LAE samples at high-$z$ typically probe objects down to stellar masses of $M_\star \sim 10^7$ \citep[][his Table 3.3]{Ouchi2019}, and the fraction of galaxies at lower masses than what is probed here is dominating in those samples.  Nevertheless, here the different kinematic classes appear rather uniformly distributed over the masses and star-formation rates of our sample.

In Figure~\ref{fig:kin_vs_msfr} we compare the relations between the kinematic parameters and $M_\star$ (left panel) or $\mathrm{SFR}_\mathrm{H\alpha}$ (right panel).  We also use Kendall's rank correlation coefficient, $\tau$, and the associated $p$-value, $p_\tau$, to reject the null-hypothesis of no correlation.  The calculation of the uncertainty on $\tau$ and the assessment of the robustness of our correlation analyses against the error-bars of the parameters is determined via a Monte-Carlo simulation; further details are given in Appendix~\ref{sec:robustn-corr}.  We recover both the $\sigma_0$ vs. $\mathrm{SFR}_\mathrm{H\alpha}$ relation and also the $v_\mathrm{shear}$ vs. $M_\star$ relation at high significance ($p_\tau < 10^{-4}$). The LARS+eLARS IFS sample thus shows the same correlations as larger samples that relate galaxy kinematics with $M_\star$ and SFR \citep[e.g.,][]{Green2014,Moiseev2015,Barat2020,Law2022}.  Moreover, we also recover a relation between $\mathrm{SFR}_\mathrm{H\alpha}$ and $v_\mathrm{shear}$.  While this correlation shows a larger scatter and decreased $|\tau|$ value than the $\mathrm{SFR}_\mathrm{H\alpha}$ vs. $\sigma_0^\mathrm{obs}$ relation it is also robust with respect to the error-bars.  The correlation between $\mathrm{SFR}_\mathrm{H\alpha}$ and $v_\mathrm{shear}$ may be interpreted as the kinematical imprint of the star-formation main sequence.  In Figure~\ref{fig:kin_vs_msfr} we can also appreciate that RDs have, unsurprisingly, overall higher $v_\mathrm{shear}/\sigma_0^\mathrm{obs}$ and higher $v_\mathrm{shear}$ values than CKs and PRs.

Summarising, we find that the eLARS galaxies shows similar star-formation rates as found in high-$z$ LAE samples, while their masses are found to be higher than the average masses such samples.  Nevertheless, we recover the known trends between those parameters and galaxy kinematics of star-forming galaxies.  We thus regard the eLARS sample as a representative probe for studying the relation between galaxy kinematics and Ly$\alpha$ observables, but being aware of potential caveats that arise from the bias towards higher masses (see Sect.~\ref{sec:disc}).

\subsection{Relations between global kinematical statistics and Ly$\alpha$
  observables}
\label{sec:corr-betw-glob}

\begin{figure*}
  \centering
  \includegraphics[width=0.95\textwidth]{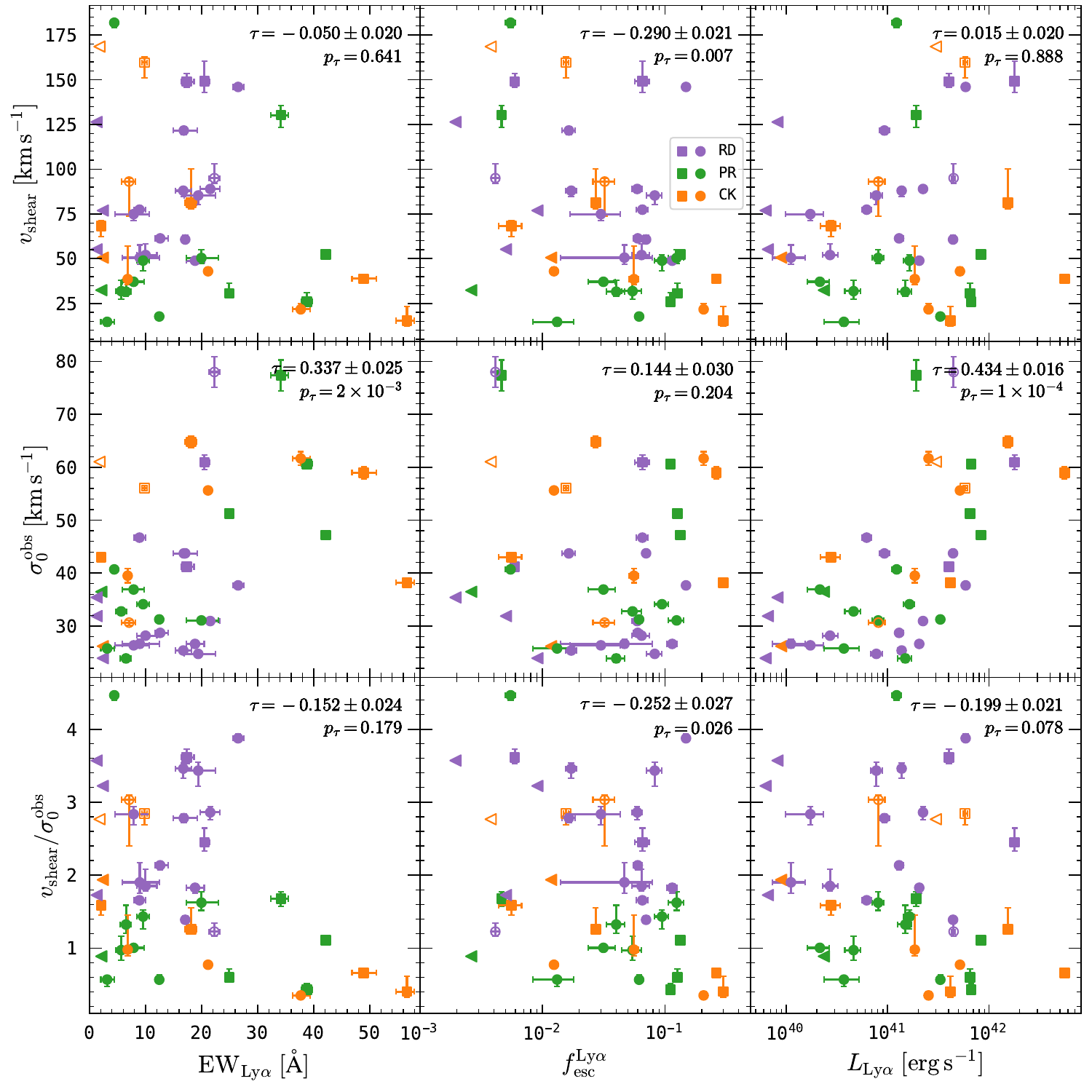}
  \vspace{-1em}
  \caption{Relations between ionised gas kinematics ($v_\mathrm{shear}$,
    $\sigma_0^\mathrm{obs}$, $v_\mathrm{shear} / \sigma_0^\mathrm{obs}$ in the
    top-, middle-, and bottom row, respectively) and Ly$\alpha$ observables
    ($\mathrm{EW}_\mathrm{Ly\alpha}$, $f_\mathrm{esc}^\mathrm{Ly\alpha}$, and
    $L_\mathrm{Ly\alpha}$, in the left-, centre-, and right column,
    respectively).  Symbols are the same as in Figure~\ref{fig:kin_vs_msfr},
    additionally here upper limits in the Ly$\alpha$ observables are shown as
    leftward pointing triangles.  Correlation coefficients (Kendall's $\tau$)
    and corresponding $p$-values are provided in the top-right corner of each
    panel.}
  \label{fig:kin_vs_lya}
\end{figure*}

When combining the observed trends between $M_\star$ or $\mathrm{SFR}_\mathrm{H\alpha}$ and kinematical trends from Sect.~\ref{sec:corr-betw-galaxy} with the fact that LAEs are predominantly found amongst low-mass galaxies on and above the star-formation main sequence, it is expected that strong LAEs exhibit high velocity dispersions and small shearing amplitudes.  This expectation is met by the LARS galaxies \citepalias{Herenz2016} and was recently also reported for $z\sim2$ and $z\sim3$ galaxies by \cite{Foran2023}.    With the now enlarged sample and more carefully determined intrinsic velocity dispersions at hand (Appendix~\ref{sec:meas-intr-disp}) we can put these trends again to the test.

Using the canonical $\mathrm{EW}_\mathrm{Ly\alpha} \geq 20$\,\AA{} boundary to separate the sample into LAEs and non-LAEs, we find that the LAEs are characterised by a on average higher $\sigma_0^\mathrm{obs}$ than the non-LAEs: $\langle \sigma_0^\mathrm{obs} \rangle_{(\mathrm{EW}_\mathrm{Ly\alpha} \geq 20\,\mathrm{\AA})} = 53 \pm 13$\,km\,s$^{-1}$ vs.  $\langle \sigma_0^\mathrm{obs} \rangle_{(\mathrm{EW}_\mathrm{Ly\alpha} < 20\,\mathrm{\AA})} = 34 \pm 9$\,km\,s$^{-1}$.  Moreover, while the $v_\mathrm{shear}$ values of LAEs and non-LAEs cover similar ranges ($\langle v_\mathrm{shear} \rangle_{(\mathrm{EW}_\mathrm{Ly\alpha} \geq 20\,\mathrm{\AA})} = 67 \pm 51$\,km\,s$^{-1}$ vs.  $\langle v_\mathrm{shear} \rangle_{(\mathrm{EW}_\mathrm{Ly\alpha} < 20\,\mathrm{\AA})} = 67 \pm 39$\,km\,s$^{-1}$), LAEs tend to exhibit lower $v_\mathrm{shear}/\sigma_0^\mathrm{obs}$ ratios ($\langle v_\mathrm{shear} / \sigma_0^\mathrm{obs} \rangle_{(\mathrm{EW}_\mathrm{Ly\alpha} \geq 20\,\mathrm{\AA})} = 1.38 \pm 1.18$) than the non-LAEs ($\langle v_\mathrm{shear} / \sigma_0^\mathrm{obs} \rangle_{(\mathrm{EW}_\mathrm{Ly\alpha} < 20\,\mathrm{\AA})} = 2.00 \pm 1.05$).  This indicates that the LAEs and non-LAEs are likely drawn from different underlying distributions in $\sigma_0^\mathrm{obs}$ and $v_\mathrm{shear} / \sigma_0^\mathrm{obs}$.  This hypothesis is confirmed by a KS-test on the sub-samples resulting from a split of LARS+eLARS at $\mathrm{EW}_\mathrm{Ly\alpha} = 20$\,\AA{} with p-values of $p_\mathrm{KS} < 10^{-3}$ for $\sigma_0^\mathrm{obs}$ and $p_\mathrm{KS} = 0.04$ for $v_\mathrm{shear} / \sigma_0^\mathrm{obs}$.  However, in $v_\mathrm{shear}$ the sub-samples appear indistinguishable ($p_\mathrm{KS} = 0.51$).  This result indicates that LAEs are more turbulent and show lower rotational support than non-LAEs.

We now investigate trends between the Ly$\alpha$ observables and the global kinematical characteristics.  We do this graphically in Figure~\ref{fig:kin_vs_lya}, where we show a 3$\times$3 scatter-plot matrix; the columns trace $\mathrm{EW}_\mathrm{Ly\alpha}$, $f_\mathrm{esc}^\mathrm{Ly\alpha}$, and $L_\mathrm{Ly\alpha}$ on the abscissa (from left to right), whereas the rows trace $v_\mathrm{shear}$, $\sigma_0^\mathrm{obs}$, and $v_\mathrm{shear} / \sigma_0^\mathrm{obs}$ on the ordinate (from top to bottom).  For each of the nine analysed relations we also compute Kendall's $\tau$ and the corresponding $p_\tau$.  The resulting $\tau$ and $p_\tau$ values are provided in each sub-panel of Figure~\ref{fig:kin_vs_lya}.  We adopt the canonical $p_\tau < 0.05$ threshold for considering the scatter not being random.  This corresponds to $|\tau| > 0.22$ ($|\tau| > 0.21$) for $N=38$ ($N=42$, for analysis involving only $v_\mathrm{shear}$; cf. Sect.~\ref{sec:line-sight-velocity}) data points (Appendix~\ref{sec:robustn-corr}).  We recover correlations for $\sigma_0^\mathrm{obs}$ vs.  $\mathrm{EW}_\mathrm{Ly\alpha}$ and $\sigma_0^\mathrm{obs}$ vs. $L_\mathrm{Ly\alpha}$ and we recover anti-correlations for $v_\mathrm{shear}$ vs. $f_\mathrm{esc}^\mathrm{Ly\alpha}$ and $v_\mathrm{shear}/\sigma_0^\mathrm{obs}$ vs. $f_\mathrm{esc}^\mathrm{Ly\alpha}$ by adopting this threshold.  All of these correlations are robust against perturbations due uncertainties on the involved measurements by the metric defined in Appendix~\ref{sec:robustn-corr}.  Moreover, the data is suggestive of a anti-correlation between $v_\mathrm{shear}/\sigma_0^\mathrm{obs}$ and $L_\mathrm{Ly\alpha}$, although the obtained $p_\tau$ is slightly above the canonical boundary of 0.05.  We will discuss the implications of our findings further in Sect.~\ref{sec:disc}.  For now we conclude that the data suggests that the kinematical state of a galaxy are in a causal relation to all three Ly$\alpha$ observables.

Since the correlation analysis tests the data for monotonicity it fails to capture trends that can not be described by a monotonic function.  In this respect the panels for $v_\mathrm{shear}$ vs. $\mathrm{EW}_\mathrm{Ly\alpha}$ or $v_\mathrm{shear}/\sigma_0^\mathrm{obs}$ vs. $\mathrm{EW}_\mathrm{Ly\alpha}$ in Figure~\ref{fig:kin_vs_lya} appear noteworthy.  Here we observe that low-$\mathrm{EW}_\mathrm{Ly\alpha}$ galaxies occupy almost the whole dynamical range sampled in $v_\mathrm{shear}$ and $v_\mathrm{shear} / \sigma_0^\mathrm{obs}$, whereas high-$\mathrm{EW}_\mathrm{Ly\alpha}$ galaxies are found predominantly at lower $v_\mathrm{shear}$ and $v_\mathrm{shear} / \sigma_0^\mathrm{obs}$ values.  More quantitatively, adopting the thresholds of $\mathrm{EW}_\mathrm{Ly\alpha} \geq 20$\,\AA{} for defining LAEs and $v_\mathrm{shear} / \sigma_0^\mathrm{obs} < 1.86$ for defining dispersion dominated systems (cf. Sect.~\ref{sec:comp-glob-kinem}), we find that 9 out of 12 LAEs are dispersion dominated (75\%).  These 9 LAEs are also characterised by $v_\mathrm{shear} < 100$\,km\,s$^{-1}$.  In this regard it appears also of interest to note that all galaxies classified as RDs in Sect.~\ref{sec:meth-class} do not qualify as dispersion dominated.  If the trends found in the LARS+eLARS sample can be generalised, then we may suspect that LAEs are preferentially dispersion dominated systems with low velocity shearing and velocity fields that are, moreover, not commensurate with an unperturbed rotating disk.

\section{Discussion}
\label{sec:disc}

\subsection{How important are the kinematical parameters for the
  Ly$\alpha$ observables?}
\label{sec:how-important-are}

\begin{table}
  \caption{LARS+eLARS galaxy parameters against which the importance of
    $v_\mathrm{shear}$, $\sigma_{0}^\mathrm{obs}$ and
    $v_\mathrm{sehar} / \sigma_0^\mathrm{obs}$ for regulating
    $\mathrm{EW}_\mathrm{Ly\alpha}$, $f_\mathrm{esc}^\mathrm{Ly\alpha}$, and
    $L_\mathrm{Ly\alpha}$ is assessed.}
  \label{tab:stdin}
  \centering
  \begin{tabular}{ccc}
    \hline \hline \noalign{\smallskip}
    \multicolumn{3}{c}{observational parameter set} \\
    Parameter & Unit  & Reference \\
    \hline
    $L_{X\in\{FUV,U,B,I\}}$ & erg\,s$^{-1}$\AA{}$^{-1}$ & \citetalias{Runnholm2020} \\
    $L_\mathrm{H\alpha}$ & erg\,s$^{-1}$ & \citetalias{Melinder2023} \\
    $L_{X\in\{\mathrm{H\beta},\mathrm{[OIII],[OII],[NII]}\}}$ & erg\,s$^{-1}$ & \citetalias{Runnholm2020} \\
    \hline \hline
    \multicolumn{3}{c}{physical parameter set} \\
    Parameter & Unit & Reference \\ \hline
    $M_\star$ & M$_\odot$ & \citetalias{Melinder2023} \\
    $s_\mathrm{UV}$ & kpc  & \citetalias{Runnholm2020} \\
    $E(B-V)$ & mag  & \citetalias{Melinder2023} \\
    $O_\mathrm{32}$ & \dots & \citetalias{Runnholm2020} \\
    $\lg (O/H)$ & $12 + \lg (O/H)$  & \citetalias{Runnholm2020} \\
    $\mathrm{SFR}_\mathrm{H\alpha}$ & M$_\odot$yr$^{-1}$ & \citetalias{Melinder2023} \\
    $v_{95}$ & km\,s$^{-1}$ &  \citetalias{Runnholm2020} \\
    $w_{90}$ & km\,s$^{-1}$ &  \citetalias{Runnholm2020} \\
    $F_\mathrm{cov}$ & km\,s$^{-1}$ & \citetalias{Runnholm2020} \\
    \hline \hline
  \end{tabular}
\end{table}

Our analysis in Sect.~\ref{sec:corr-betw-glob} shows the potential importance of the global kinematical state of the ionised gas in a galaxy regarding its Ly$\alpha$ observables.  We found significant correlations of $\sigma_0^\mathrm{obs}$ with $\mathrm{EW}_\mathrm{Ly\alpha}$ and $L_\mathrm{Ly\alpha}$, as well as the anti-correlations of $v_\mathrm{shear}$ and $v_\mathrm{shear}/\sigma_0^\mathrm{obs}$ with $f_\mathrm{esc}$ (Figure~\ref{fig:kin_vs_lya}).  We reasoned that such relations are expected as they are the kinematical imprints of the fact that strong Ly$\alpha$ emitters are preferentially lower-mass systems with higher star-formation rates in comparison to weaker Ly$\alpha$ emitters or even absorbers.  Of course, higher star-formation rates result in a larger amount of intrinsically produced Ly$\alpha$ photons, but we may further reason that feedback from star formation is creating an environment that facilitates easy escape of those photons.  This may be especially the case in low-mass galaxies, where stellar winds and supernovae explosions can remove gas more efficiently because of their shallower gravitational potentials.

We recall that other galaxy characteristics are known to significantly affect the Ly$\alpha$ observability.  In particular, the size and the dust content are of high relevance.  Studies both at high-$z$ and low-$z$ show that LAEs are preferentially compact systems with only small amounts of dust \citep[e.g.,][]{Guaita2010,Paulino-Afonso2018,Marchi2019,Paswan2022,Napolitano2023}.  The low dust content of LAEs is also reflected in their young ages and low metallicities.  Moreover, the outflow kinematics and covering fraction of the scattering medium, probed by low-ionisation absorption lines in the rest-frame UV, are important;  Ly$\alpha$ emitting galaxies are found to exhibit lower covering fractions and velocity offsets indicative of outflows \cite[e.g.,][]{Wofford2013,Rivera-Thorsen2015,Trainor2015,Ostlin2021,Reddy2022,Hayes2023}.  Lastly, Ly$\alpha$ emitters exhibit higher degrees of ionisation and overall a harder UV radiation field \citep{Trainor2016,Hayes2023b}.  Clearly, Ly$\alpha$ production and the subsequent escape depend on a multitude of factors, and it is not at all intuitive which are the most important.  Thus, we now want to quantify the importance of the global ionised gas kinematics regarding $f_\mathrm{esc}^\mathrm{Ly\alpha}$, $\mathrm{EW}_\mathrm{Ly\alpha}$, and $L_\mathrm{Ly\alpha}$ (hereafter responses) relative to the other global observational and observationally inferred physical galaxy characteristics (hereafter parameters) of LARS+eLARS.

Assessing the relative parameter importance from a set of parameters with respect to a response is a difficult statistical problem.  It is an area of active and controversial research and there exists no general agreed upon solution \citep[see review by][]{Groemping2015}.  A main problem are intra-correlations between parameters, that lead to fundamental problems in defining the concept of importance.  This hampers the robustness of the interpretation of heuristic approaches.  Moreover, the more robust approaches implicitly assume that measurement errors are negligible, a situation unfortunately not encountered for the here considered responses and parameters.  We thus opt for using rather simple metrics that were already applied in the context of understanding Ly$\alpha$ emission from galaxies (\citealt{Runnholm2020}, hereafter \citetalias{Runnholm2020}; \citealt{Napolitano2023}; \citealt{Hayes2023}).  Nevertheless, we also need to be aware of their shortcomings.

First, we will use the absolute values of Kendall's $\tau$ for each parameter on its own (Sect.~\ref{sec:mp}).  This so-called ``marginal perspective'' was recently used by \cite{Napolitano2023} alongside a Random Forrest classifier for identifying the important parameters that regulate $\mathrm{EW}_\mathrm{Ly\alpha}$ in high-$z$ galaxies.  The downside of the marginal perspective is that seemingly unimportant parameters may help each other in influencing the response variable \citep[e.g.,][]{Guyon2003}.  For example, it may not be sufficient that a galaxy shows high $O_{32}$ for having a large $\mathrm{EW}_\mathrm{Ly\alpha}$, but the galaxy is also required to have a low $F_\mathrm{cov}$ -- both parameters together may thus be of high-importance.  Thus, we also need to analyse the problem from a ``conditional perspective'' \citep{Groemping2015}.  To this aim we first analyse all intra-parameter correlations in our sample before we follow \citetalias{Runnholm2020} and \cite{Hayes2023b} and adopt a procedure known as stepwise-regression in a multi-variable linear regression framework (Sect.~\ref{sec:cp}).

As in \citetalias{Runnholm2020} we define a purely observational parameter set and a physical parameter that is derived from the observational parameters. The observational parameter set consists of luminosities in the FUV, U, B, and I-band (expressed as luminosity densities $L_X$, with ${X\in\{FUV,U,B,I\}}$), emission line luminosities of the lines H$\alpha$, H$\beta$, [\ion{O}{III}] $\lambda5007$, [\ion{O}{II}] $\lambda3727 + \lambda3729$, [\ion{N}{II}] $\lambda6584$ ($L_Y$, with $Y\in\{\mathrm{H\alpha, H\beta, [OII], [OIII] [NII]}\}$).  The physical parameter set consists of stellar mass $M_\star$, the flux ratio between the [\ion{O}{III}] and [\ion{O}{II}] lines, $O_{32} = F_{\mathrm{[OIII]},\lambda5007}/(F_{\mathrm{[OII],\lambda3727}} + F_{\mathrm{[OII]},\lambda3729})$ as proxy for the degree of ionisation, the nebular dust attenuation, $E(B-V)$, the star-formation rate, $\mathrm{SFR}_\mathrm{H\alpha}$, the nebular oxygen abundance, $12 + \log (\mathrm{O/H})$, the UV size $s_\mathrm{UV}$, as well as three parameters from low-ionisation absorption line measurements from HST/COS spectroscopy, namely the covering fraction $F_\mathrm{cov}$, the line width, $w_{90}$, and the velocity offset, $v_{95}$, with respect to the systemic redshift determined from the nebular emission lines.  All parameters are taken from \citetalias{Runnholm2020} and \citetalias{Melinder2023}; these publications detail the measurements of most of those parameters and we also refer to \cite{Rivera-Thorsen2015} and \cite{Hayes2023b} for a description of the HST/COS based parameters.  Both parameter sets consist of 9 parameters, which are again summarised in Table~\ref{tab:stdin}.

Lastly, in the following analyses we remove all six galaxies without detected Ly$\alpha$ emission from the statistical analyses presented below, since methods of the conditional perspective can not deal with upper limits.  As mentioned in Sect.~\ref{sec:line-sight-velocity}, galaxies with double component H$\alpha$ profiles are also not considered in our statistical analyses (i.e., we also have to remove additionally the three galaxies LARS 9, eLARS 7, and eLARS 24; LARS 13 has both double components and upper limits in the Ly$\alpha$ observables).  Thus, the following analyses consider a sample of $N=33$ galaxies.

\subsubsection{Importance of kinematics -- marginal perspective}
\label{sec:mp}

\begin{figure*}
  \centering
  \includegraphics[width=0.33\textwidth]{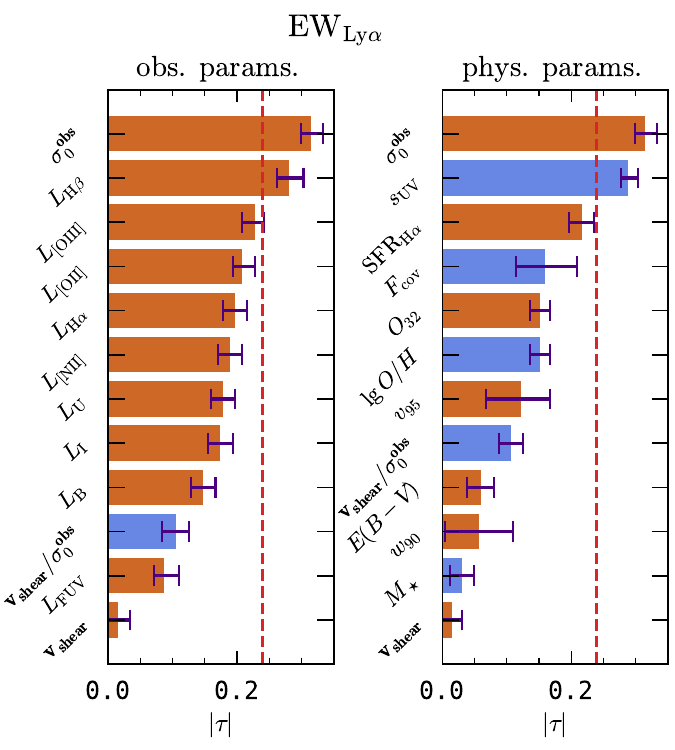}
  \includegraphics[width=0.33\textwidth]{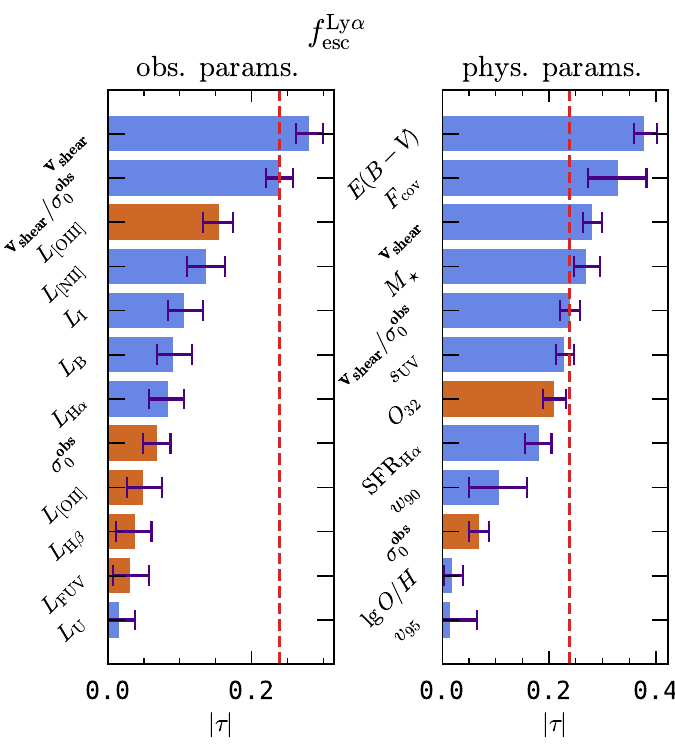}
  \includegraphics[width=0.33\textwidth]{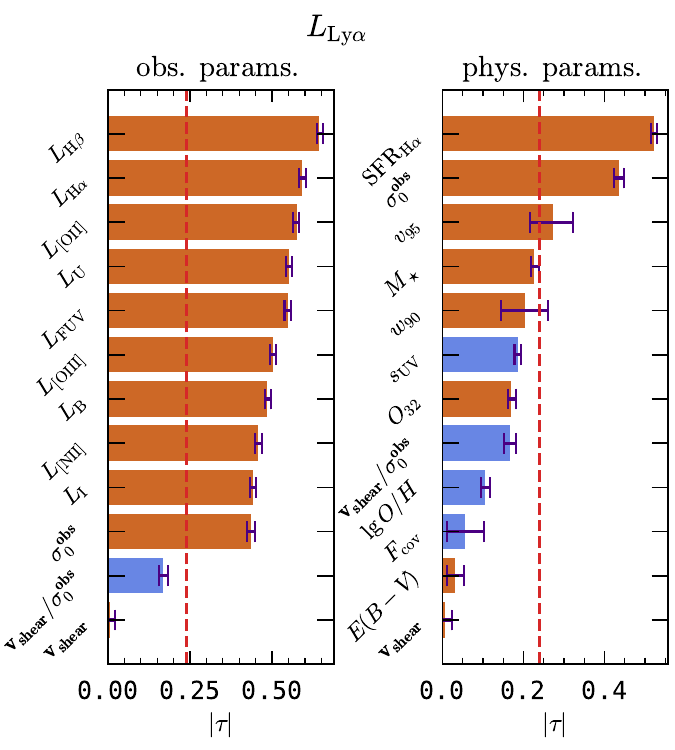}
  \caption{Importance of observational and physical parameters for $f_\mathrm{esc}^\mathrm{Ly\alpha}$, $\mathrm{EW}_\mathrm{Ly\alpha}$, and $L_\mathrm{Ly\alpha}$ using the absolute value of Kendall's $\tau$ (marginal perspective; Sect.~\ref{sec:mp}).  In each panel the parameters are ranked according to $|\tau|$ from top to bottom.  The length of each bar indicates the absolute value of $\tau$ and the error-bars indicate the upper- and lower quartile from the robustness analysis presented in Appendix~\ref{sec:robustn-corr}.  Each bar is coloured in red or blue depending on whether $\tau > 0$ (correlated trend between parameter and response) or $\tau < 0$ (anti-correlated trend between parameter and response), respectively.  The vertical dashed red line indicates the threshold $|\tau| > 0.239$ that corresponds to $p_\tau < 0.05$ for $N=33$ (Appendix~\ref{sec:robustn-corr}).  The \emph{left-}, \emph{centre-}, and \emph{right panel} show the results for the response $\mathrm{EW}_\mathrm{Ly\alpha}$, $f_\mathrm{esc}^\mathrm{Ly\alpha}$, and $L_\mathrm{Ly\alpha}$, respectively, where each left- and right sub-panel show the rankings for the observational- and physical parameters, respectively.}
  \label{fig:mp}
\end{figure*}

The marginal perspective regarding the importance of observational and physical parameters with respect to the responses $EW_\mathrm{Ly\alpha}$, $f_\mathrm{esc}^\mathrm{Ly\alpha}$, and $L_\mathrm{Ly\alpha}$ in LARS+eLARS is presented graphically in Figure~\ref{fig:mp}.  We include the global kinematical statistics $v_\mathrm{shear}$, $\sigma_0^\mathrm{obs}$, and $v_\mathrm{shear} / \sigma_0^\mathrm{obs}$ in each of the two analysed parameter sets to asses their relative importance regarding the responses in comparison to the other parameters.
In  Figure \ref{fig:mp} we rank the parameters by the absolute values of
Kendall's $\tau$.  The rationale here is, that $|\tau|$ is a direct measure of
the frequency at which the change of a parameter leads to a parallel or
anti-parallel change of the response.  The higher the frequency of changes in
the same direction, the higher the value of $|\tau|$.  Thus, an ordering of the
parameters by $|\tau|$ can be interpreted as a ranking of their importance in
influencing the response.  As before, we deem a correlation significant if
$p_\tau <0.05$, which corresponds to $|\tau| > 0.239$ (indicated by a vertical
dashed line in Figure~\ref{fig:mp}) for $N = 33$.  Moreover,  the correlation is deemed
robust if the inter-quartile range of $\tau$'s from a Monte-Carlo simulation,
that takes errors on the parameters and observables into account, does not overlap with this threshold (Appendix~\ref{sec:robustn-corr}).

Focusing first on the response $\mathrm{EW}_\mathrm{Ly\alpha}$ we find that $\sigma_0^\mathrm{obs}$ dominates in importance over all parameters in the observational- and physical parameter sets.  On the contrary, $v_\mathrm{shear}$ is found to be least influential and, moreover, $v_\mathrm{shear} / \sigma_0^\mathrm{obs}$ appears also irrelevant.  Formally, the emission line luminosities are ranked directly below $\sigma_0^\mathrm{obs}$, followed by the broad-band luminosity densities.  However, here only the correlation with $L_\mathrm{H\beta}$ is statistically significant and robust.  Lastly, consistent with the literature \citep[e.g.,][]{Law2012b,Paulino-Afonso2018}, we find an anti-correlation between $\mathrm{EW}_\mathrm{Ly\alpha}$ and $s_\mathrm{UV}$ .  Taken together, these trends suggest that compact galaxies that are bright in the rest-frame optical emission lines, and which are characterised by a highly turbulent interstellar medium, are more likely to show higher $\mathrm{EW}_\mathrm{Ly\alpha}$.

Next we focus on the response $f_\mathrm{esc}^\mathrm{Ly\alpha}$.  The kinematical parameters $v_\mathrm{shear}$ and $v_\mathrm{shear} / \sigma_0^\mathrm{obs}$, while not being highly-ranked for $\mathrm{EW}_\mathrm{Ly\alpha}$, emerge as highly influential parameters for $f_\mathrm{esc}$.  Both anti-correlations dominate over all other observational parameters, albeit the correlation with $v_\mathrm{shear} / \sigma_0^\mathrm{obs}$ is not robust (Appendix~\ref{sec:robustn-corr}.  Notably, except for $L_\mathrm{[OIII]}$ and $L_\mathrm{[NII]}$, all observational parameters show even $|\tau| < 0.1$ and appear thus individually completely irrelevant with respect to $f_\mathrm{esc}^\mathrm{Ly\alpha}$.  This is also the case for $\sigma_0^\mathrm{obs}$.  Regarding relations between physical parameters and $f_\mathrm{esc}^\mathrm{Ly\alpha}$ we find that the anti-correlations with $E(B-V)$ and $F_\mathrm{cov}$ are ranked higher than the anti-correlations with $v_\mathrm{shear}$.  Moreover, the value of $|\tau|$ for $v_\mathrm{shear}$ appears on par with the coefficient recovered for $M_\star$.

The high importance of $\sigma_0^\mathrm{obs}$ and simultaneous irrelevance of $v_\mathrm{shear}$ or $v_\mathrm{shear} / \sigma_0$ for regulating $\mathrm{EW}_\mathrm{Ly\alpha}$ appears noteworthy, given that the situation is reverse for $f_\mathrm{esc}^\mathrm{Ly\alpha}$.  This situation is also evident in Figure~\ref{fig:kin_vs_lya}.  Taken at face value, this is suggestive of different mechanisms regulating $\mathrm{EW}_\mathrm{Ly\alpha}$ and $f_\mathrm{esc}^\mathrm{Ly\alpha}$.  This appears intriguing, because a tight linear relation between $f_\mathrm{esc}^\mathrm{Ly\alpha}$ and $\mathrm{EW}_\mathrm{Ly\alpha}$ is found in our sample (\citetalias{Melinder2023}), as well as in samples at higher redshifts \citep{Sobral2019,Roy2023,Tang2023}.  However, in the following Sect.~\ref{sec:cp} we show that $\sigma_0^\mathrm{obs}$ also emerges as an important parameter for regulating $f_\mathrm{esc}^\mathrm{Ly\alpha}$ from the conditional perspective, but only when taken in concert with other parameters that are known to influence the Ly$\alpha$ radiative transfer.  

Here we recover another intriguing discrepancy regarding the influence of individual parameters with respect to the responses $f_\mathrm{esc}^\mathrm{Ly\alpha}$ and $\mathrm{EW}_\mathrm{Ly\alpha}$.  Namely, we find that $E(B-V)$ and $M_\star$ are not correlated with $\mathrm{EW}_\mathrm{Ly\alpha}$, whereas both parameters anti-correlate with $f_\mathrm{esc}^\mathrm{Ly\alpha}$.  Kinematically, the presence or absence of an anti-correlation between $M_\star$ and $f_\mathrm{esc}^\mathrm{Ly\alpha}$ or $\mathrm{EW}_\mathrm{Ly\alpha}$, respectively, is reflected in our sample by the different behaviour of both parameters with respect to $v_\mathrm{shear}$, which we just discussed.  Moreover, this discrepancy does not vanish under the conditional perspective in Sect.~\ref{sec:cp} and it is appears also in contrast to some literature results \citep{Marchi2019,Napolitano2023}.  Nevertheless, other studies also do not report monotonic relations between $\mathrm{EW}_\mathrm{Ly\alpha}$ and $E(B-V)$ or $M_\star$ \citep{Atek2014,Bolan2024}.  

Finally, we focus on the response $L_\mathrm{Ly\alpha}$.  Here we find that all the other observational parameters are ranked above the kinematical parameters.  However, $\sigma_0^\mathrm{obs}$ is ranked in second place amongst the physical parameters below $\mathrm{SFR}_\mathrm{H\alpha}$; $v_\mathrm{shear}$ and $v_\mathrm{shear} / \sigma_0^\mathrm{obs}$ appear in the bottom half of the ranks and are thus not important in influencing $L_\mathrm{Ly\alpha}$.  The strong relation between the emission line luminosities of H$\beta$ and H$\alpha$ for $L_\mathrm{Ly\alpha}$ indicates that the production of intrinsic Ly$\alpha$ is a dominating cause in regulating the total Ly$\alpha$ photon output.  This normalisation of Ly$\alpha$ luminosity by recombination rate is also reflected in the physical parameter set, where $\mathrm{SFR}_\mathrm{H\alpha}$ ($\propto L_\mathrm{H\alpha}$) is ranked first.

As already mentioned in Sect.~\ref{sec:how-important-are}, we need to refrain from over-interpreting the rankings based on the marginal perspective for getting a grip on the physics that regulate the escape and observability of Ly$\alpha$ emission.  For now we only conclude that our findings here are not inconsistent with the hypothesis that the global gas kinematics play a very important role in regulating the Ly$\alpha$ observables of a galaxy.

\subsubsection{Importance of kinematics -- conditional perspective}
\label{sec:cp}

\begin{figure*}
  \centering
  \includegraphics[width=0.49\textwidth,trim=20 0 0 0,clip=true]{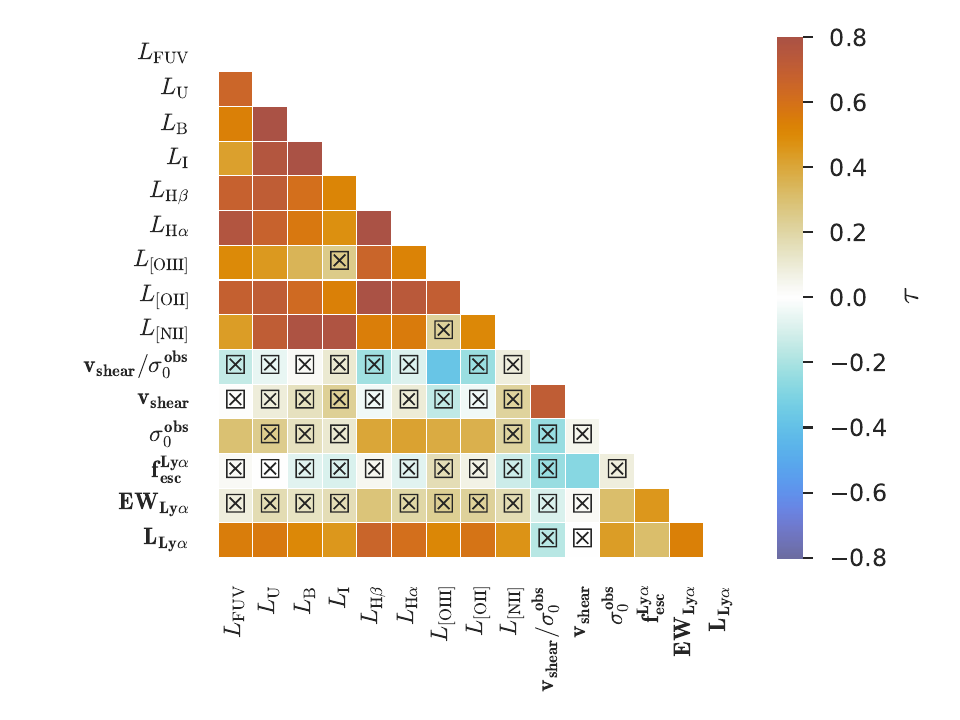} 
  \includegraphics[width=0.49\textwidth,trim=20 0 0 0,clip=true]{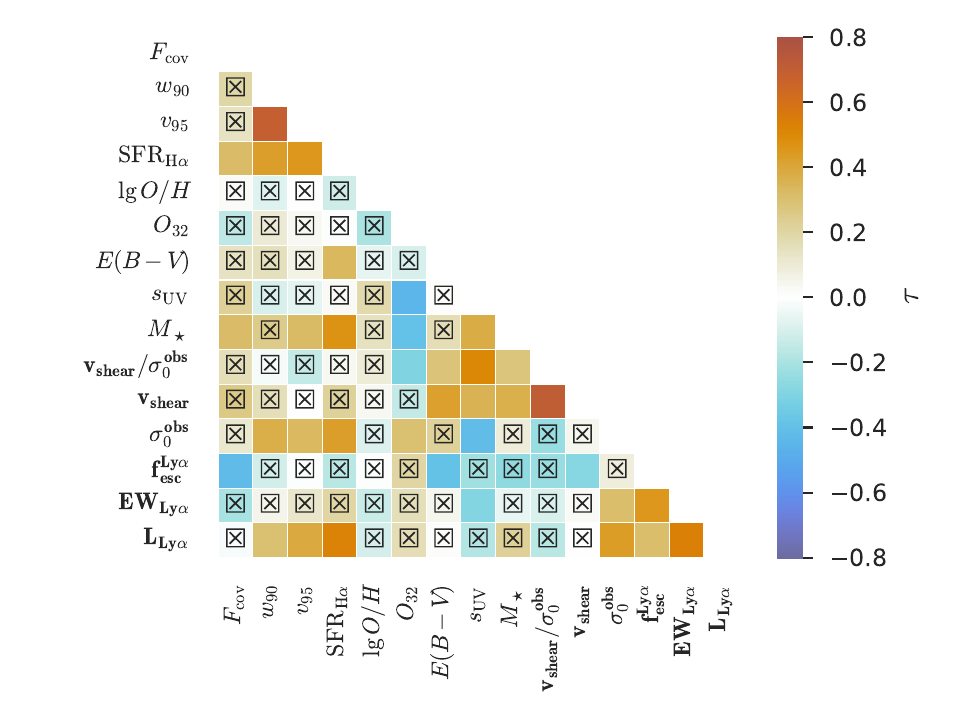}
  \caption{Correlation matrix heatmap visualisation (Kendall's $\tau$) for observational parameters (\emph{left panel}) and physical parameters (\emph{right panel}), including $v_\mathrm{shear}$, $\sigma_0^\mathrm{obs}$, and $v_\mathrm{shear} / \sigma_0^\mathrm{obs}$, as well as the Ly$\alpha$ observables $f_\mathrm{esc}^\mathrm{Ly\alpha}$, $\mathrm{EW}_\mathrm{Ly\alpha}$, and $L_\mathrm{Ly\alpha}$.  Statistical insignificant trends, i.e. where $p_0 > 0.05$ or, correspondingly $|\tau| > 0.239$ for $N=33$, are crossed out.}
  \label{fig:heat}
\end{figure*}

\begin{figure*}
  \centering
  \includegraphics[width=0.49\textwidth]{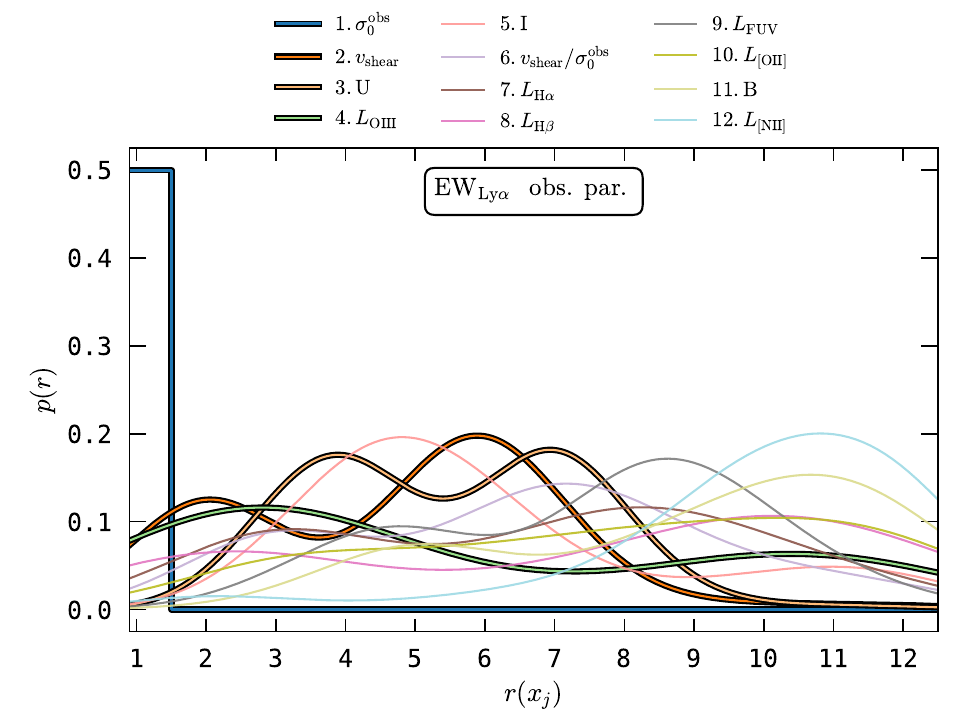}
  \includegraphics[width=0.49\textwidth]{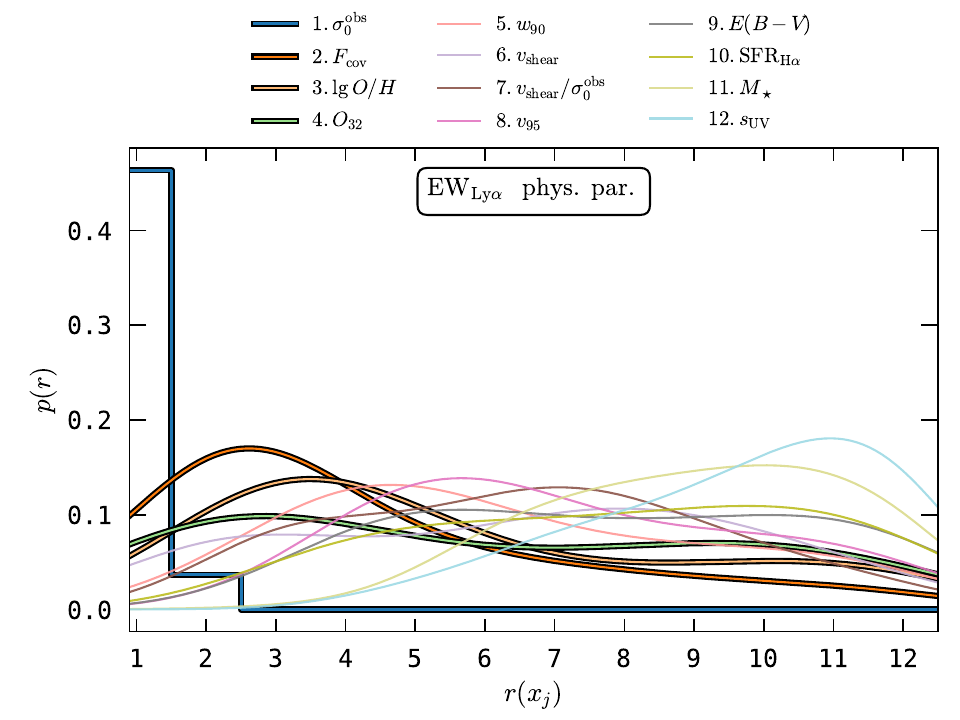} \\ \vspace{0.3em}
  \includegraphics[width=0.49\textwidth]{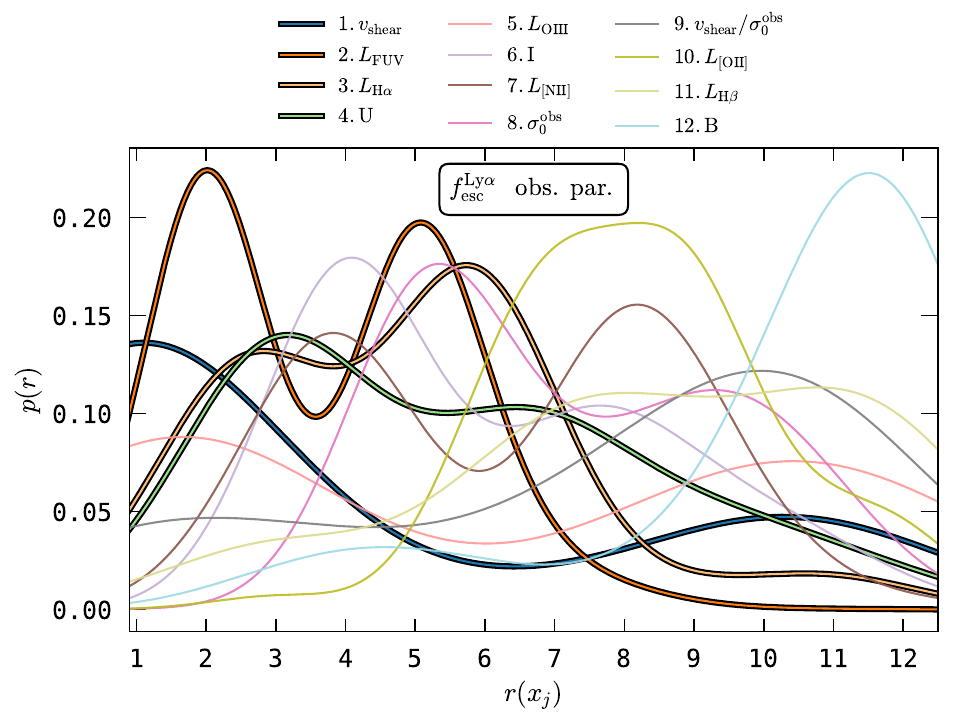}
  \includegraphics[width=0.49\textwidth]{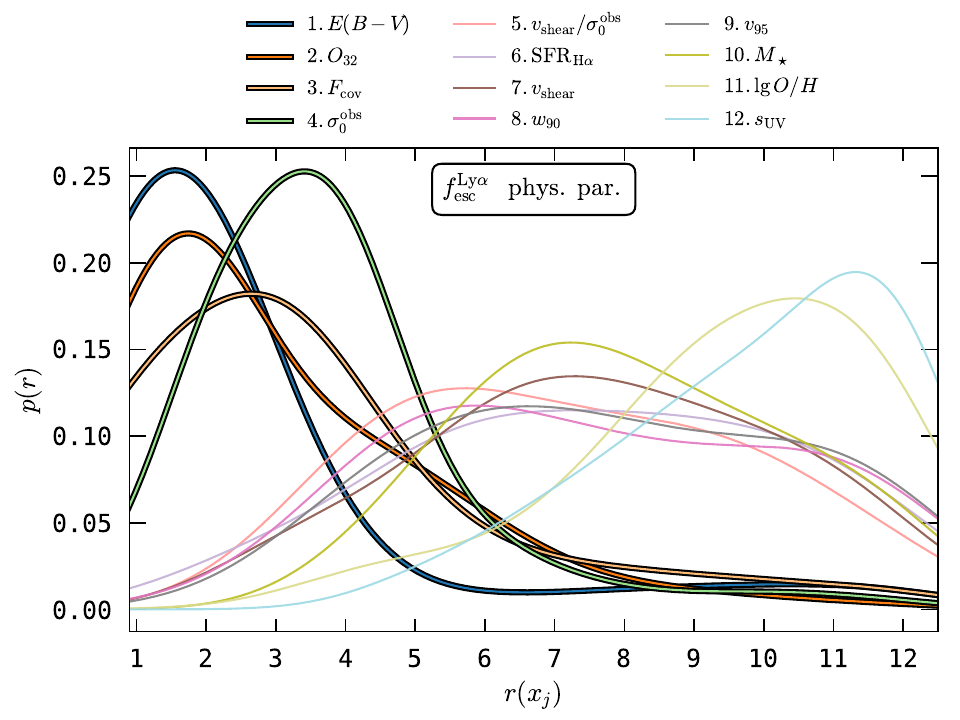} \\ \vspace{0.3em}
  \includegraphics[width=0.49\textwidth]{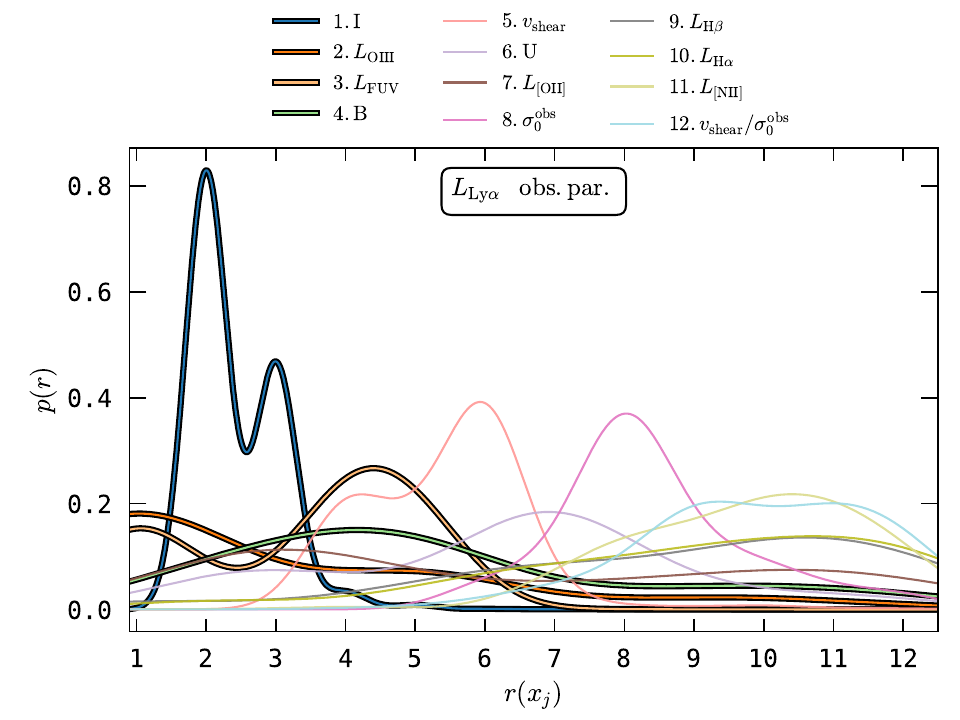}
  \includegraphics[width=0.49\textwidth]{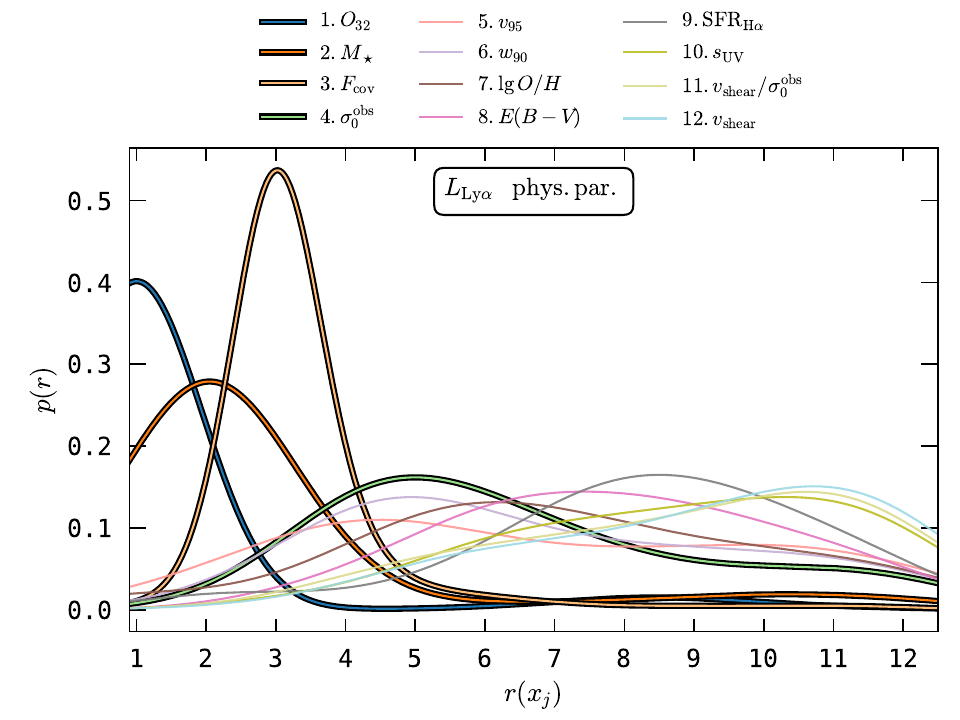}
  \vspace{1em}
  \caption{Visualisation of the parameter importance for responses $\mathrm{EW}_\mathrm{Ly\alpha}$ (\emph{top panels}), $f_\mathrm{esc}^\mathrm{Ly\alpha}$ (\emph{middle panels}), and $L_\mathrm{Ly\alpha}$ (\emph{bottom panels}) from a conditional perspective using step-wise regression in a multi-variate linear regression framework (see Sect.~\ref{sec:cp} for details).  The panels on the left- and right-hand side visualise the distribution of rankings $p(r(x_i))$ for the parameters $x_i$ from the observational- and physical parameter set, respectively, where each set includes the kinematical parameters that have been derived in this work.  Low values of $r(x_i)$ indicate high importance of $x_i$, whereas high values indicate low importance.  All except two curves are obtained by using a kernel-density estimator (Gaussian kernel, width = 0.5) on the ranks from $N_\mathrm{MC} = 10^3$ Monte-Carlo realisations.  The exceptions are the $\sigma_0^\mathrm{obs}$ curves for the response $\mathrm{EW}_\mathrm{Ly\alpha}$ (see footnote \ref{fn:1}).  The four most important parameters according to Eq.~\eqref{eq:4} are shown with bold lines, whereas the remaining ranks are shown with thin lines and in subdued colours.  The legend atop of each panel is sorted according to the ranks from Eq.~\eqref{eq:4} that are also tabulated in Table~\ref{tab:cp}.}
  \label{fig:cp}
\end{figure*}

\begin{table*}
  \caption{Importance of observational and physical parameters for
    $f_\mathrm{esc}^\mathrm{Ly\alpha}$, $\mathrm{EW}_\mathrm{Ly\alpha}$, and
    $L_\mathrm{Ly\alpha}$ from a conditional perspective using the rank score
    defined in Eq.~\eqref{eq:4} over $N_{MC}=10^3$ Monte-Carlo realisations
    (see Sect.~\ref{sec:cp} for details). }
  \label{tab:cp}
  \centering 
  \addtolength{\tabcolsep}{-0.3em}
  \begin{tabular}{c|cr|cr|cr|cr|cr|cr}
    \hline \hline 
    \multirow{2}{*}{Rank} & \multicolumn{4}{c|}{$\mathrm{EW}_\mathrm{Ly\alpha}$}   & \multicolumn{4}{c|}{$f_\mathrm{esc}^\mathrm{Ly\alpha}$} & \multicolumn{4}{c}{$L_\mathrm{Ly\alpha}$} \\ 
    {}     & obs. par. & \multicolumn{1}{c}{$r(x_i)$} & phys. par. & \multicolumn{1}{c|}{$r(x_i)$} & obs. par. & \multicolumn{1}{c}{$r(x_i)$} & phys. par. & \multicolumn{1}{c| }{$r(x_i)$}  &  obs. par. & \multicolumn{1}{c}{$r(x_i)$} & phys. par. & \multicolumn{1}{c }{$r(x_i)$} \\  \hline
    1 & $\sigma_0^\mathrm{obs}$ & 1.002 & $\sigma_0^\mathrm{obs}$ & 1.131 & $L_\mathrm{FUV}$ & 3.652 & $E(B-V)$ & -2.294 & $\mathrm{I}$ & 2.446 & $O_{32}$ & 1.600 \\
    2 & $v_\mathrm{shear}$ & -4.817 & $F_\mathrm{cov}$ & -4.391 & $v_\mathrm{shear}$ & -3.806 & $O_{32}$ & 3.049 & $L_\mathrm{OIII}$ & 2.834 & $M_\star$ & 3.066 \\
    3 & $\mathrm{U}$ & -5.553 & $\lg O/H$ & -5.441 & $L_\mathrm{H\alpha}$ & -4.788 & $\sigma_0^\mathrm{obs}$ & 3.609 & $L_\mathrm{FUV}$ & 3.269 & $F_\mathrm{cov}$ & -3.346 \\
    4 & $L_\mathrm{OIII}$ & -5.641 & $O_{32}$ & 6.022 & $\mathrm{U}$ & -5.535 & $F_\mathrm{cov}$ & -3.611 & $\mathrm{B}$ & -5.334 & $\sigma_0^\mathrm{obs}$ & 6.540 \\
    5 & $\mathrm{I}$ & 6.165 & $w_{90}$ & -6.167 & $L_\mathrm{OIII}$ & 5.942 & $v_\mathrm{shear} / \sigma_0^\mathrm{obs}$ & 7.083 & $v_\mathrm{shear}$ & -5.452 & $v_{95}$ & 6.854 \\
    6 & $v_\mathrm{shear} / \sigma_0^\mathrm{obs}$ & 6.545 & $v_\mathrm{shear}$ & -6.528 & $\mathrm{I}$ & 6.128 & $w_{90}$ & -7.394 & $\mathrm{U}$ & -6.254 & $w_{90}$ & -6.960 \\
    7 & $L_\mathrm{H\alpha}$ & -6.596 & $v_\mathrm{shear} / \sigma_0^\mathrm{obs}$ & 6.697 & $L_\mathrm{[NII]}$ & 6.223 & $v_\mathrm{shear}$ & -7.547 & $L_\mathrm{[OII]}$ & 6.437 & $\lg O/H$ & -7.019 \\
    8 & $L_\mathrm{H\beta}$ & 7.279 & $v_{95}$ & 7.176 & $\sigma_0^\mathrm{obs}$ & 7.261 & $v_{95}$ & 7.850 & $\sigma_0^\mathrm{obs}$ & 8.357 & $E(B-V)$ & -7.811 \\
    9 & $L_\mathrm{FUV}$ & 7.534 & $E(B-V)$ & -7.742 & $v_\mathrm{shear} / \sigma_0^\mathrm{obs}$ & 7.474 & $\mathrm{SFR}_\mathrm{H\alpha}$ & 7.883 & $L_\mathrm{H\beta}$ & -8.900 & $\mathrm{SFR}_\mathrm{H\alpha}$ & 8.124 \\
    10 & $L_\mathrm{[OII]}$ & -8.071 & $\mathrm{SFR}_\mathrm{H\alpha}$ & 7.920 & $L_\mathrm{[OII]}$ & 8.187 & $M_\star$ & -8.239 & $L_\mathrm{H\alpha}$ & -9.102 & $s_\mathrm{UV}$ & 8.762 \\
    11 & $\mathrm{B}$ & 8.951 & $M_\star$ & -9.183 & $L_\mathrm{H\beta}$ & 8.584 & $\lg O/H$ & 9.459 & $L_\mathrm{[NII]}$ & 9.777 & $v_\mathrm{shear} / \sigma_0^\mathrm{obs}$ & -8.804 \\
    12 & $L_\mathrm{[NII]}$ & 9.847 & $s_\mathrm{UV}$ & 9.604 & $\mathrm{B}$ & -10.419 & $s_\mathrm{UV}$ & 9.982 & $v_\mathrm{shear} / \sigma_0^\mathrm{obs}$ & 9.837 & $v_\mathrm{shear}$ & 9.114 \\ \hline \hline
  \end{tabular}
  \tablefoot{A minus sign in front of $r(x_i)$ indicates that this parameter is
    contributing with a negative $c_i$ in Eq.~\eqref{eq:2}, otherwise the
  contribution is positive.}
\end{table*}

In order to move away from the marginal perspective, we first analyse intra-parameter correlations in our parameter sets from Table~\ref{tab:stdin}.  To this aim we use heat-map visualisations in Figure~\ref{fig:heat}.  Using Kendall's $\tau$ as a correlation measure Figure~\ref{fig:heat} visualise the strength and directions of possible monotonic relations between all possible parameter combinations.  The aim of this analysis is primarily to illustrate possible caveats when analysing only single parameter correlations with respect to the Ly$\alpha$ observables.  We therefore also refrained from carrying out a robustness analysis on $\tau$, instead we choose a more stringent $p$-value of $p_\tau = 0.03$ or, correspondingly, $|\tau| > 0.265$ for $N=33$ (Eq.~\eqref{eq:30} in Appendix~\ref{sec:robustn-corr}).  Intra-parameter relations resulting in a smaller $|\tau|$ are crossed out in Figure~\ref{fig:heat}.

Considering the observational parameters in Figure~\ref{fig:heat} (left panel) it becomes apparent that galaxies brighter in the continuum (irrespective of the band) are also almost always brighter in the rest-frame optical emission lines ($\tau \gtrsim 0.5$, $p_\tau \lesssim 10^{-3}$) and also in the Ly$\alpha$ line.  This ``more is more'' effect regarding $L_\mathrm{Ly\alpha}$ might be specific to the here analysed sample that excludes Ly$\alpha$ absorbers.  Regarding our kinematical parameters, only correlations between $\sigma_0^\mathrm{obs}$ and the emission lines, including Ly$\alpha$, appear significant\footnote{The relation between $\sigma_0^\mathrm{obs}$ (and also $\sigma_\mathrm{tot}$) and optical emission lines in star-forming galaxies is well known \citep{Terlevich1981}, albeit its physical origins are debated. Nevertheless, these trends have been suggested as an alternative cosmological standard candle \citep[e.g.,][and references therein]{Gonzalez-Moran2019}.}.

Focusing on the physical parameters in Figure~\ref{fig:heat}, we see that $M_\star$ shows the highest number of intra-correlations with other parameters; only $\lg (O/H)$ and $E(B-V)$ are found not to be correlated with $M_\star$.  Numerous other significant intra-correlations exist amongst the physical parameters.  Amongst the kinematical parameters $v_\mathrm{shear}/\sigma_0^\mathrm{obs}$ correlates with $v_\mathrm{shear}$ and anti-correlates with $\sigma_0^\mathrm{obs}$.  We also note that all Ly$\alpha$ observables correlate with each other.  While some of the uncovered monotonic relations are certainly imprints of astrophysical scaling relations, other correlations or non-correlations might be specific to the sample at hand.  We will not discuss this further here, as this is not the main aim of our study.  For us of interest is, that the high number of intra-correlations highlight the complex interdependence between the individual parameters in the observational and in the physical parameter sets.  This complex network of relationships amongst the analysed parameters complicates the problem of assessing parameter importance, but it also highlights the need for a more sophisticated approach than just looking at parameter pairings.

We opt for a simple heuristic approach based on multi-variate regression,
\begin{equation}
  \label{eq:2}
  y_i = \sum_{i=1}^{N_p} c_i x_i \;\text{,}
\end{equation}
where the $x_i$ are the $N_p=12$ parameters from the observational- and physical parameter sets, respectively, including the kinematical parameters.  The $y_i$ in Eq.~\eqref{eq:2} are the responses, i.e. here the Ly$\alpha$ observables, and the $c_i$ denote the coefficients maximising the coefficient of determination,
\begin{equation}
  \label{eq:3}
  R^2 = 1 - \frac{\sum_{i=1}^N (y_i - f_i)^2}{\sum_{i=1}^N (y_i - \langle y_i \rangle)^2} \;\text{,}
\end{equation}
where the $f_i$ are the predictions of the observable $y_i$ and $\langle y_i \rangle$ is the mean over all $y_i$.  Formally, $R^2$ encapsulates the variance explained by the empirical model of Eq.~\eqref{eq:2}.  If $R^2 \sim 0$, the linear relation in $x_i$ is not usable to describe and predict the $y_i$, whereas for $R^2 \sim 1$ Eq.~\eqref{eq:2} is a nearly perfect empirical model.

In above framework we apply a forward and backward selection technique (also known as step-wise regression) to establish importance rankings for the parameters.  Here we iteratively include (forward selection) or exclude (backward selection) parameters from Eq.~\eqref{eq:2}, such that the included or excluded parameter leads to the largest increase or smallest decrease of $R^2$, respectively.  Moreover, this procedure requires that the parameters are standardised to a comparable scale.  As in \cite{Hayes2023b} we here use Z-score normalisation i.e., we subtract the mean over all values for each parameter and then we divide the mean shifted values by their standard deviation.  Finally, we account for the statistical errors on the parameters and responses by performing $N_\mathrm{MC}=10^3$ Monte-Carlo realisations of the above procedure.  In each realisation we perturb both the response and the parameters according to their statistical uncertainties.  This procedure provides us then with $N_\mathrm{MC}$ ranks for each parameter $x_i$ under forward- and backward selection; we write for these $j$ ranks $r_j^\mathrm{FW}(x_i)$ and $r_j^\mathrm{BW}(x_i)$, respectively, where $j=1,\dots,N_\mathrm{MC}$.

We find that in almost all realisations the rankings obtained under forward selection are not completely consistent with the rankings obtained under backward selection.  This so called ``violation of anonymity'' of the parameters regarding their positions in the regression is driven by the parameter intra-correlations (Figure~\ref{fig:heat}).  It is a fundamental flaw of step-wise regression \citep{Groemping2015}.  This flaw also remains in bi-directional step-wise regression, where parameters are included and excluded in each iteration \citep{Smith2018}.  A potential alternative are more computational intensive methods related to game theory, which assign relative importance to a parameter based on all possible permutations that a parameter can have with all other parameters \citep{Groemping2015}.  However, given that we also require to include the measurement uncertainties via Monte Carlo simulations in our importance estimate, this approach is computationally expensive and will be the subject of future work.  Here we instead opt for an ad-hoc definition of a final ranking score.  With
\begin{equation}
  \label{eq:4}
  r(x_i) = \frac{1}{2 N_\mathrm{MC}} \sum_{j=1}^{N_\mathrm{MC}}
  r_j^\mathrm{FW}(x_i) + r_j^\mathrm{BW}(x_i)
\end{equation}
we define the rank importance $r(x_i)$ of parameter $x_i$ as the average over all ranks obtained under forward and backward selection for a particular Monte Carlo realisation.  The smaller $r(x_i)$, the more important is $x_i$.  The so obtained order of importance for the responses $\mathrm{EW}_\mathrm{Ly\alpha}$, $f_\mathrm{esc}^\mathrm{Ly\alpha}$, and $L_\mathrm{Ly\alpha}$ are provided in Table~\ref{tab:cp}.

Finally, following \cite{Hayes2023b}, we visualise in Figure~\ref{fig:cp} the probability distributions of all ranks $r(x_i)$, $p\left(r(x_i)\right)$ by means of a Gaussian kernel-density estimator (kernel size 0.5).  We highlight the four curves with the smallest $r(x_i)$ in Figure~\ref{fig:cp}, as these are formally the top ranked parameters according to Eq.~\eqref{eq:4}.  Moreover, we deem a high ranking of a parameter as robust if the $p\left(r(x_i)\right)$ curve for a given parameter peaks at a low $r$.  Contrariwise, the ranking of a parameter is not considered as robust if the curve shows a flat distribution that encompasses a larger number of possible ranks.  As can be seen from Figure~\ref{fig:cp}, and as we will discuss below, for most of the parameters no clear order of importance can be established, even if a parameter is formally ranked high according to Eq.~(\ref{eq:4}).  However, for some parameters a clear picture emerges.

The most notable and robust result from this statistical experiment is, that $\sigma_0$ is found as the most important parameter in determining $EW_\mathrm{Ly\alpha}$ (top panels of Figure~\ref{fig:cp}).  In fact, in almost all Monte-Carlo realisations $\sigma_0^\mathrm{obs}$ emerges as the top ranked parameter, both under forward and backward selection\footnote{For this reason we have to visualise the probability density $p(r(\sigma_0^\mathrm{obs}))$ in Figure~\ref{fig:cp} as a histogram, since the kernel density estimator would produce a highly peaked narrow curve, whose visualisation would render the other curves indistinguishable. \label{fn:1}}.  The dominating importance of $\sigma_0$ in regulating $\mathrm{EW}_\mathrm{Ly\alpha}$ is thus found both from the marginal perspective (Sect.~\ref{sec:mp}) and from the conditional perspective.  We will discuss this result in Sect.~\ref{sec:astr-interpr} below.  Moreover, the distribution for the parameters ranked formally below $\sigma_0^\mathrm{obs}$ do not show significant peaks.  We thus regard the order of those parameters obtained by Eq.~\eqref{eq:4} not as robust indications of their relative importance with respect to $\mathrm{EW}_\mathrm{Ly\alpha}$.  The average $R^2$ for all 12 parameters regarding the response $\mathrm{EW}_\mathrm{Ly\alpha}$ is $0.55\pm 0.04$ and $0.50 \pm 0.04$ for the observational- and physical parameter set, respectively.

The distributions that we recover for the observational parameters with respect to $f_\mathrm{esc}^\mathrm{Ly\alpha}$ appear completely nondescript (centre left panel in Figure~\ref{fig:cp}).  However, here a clearer picture emerges regarding the role of the physical parameter with respect to $f_\mathrm{esc}^\mathrm{Ly\alpha}$ (centre right panel in Figure~\ref{fig:cp}).  Here the four top ranked parameters -- $E(B-V)$, $O_{32}$, $F_\mathrm{cov}$, and $\sigma_\mathrm{obs}$ -- are characterised by single peaked distributions in the left side of the diagram, whereas the remaining parameters show flattened out distributions towards the right side.  For both parameter sets we recover an average $R^2 = 0.60 \pm 0.05$.

Attending the bottom panels of Figure~\ref{fig:cp} we see that for $L_\mathrm{Ly\alpha}$ the $I$-band luminosity arises as a top ranked observational parameter, with the other parameters not yielding conclusive rankings.  Here $R^2 = 0.966 \pm 0.003$, but this high $R^2$ is a result of the strong positive correlations of almost all parameters with $L_\mathrm{Ly\alpha}$.  In the physical parameter set the top three ranked parameters -- $O_{32}$, $M_\star$, and $F_\mathrm{cov}$ -- separate themselves distinctly from the remaining parameters.  The $R^2$ for the physical parameter set is $0.64 \pm 0.05$.

\subsection{Astrophysical Interpretation}
\label{sec:astr-interpr}

Our analysis above leads to the conclusion, that $\sigma_0^\mathrm{obs}$ appears as a dominating parameter in regulating $\mathrm{EW}_\mathrm{Ly\alpha}$ when compared to other observational and physical parameters.  This result is found both from the marginal- (Sect.~\ref{sec:mp}) and from the conditional perspective (Sect.~\ref{sec:cp}).  A slightly different picture emerges regarding the role of $\sigma_0^\mathrm{obs}$ with respect to $f_\mathrm{esc}^\mathrm{Ly\alpha}$.  Here no direct correlation is found (Sect.~\ref{sec:corr-betw-glob} and Sect.~\ref{sec:mp}), i.e. $\sigma_0^\mathrm{obs}$ appears unimportant for $f_\mathrm{esc}^\mathrm{Ly\alpha}$ when taken on its own, however it is found as an important parameter in regulating $f_\mathrm{esc}^\mathrm{Ly\alpha}$ when considered in concert with other parameters (Sect.~\ref{sec:cp}).  These parameters are the neutral gas covering along the line of sight traced by $F_\mathrm{cov}$, the ionisation state traced by $O_{32}$, and the dust extinction traced by $E(B-V)$.  That these three other parameters are also expected and known to influence the escape of Ly$\alpha$ photons provides credibility to our method and result.

Unmistakably, Eq.~\eqref{eq:2} from Sect.~\ref{sec:cp} can not replace an astrophysical model.  This statistical method just helps to uncover trends in the data beyond the conventional two-parameter correlation analyses.  The result from our statistical analysis requires us now to understand the dominant role of $\sigma_0^\mathrm{obs}$ for $\mathrm{EW}_\mathrm{Ly\alpha}$ and the role of $\sigma_0^\mathrm{obs}$ in modulating $f_\mathrm{esc}^\mathrm{Ly\alpha}$.

\subsubsection{The role of the age of the stellar population}
\label{sec:sigma_vs_age}

\begin{figure*}
  \centering
  \includegraphics[width=\textwidth]{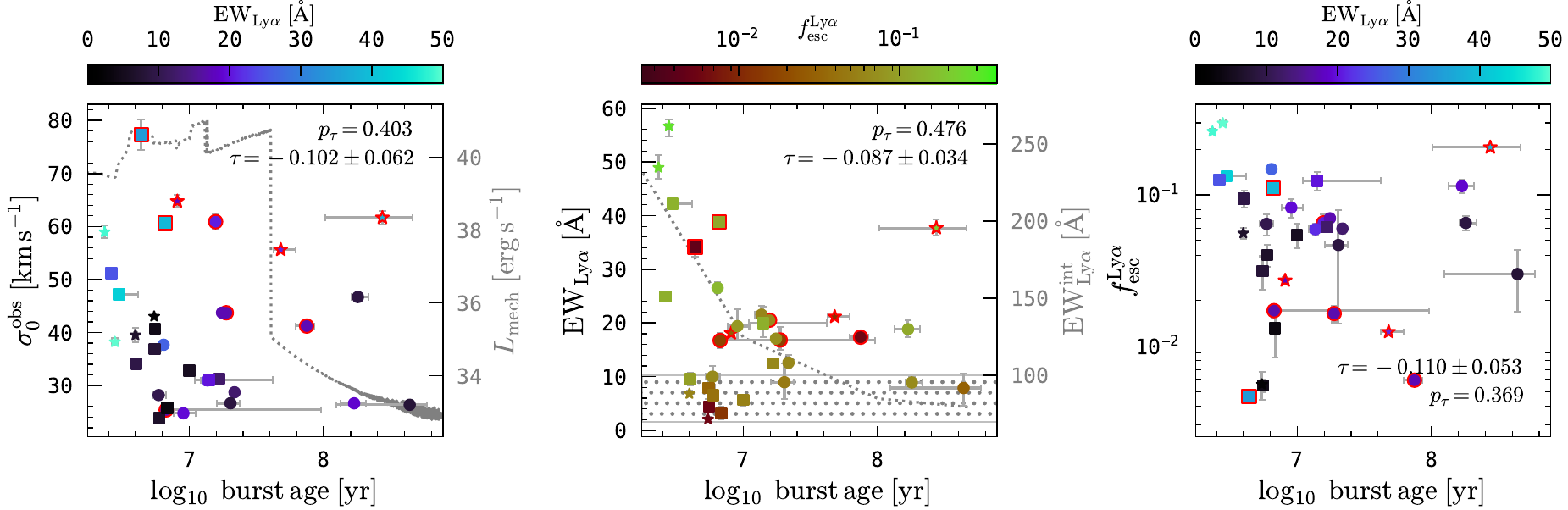}
  \caption{Relations between $\sigma_0^\mathrm{obs}$ (\emph{left panel}), $\mathrm{EW}_\mathrm{Ly\alpha}$ (\emph{middle panel}), as well as $f_\mathrm{esc}^\mathrm{Ly\alpha}$ (\emph{bottom panel}) with the dust-corrected burst age.  Kendall's $\tau$ and the respective $p$-Value are provided in the bottom left of each panel. Different symbols are used for rotating disks (circles), perturbed rotators (squares), and galaxies exhibiting complex kinematics (stars).
We colour code the points by $\mathrm{EW}_\mathrm{Ly\alpha}$ (left and right panel) or $f_\mathrm{esc}^\mathrm{Ly\alpha}$ (middle panel).  We also highlight points with red outlines that show $L_\mathrm{Ly\alpha} / L_\mathrm{H\alpha}$ above the maximum dust screen expectation from Eq.~\eqref{eq:5}. As visual aid for the interpretation we show model calculations for the mechanical luminosity, $L_\mathrm{mech}$, and the intrinsic Ly$\alpha$ equivalent width, $\mathrm{EW}_\mathrm{Ly\alpha}^\mathrm{int}$, as a function of burst age with grey dotted lines.  We note that the ticks on the right side of each panel provide the scale for these model calculations.  In particular, we show the $L_\mathrm{mech}$ of a $Z = 0.008$ starburst normalised to 1\,M$_\odot$ in the left panel and we indicate the temporal evolution of  $\mathrm{EW}_\mathrm{Ly\alpha}^\mathrm{int}$ for a $Z=0.008$ burst and the $\mathrm{EW}_\mathrm{Ly\alpha}^\mathrm{int}$ range for constant star-formation at $Z = 0.001 - 0.040$ in the middle panel (see text for details).}
  \label{fig:age}
\end{figure*}

The $\sigma_0^\mathrm{obs}$ values in our sample are all significantly above the expected combined velocity dispersions due to thermal and turbulent motions of the individual \ion{H}{ii} regions ($\sigma_\mathrm{therm} = \sqrt{2 k_\mathrm{B} T / m_\mathrm{H}} = 12.85$\,km\,s$^{-1}$ for $T=10^4$\,K).  Therefore, the observed large values of $\sigma_0^\mathrm{obs}$ are dominated by the turbulence of the cold and/or warm neutral medium in which the unresolved \ion{H}{II} regions are embedded.  The monotonic relation between SFR and $\sigma_0$ (Sect.~\ref{sec:corr-betw-galaxy}) thus indicates a causal connection between star-formation and turbulence of the interstellar medium.

There are competing scenarios regarding the main physical mechanism that are responsible for the SFR vs. $\sigma_0$ relation.  One the one hand, it is suggested that gravitational instabilities, which are required to form stars, may stir-up the interstellar medium and decay as turbulence \citep[e.g.,][]{Krumholz2010}.  On the other hand, feedback from star-formation (i.e., stellar winds and especially supernovae) may act as the main source of energy driving this turbulence \citep[e.g.,][]{Bacchini2020}.  Observational results are often used to argue in favour of latter scenario \citep{Moiseev2015,Law2022}, but there are also results that favour the former scenario \citep[e.g.][]{Krumholz2016,Yu2021}.  Recent modelling by \cite{Ejdetjarn2022} indicates that the warm-ionised phase traced by H$\alpha$ may actually also effectively couple to kinematic energy injected by feedback.

Adopting feedback as the driving force beyond $\sigma_0$ as a working hypothesis we recall that for a simple stellar population the amount of energy injected by stellar winds and supernovae is related to the duration since the star-burst event (burst age, e.g., \citealt{Leitherer1999}).  Furthermore, we recall that the intrinsic equivalent width, $\mathrm{EW}_\mathrm{Ly\alpha}^\mathrm{int}$, is controlled by the ionising photon production efficiency, $\xi_\mathrm{ion}$, and both $\xi_\mathrm{ion}$ and $\mathrm{EW}_\mathrm{Ly\alpha}$ are controlled by the burst age and metallicity \citep{Charlot1993,Schaerer2003,Raiter2010,Inoue2011}.  Thus, the evolutionary state of the stellar population is inherently linked to Ly$\alpha$ production, and given our working hypothesis and the results from Sect.~\ref{sec:corr-betw-glob} and Sect.~\ref{sec:cp}, should also affect the escape of Ly$\alpha$ photons and modulate the observed equivalent width.  Indeed, this idea finds support by the recent work of \cite{Hayes2023b}, who report a strong anti-correlation between $\mathrm{EW}_\mathrm{Ly\alpha}$ and burst age in their sample of 87 low-$z$ galaxies observed with HST/COS.  These authors also report an equally strong anti-correlation between burst age and $f_\mathrm{esc}^\mathrm{Ly\alpha}$.  If our hypothesis is true, then the effect that burst age has on $\mathrm{EW}_\mathrm{Ly\alpha}$ and $f_\mathrm{esc}^\mathrm{Ly\alpha}$ should also be identifiable as an effect on $\sigma_0^\mathrm{obs}$.

In order to analyse the dependence of $\sigma_0^\mathrm{obs}$, $\mathrm{EW}_\mathrm{Ly\alpha}$, and $f_\mathrm{esc}^\mathrm{Ly\alpha}$ on stellar age, we require estimates of the latter.  The work by \citetalias{Melinder2023} provided age estimates for each Voronoi cell from the HST imaging data.  We here use an average of those ages by weighing with the extinction corrected UV continuum.  We note that our estimate differs from the age averages tabulated in \citetalias{Melinder2023}, as these were only weighted by the extinction uncorrected UV continuum.  As shown in Appendix~\ref{sec:extinct-corr-starb}, where we also provide a table of the extinction corrected ages, for the majority of galaxies we now obtain significantly younger ages since the dust-uncorrected ages did not account for the very young populations enshrouded in dust.

We visualise the dependence of $\sigma_0^\mathrm{obs}$, $\mathrm{EW}_\mathrm{Ly\alpha}$, and $f_\mathrm{esc}^\mathrm{Ly\alpha}$ on the so derived burst ages in Figure~\ref{fig:age}.  None of the three parameters are found to correlate with burst age.  Hence, the anti-correlations between burst-age and $\mathrm{EW}_\mathrm{Ly\alpha}$ or $f_\mathrm{esc}^\mathrm{Ly\alpha}$, that were recovered in the sample of \cite{Hayes2023b}, are not found in our sample.  However, a comparison with this sample is not on an equal footing.  The crucial methodological difference is, that \cite{Hayes2023b} estimated stellar ages directly from fitting stellar templates to UV and optical spectra, whereas our ages are derived photometrically.  Especially the UV spectra contain many age sensitive features \citep[e.g.,][]{Chisholm2019}.  Nevertheless, the lack of a dirrect correlation between starburst age and $\sigma_0^\mathrm{obs}$ may indicate that kinetic energy released from the stellar population and the conversion to turbulent energy follows more complex patterns.

To analyse this further we plot in Figure~\ref{fig:age} (upper panel) the evolution of mechanical luminosity (normalised to 1\,M$_\odot$) according to the STARBURST99 stellar population model \citep{Leitherer1999} for a $Z=0.008$ population (with a power law initial mass function of slope $\alpha = -2.35$ and upper mass cut-off at 100\,M$_\odot$).  To exemplify how $\mathrm{EW}_\mathrm{Ly\alpha}^\mathrm{int}$ depends on the burst-age we plot in the in the middle panel of Figure~\ref{fig:age} the $\mathrm{EW}_\mathrm{Ly\alpha}^\mathrm{int}$ evolution with time for a star-burst with $Z=0.008\,Z_\odot$ according to the calculations by \cite{Inoue2011}.  We also show the range of $\mathrm{EW}_\mathrm{Ly\alpha}^\mathrm{int}$ for similar types of low-$Z$ populations with constant star formation from the calculations of \cite{Schaerer2003}.  Considering the idealised star-burst, where all stars form instantaneously, initally winds from the highest mass stars (O stars) dominate the mechanical energy output.  During these first few $10^6$ years after the burst, the intrinsic Ly$\alpha$ equivalent widths are expected to be very high, with $\mathrm{EW}_\mathrm{Ly\alpha}^\mathrm{int} \sim 300$\,\AA{} at the typical metallicities ($Z\sim10^{-2}$) of our sample.  After $\approx 5 \times 10^6$ years the injection of kinematic energy will increase substantially due to the most massive stars going supernova.  The absence of ionising photons from these stars in otherwise very blue continua will lead to a gradual decrease of $\mathrm{EW}_\mathrm{Ly\alpha}^\mathrm{int}$.  At $\approx 10^7$ years the O-stars are gone, the ionising luminosity drops by a factor of $\sim 100$ compared to the initial phase of the burst, and the $\mathrm{EW}_\mathrm{Ly\alpha}^\mathrm{int}$ decreases even further.  At this phase the now the injected kinetic energy stems only from stellar winds of lower mass stars.

Of course, the galaxies here consist of multiple clusters of different ages. Moreover, turbulent energy dissipates over varying timescales \citep[$\sim 10 - 100$\,Myr; e.g.,][]{Bacchini2020}, so the link between mechanical luminosity or $\mathrm{EW}_\mathrm{Ly\alpha}^\mathrm{int}$ with burst age in Figure~\ref{fig:age} serves only as a schematic illustration.  Still, in such a simple scenario, we would expect the lowest $\sigma_0^\mathrm{obs}$ values for the oldest population, but this is not what we observe.  Also the trend for the EWs appears more nuanced.  Except for one outlier, this plot is characterised by a zone of avoidance of old objects with high-EWs.  Moreover, the highest EWs indeed appear at the youngest ages.  Lastly, no discernible trend appears in the $f_\mathrm{esc}^\mathrm{Ly\alpha}$ plot.  We conclude, that we do not recover the anticipated anti-correlation of burst-age on $\mathrm{EW}_\mathrm{Ly\alpha}^\mathrm{int}$ and $f_\mathrm{esc}^\mathrm{Ly\alpha}$, nor can we establish a causal relation between burst-age and $\sigma_0^\mathrm{obs}$.

\subsubsection{Gas kinematics can shape dust geometry}
\label{sec:interpl-betw-kinem}

\begin{figure}
  \centering
  \includegraphics[width=0.45\textwidth]{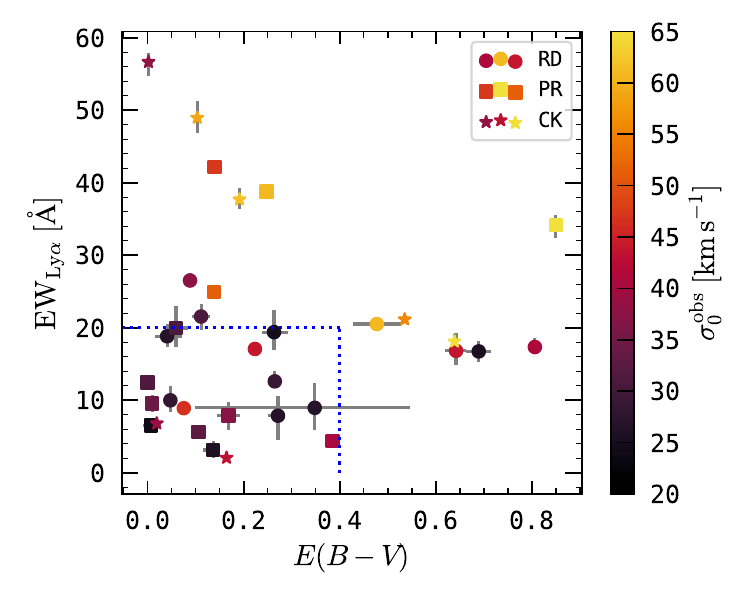}
  \caption{Ly$\alpha$ equivalent width, $\mathrm{EW}_\mathrm{Ly\alpha}$, vs. nebular extinction, $E(B-V)$.  Points are coloured according to $\sigma_0^\mathrm{obs}$ as indicated by the colour bar on the right.  Different symbols indicate the kinematical classes of the galaxies introduced in Sect.~\ref{sec:visu-class-veloc}.  The dotted blue line indicates the zone of $\mathrm{EW}_\mathrm{Ly\alpha} \leq 20$\,\AA{} and $E(B-V) \leq 0.4$ discussed in the text.   There is no correlation between both quantities ($\tau = -0.049$ / $p_\tau = 0.7$).}
  \label{fig:ewebv}
\end{figure}

\begin{figure}
  \centering
  \includegraphics[width=0.49\textwidth]{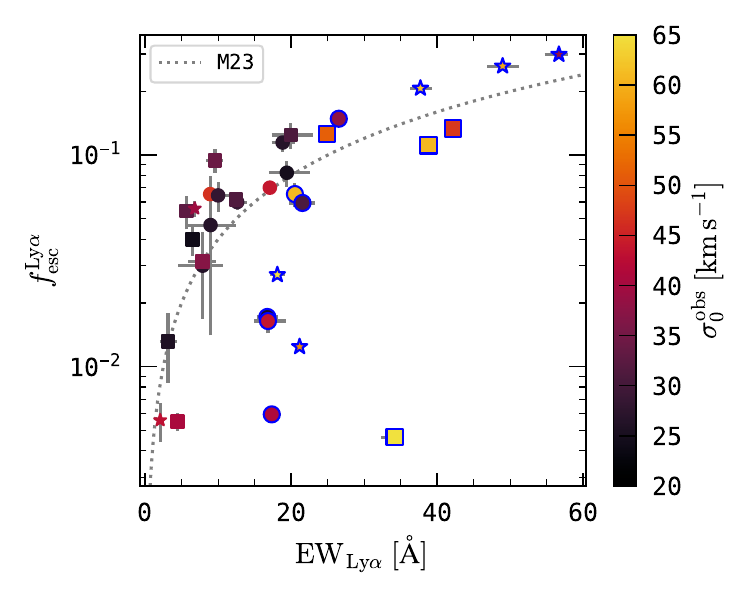}
  \caption{Ly$\alpha$ escape fraction, $f_\mathrm{esc}^\mathrm{Ly\alpha}$ vs. Ly$\alpha$ equivalent width, $\mathrm{EW}_\mathrm{Ly\alpha}$ for the in Sect.~\ref{sec:disc} analysed sub-sample of 33 galaxies as in Figure~\ref{fig:ewebv}.  Both quantities are highly correlated ($\tau = 0.453$ / $p_\tau \sim 10^{-3}$), even in a linear sense (Pearson's $r=0.78$ / $p_r \simeq 10^{-7}$).  The dotted line shows the linear fit to the whole sample from \citetalias{Melinder2023}.  Points are coloured according to $\sigma_0^\mathrm{obs}$ as indicated by the colour bar on the right.  Symbols are the same as in Fig.~\ref{fig:ewebv}, but here we highlight by blue outlines the objects outside of the $\mathrm{EW}_\mathrm{Ly\alpha}$ and $E(B-V) \leq 0.4$ zone in Figure~\ref{fig:ewebv}.}
  \label{fig:ewfesc}
\end{figure}

\begin{figure}
  \centering
  \includegraphics[width=0.4\textwidth]{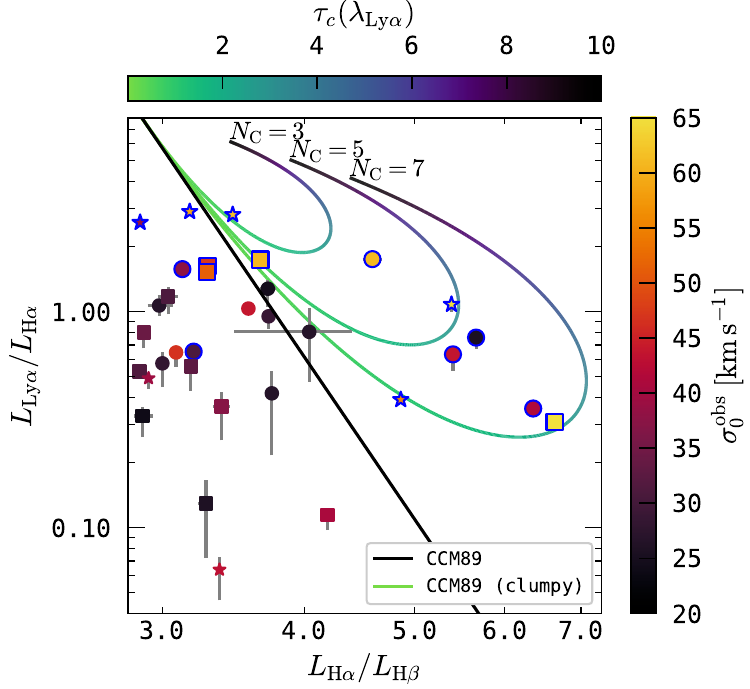}
  \caption{Ly$\alpha$/H$\alpha$ ratio against the H$\alpha$/H$\beta$ luminosity ratio for the in Sect.~\ref{sec:disc} analysed sub-sample of 33 galaxies as.  there is no correlation between both quantities ($\tau = -0.09$ / $p_\tau = 0.47$).  Symbols, colouring, and highlighting of the symbols is the same as in Figure~\ref{fig:ewfesc}.  The expectation of Ly$\alpha$/H$\alpha$ as a function of H$\alpha$/H$\beta$ for a homogeneous dust-screen without radiative transfer effects in front of the line emission is shown as a solid line (Eq.~\ref{eq:5}).  The curves with a colour gradient indicate possible values for Ly$\alpha$/H$\alpha$ and H$\alpha$/H$\beta$ for a clumpy dust screen, where on average $N_\mathrm{C}=\{3, 7, 9\}$ clumps block the sight lines (Eq.~\ref{eq:8}).  These lines are drawn with a colour gradient that indicates the dust optical depth $\tau_C$ at 1216\,\AA{} of a single clump (colour bar at the top of the plot).}
  \label{fig:lyahahahb}
\end{figure}

We analyse in Figure~\ref{fig:ewebv} the distribution of our sample in the $\mathrm{EW}_\mathrm{Ly\alpha}$ vs. $E(B-V)$ plane.  In this diagram we find a zone of avoidance: the five objects with the highest $\mathrm{EW}_\mathrm{Ly\alpha}$ measurements are found at low extinction ($E(B-V) < 0.4$), whereas the objects with lower $\mathrm{EW}_\mathrm{Ly\alpha}$'s occupy the full range in $E(B-V)$.  The distribution of galaxies in the $\mathrm{EW}_\mathrm{Ly\alpha}$ vs. $E(B-V)$ plane appears similar as in the compilation of low-$z$ starbursts presented by \cite{Atek2014}.  We colour code the points in Figure~\ref{fig:ewebv} by $\sigma_0^\mathrm{obs}$ to visualise the influence of this parameter on the observed $\mathrm{EW}_\mathrm{Ly\alpha}$ vs. $E(B-V)$ distribution.  While $\mathrm{EW}_\mathrm{Ly\alpha}$ is strongly correlated with $\sigma_0^\mathrm{obs}$ (Sect.~\ref{sec:corr-betw-glob} and Sect.~\ref{sec:mp}), we formally have to accept the null-hypothesis that $\sigma_0^\mathrm{obs}$ and $E(B-V)$ are not correlated (see also Figure~\ref{fig:heat}, numerically $\tau = 0.223$ / $p_0 = 0.07$).  Still, in the $E(B-V)$ vs. $\mathrm{EW}_\mathrm{Ly\alpha}$ plane it is apparent, how the corner of low extinction objects ($0 \leq E(B-V) \leq 0.4$) with low Ly$\alpha$ equivalent widths ($\mathrm{EW}_\mathrm{Ly\alpha} \leq 20$\,\AA{}) appears to be dominated by galaxies with lower $\sigma_0^\mathrm{obs}$, whereas objects outside of this corner appear to be characterised by higher $\sigma_0^\mathrm{obs}$.  These boundaries may appear somewhat arbitrary, but $\mathrm{EW}_\mathrm{Ly\alpha} \geq 20$\,\AA{} represents the canonical cut to define LAEs, while $E(B-V) = 0.4$ halves the $E(B-V)$ range probed by our sample.   We find that 18 objects with $\mathrm{EW}_\mathrm{Ly\alpha} \leq 20$\,\AA{} and $0 \leq E(B-V) \leq 0.4$ show a mean (median) $\sigma_0^\mathrm{obs}$ of 32.8\,km\,s$^{-1}$ (31.2\,km\,s$^{-1}$), whereas the other 15 galaxies are characterised by a mean (median) $\sigma_0^\mathrm{obs}$ of 50.3\,km\,s$^{-1}$ (51.2\,km\,s$^{-1}$).  Thus, excluding objects in this low-$\sigma_0^\mathrm{obs}$ and low-$E(B-V)$ region leaves us with objects that show simultaneously $\mathrm{EW}_\mathrm{Ly\alpha} > 20$\,\AA{} and high velocity dispersions, but we also find galaxies with higher dust extinction that still show appreciable Ly$\alpha$ equivalent widths ($\mathrm{EW}_\mathrm{Ly\alpha} \approx 20$\,\AA{}) and, intriguingly, their intrinsic velocity dispersions are also high.

Next, we analyse in Figure~\ref{fig:ewfesc} the distribution of our sample in the $f_\mathrm{esc}^\mathrm{Ly\alpha}$ vs. $\mathrm{EW}_\mathrm{Ly\alpha}$ plane.  There we also show the linear relation $f_\mathrm{esc}^\mathrm{Ly\alpha} \propto 4\times10^{-3} \, \mathrm{EW}_\mathrm{Ly\alpha}$ from \citetalias{Melinder2023}\footnote{The slope from \citealt{Melinder2023} is identical to the slope of the $z \in \{0,0.3\}$ sample from \cite{Sobral2019}, but slightly shallower than then the slope found by these authors at higher redshifts ($f_\mathrm{esc}^\mathrm{Ly\alpha} \propto 5\times10^{-3} \, \mathrm{EW}_\mathrm{Ly\alpha}$).}.  As in Figure~\ref{fig:ewebv}, we colour code the points by $\sigma_0^\mathrm{obs}$.  We furthermore now highlight objects outside of the low-$\mathrm{EW}_\mathrm{Ly\alpha}$ -- low-$E(B-V)$ region defined in Figure~\ref{fig:ewebv}.  We see that the dominant fraction of objects that are below the $f_\mathrm{esc}^\mathrm{Ly\alpha} \propto 4\times 10^{-3} \mathrm{EW}_\mathrm{Ly\alpha}$ line are outside this $E(B-V) \leq 0.4$ and $\mathrm{EW}_\mathrm{Ly\alpha} \leq 20$\,\AA{} box of Figure~\ref{fig:ewebv} (9 out of 12 objects).  Thus, galaxies that have low $f_\mathrm{esc}^\mathrm{Ly\alpha}$ but higher $\mathrm{EW}_\mathrm{Ly\alpha}$ are characterised by high $\sigma_0^\mathrm{obs}$.

Interestingly, almost all of the outliers below the $f_\mathrm{esc} \propto 4 \times 10^{-3} \mathrm{EW}_\mathrm{Ly\alpha}$ relation appear incompatible with extinction by a homogeneous dust screen.  This can be appreciated in Figure~\ref{fig:lyahahahb} where we plot the Ly$\alpha$/H$\alpha$ luminosity ratio against the Balmer decrement, H$\alpha$/H$\beta$, for our sample.  In Figure~\ref{fig:lyahahahb} we also indicate the expectation of Ly$\alpha$/H$\alpha$ as a function of H$\alpha$/H$\beta$ for extinction by a homogeneous dust-screen in the absence of radiative transfer effects:
\begin{equation}
  \label{eq:5}
  \frac{L_\mathrm{Ly\alpha}}{L_\mathrm{H\alpha}} =  \left(\frac{L_\mathrm{Ly\alpha}}{L_\mathrm{H\alpha}}\right)_\mathrm{int} \cdot \frac{e^{-\tau_{d,\mathrm{Ly\alpha}}}}{e^{-\tau_{d,\mathrm{H\alpha}}}}\;\text{.}
\end{equation}
Here
\begin{equation}
  \label{eq:6}
  \tau_{d,\mathrm{Ly\alpha}} = 0.4 \cdot \ln(10) \cdot k(\lambda_\mathrm{Ly\alpha}) \cdot E(B-V)
\end{equation}
and
\begin{equation}
  \label{eq:7}
  \tau_{d,\mathrm{H\alpha}} = \tau_\mathrm{Ly\alpha} \cdot \left(k(\lambda_\mathrm{H\alpha})/k(\lambda_\mathrm{Ly\alpha})\right)
\end{equation}
are the dust optical depths at $\lambda_\mathrm{Ly\alpha}$ and $\lambda_\mathrm{H\alpha}$, respectively; the definitions for $(L_\mathrm{Ly\alpha} / L_\mathrm{H\alpha})_\mathrm{int}$, $k(\lambda)$, and $E(B-V)$ were given in Sect.~\ref{sec:glob-meas-from}.

As discussed extensively in the literature, the intuition that the observed Ly$\alpha$/H$\alpha$ is always below the ratio given in Eq.~\eqref{eq:5} due to the loss of Ly$\alpha$ photons by radiative transfer effects proves often to be wrong (\citealt{Scarlata2009}; \citealt{Atek2014}; \citealt{Hayes2014}; \citealt{Bridge2018}; \citetalias{Melinder2023}).  As can be seen from Figure~\ref{fig:lyahahahb}, here we have 9 objects in our sample of 33 galaxies that are significantly above the pure dust attenuation expectation of Eq.~\eqref{eq:5}.  We also see in this figure, that these 9 galaxies all are outside the $\mathrm{EW} \leq 20$\,\AA{} and $E(B-V) \leq 0.4$ box from Figure~\ref{fig:ewebv}.  Moreover, 8 of those 9 galaxies are below the $f_\mathrm{esc}^\mathrm{Ly\alpha} \propto 10^{-3} \mathrm{EW}_\mathrm{Ly\alpha}$ line from Figure~\ref{fig:ewfesc}.  Thus, galaxies that are not compatible with extinction by a homogeneous dust screen are characterised by high intrinsic velocity dispersions.

\cite{Scarlata2009} point out that galaxies where the measured Ly$\alpha$/H$\alpha$ is higher than the expectation of Eq.~\eqref{eq:5}, can be explained by obscuration by a system of dusty clumps, where on average $N_\mathrm{C}$ clumps obscure the lines of sight.  For this geometrical scenario \cite{Natta1984} derive the effective dust optical depth
\begin{equation}
  \label{eq:8}
  \tau_\mathrm{eff}(\lambda) = N_C \cdot \left ( 1 - e^{-\tau_c(\lambda)} \right ) \; \text{,}
\end{equation}
where $\tau_c(\lambda)$ is the dust optical depth of a single clump.  In Figure~\ref{fig:lyahahahb} we plot the curves from this scenario parameterised by $\tau_c(\lambda_\mathrm{Ly\alpha})$ for $N_C = \{3, 5, 7\}$.  Here we assume that the dust-extinction of the individual clumps obey also the \cite{Cardelli1989} law that was assumed for homogeneous screen.  The notable feature of these curves is that they describe loops:  With increasing dust optical depth the individual clumps start to significantly block emission over all involved lines and  while the overall intensities decrease, with $\lim_{\tau_c \rightarrow \infty} \tau_\mathrm{eff}(\lambda) = N_C$, the ratio starts to become dominated by the sightlines that are not obscured by the dusty clumps.  As evident  from Figure~\ref{fig:lyahahahb} all 9 galaxies above the expectation from Eq.~\eqref{eq:5} show Ly$\alpha$/H$\alpha$ ratios that are on or close to the loci associated with clumpy dust.  Our analysis thus implies a connection between the small scale distribution of dust and ionised gas kinematics as traced by $\sigma_0^\mathrm{obs}$.

\subsubsection{The influence of turbulence on $f_\mathrm{esc}^\mathrm{Ly\alpha}$}
\label{sec:lyalpha-radi-transf}

\begin{figure}
  \centering
  \includegraphics[width=0.5\textwidth]{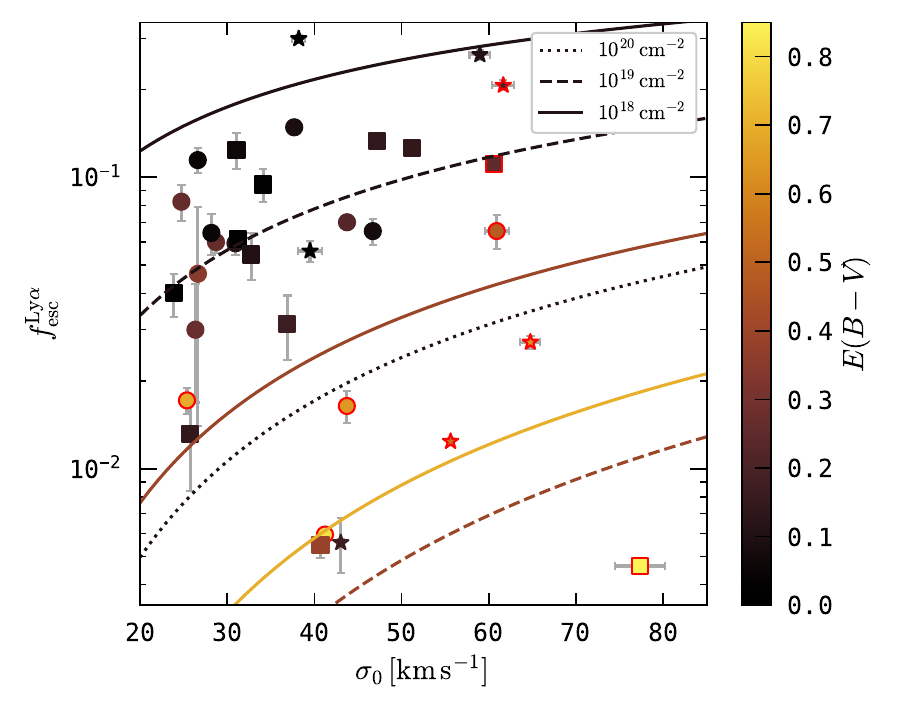}
  \caption{Ly$\alpha$ escape fraction, $f_\mathrm{esc}^\mathrm{Ly\alpha}$, vs. intrinsic velocity dispersion $\sigma_0$.  Solid, dashed, and dotted lines indicate the analytically derived approximate solution, our Eq.~\eqref{eq:11}, for monochromatic Ly$\alpha$ radiation escaping a plane parallel slab of $N_\mathrm{H} = 10^{18}$\,cm$^{-2}$ (solid lines), $10^{19}$\,cm$^{-2}$ (dashed lines), and $10^{20}$\,cm$^{-2}$ (dotted line), respectively.  For $N_\mathrm{H} = 10^{18}$\,cm$^{-2}$ we show curves with $E(B-V) = \{0.1, 0.4, 0.7\}$, for $N_\mathrm{H} = 10^{19}$\,cm$^{-2}$ curves with $E(B-V) = \{0.1, 0.4\}$, and for $N_\mathrm{H} = 10^{20}$\,cm$^{-2}$ the curve for $E(B-V) = 0.1$.  We also show the $f_\mathrm{esc}^\mathrm{Ly\alpha}$ vs. $\sigma_0^\mathrm{obs}$ pairings of the discussed sample.  Scatter points and curves are colour coded by $E(B-V)$ as indicated by the colour bar on the right.  As in Figure~\ref{fig:age} we highlight points by red outlines that show $L_\mathrm{Ly\alpha}/L_\mathrm{H\alpha}$ ratios above the maximum dust screen expectation from Eq.~\eqref{eq:5}}.
  \label{fig:fesigth}
\end{figure}

We close by discussing the anticipated effects of turbulence on Ly$\alpha$ radiative transfer in the very idealised set-up of a homogeneous static medium \citep[e.g.,][]{Neufeld1990,Verhamme2006,Laursen2009}.  This idealised scattering medium is characterised by the density of neutral hydrogen, $n_\mathrm{H}$, or a corresponding neutral column, $N_\mathrm{H} = n_\mathrm{H} s$, along the path length $s$.  Without turbulence the velocities of the atoms follow a Maxwellian distribution with the corresponding thermal velocity dispersion $\sigma_\mathrm{therm} = 12.849$\,km\,s$^{-1} \,\times \sqrt{T_4}$ for a given temperature $T_4$ in $10^4$\,K.  Turbulent motions, characterised by a velocity dispersion $\sigma_\mathrm{turb}$, broaden the velocity dispersion.  In the framework of micro-turbulence this broadening is expressed by a convolution of the velocity distribution of the thermal motions with the velocity distribution of the turbulent motions \citep[e.g.,][]{Verhamme2006,Smith2022b}.  The intrinsic velocity dispersion is thus given by $\sigma_0 = \sqrt{\sigma_\mathrm{therm}^2 + \sigma_\mathrm{turb}^2}$.  Adopting Eq.~(9) from \cite{Verhamme2006} we can express the monochromatic optical depth of this medium at the Ly$\alpha$ line centre as
\begin{equation}
  \label{eq:9}
  \tau_{\lambda_\mathrm{Ly\alpha}} =  4.253 \times 10^{-13} \, \times \, \frac{N_\mathrm{H} \,[\mathrm{cm}^{-2}]}{\sigma_0 \, [\mathrm{km}\,\mathrm{s}^{-1}]}
 \;\text{.}
\end{equation}
Hence, a highly turbulent medium will naturally become less opaque for resonant absorption of Ly$\alpha$ photons.  The reduced optical depth will decrease the number of scatterings before a Ly$\alpha$ photon can escape the medium, and thereby reduce the probability that it will be absorbed by dust.

In fact, there exists an analytically derived approximate solution for $f_\mathrm{esc}^\mathrm{Ly\alpha}$ for the homogeneous static medium, if we assume a monochromatic Ly$\alpha$ source at the centre of a plane-parallel slab and if we assume that this medium is mixed with dust of optical depth $\tau_{d,\mathrm{Ly\alpha}}$ \citep[][]{Neufeld1990}.  Following \cite{Verhamme2006} this solution can be written as
\begin{equation}
  \label{eq:10}
  f_\mathrm{esc}^\mathrm{Ly\alpha} \approx 1 \Bigg / \cosh \left ( \frac{\sqrt{3}}{\pi^{5/12} \zeta } \left [ \left (  (a \tau_{\lambda_\mathrm{Ly\alpha}})^{1/3} \tau_{d,\mathrm{Ly\alpha}} \right )^{1/2} \right ]  \right ) \; \text{.}
\end{equation}
Here $a = A_{21} / 4 \pi \Delta \nu_D$, with
$A_{21} = 6.265\times10^8$\,s$^{-1}$ being the Einstein coefficient of the
Ly$\alpha$ transition and
$\Delta \nu_D = (\sigma_0 / c) \times \nu_\mathrm{Ly\alpha} $ being the Doppler
frequency broadening due to the thermal and turbulent motions
($\nu_\mathrm{Ly\alpha} = 2.466 \times 10^{15}$\,Hz).  Moreover, $\zeta$ in
Eq.~\eqref{eq:10} is a fitting parameter that \cite{Neufeld1990} derive to be
$\zeta \approx 0.525$.  Using Eq.~\eqref{eq:6} and Eq.~\eqref{eq:9} we can rewrite Eq.~\eqref{eq:10} as
\begin{equation}
  \label{eq:11}
  f_\mathrm{esc}^\mathrm{Ly\alpha} \approx  1 \Bigg /
  \cosh \left ( 2.4 \times 10^{-2} \frac{( N_\mathrm{H} \,[\mathrm{cm}^{-2}] )^{1/6} E(B-V)^{1/2}}{( \sigma_0 \, [\mathrm{km\,s}^{-1}])^{2/6}}  \right ) \; \text{.}
\end{equation}
This analysis shows that already in a very idealised set-up no simple monotonic relation between $f_\mathrm{esc}^\mathrm{Ly\alpha}$ and $\sigma_0^\mathrm{obs}$ is expected since additionally variations in neutral column and dust content will factor into the solution in a non-trivial manner.  This result is in line with our finding that $\sigma_0^\mathrm{obs}$ on its own does not affect $f_\mathrm{esc}^\mathrm{Ly\alpha}$ (Sect.~\ref{sec:mp}), but that it emerges as an important parameter in concert with $E(B-V)$ and two other parameters ($F_\mathrm{cov}$ and $O_{32}$ (Sect.~\ref{sec:cp}) that are related to the neutral column.

In Figure~\ref{fig:fesigth} we show curves according to Eq.~\eqref{eq:11} in the $f_\mathrm{esc}^\mathrm{Ly\alpha}$ vs. $\sigma_0$ parameter space for $f_\mathrm{esc}^\mathrm{Ly\alpha}$, $\sigma_0^\mathrm{obs}$, and $E(B-V)$ values that are encountered in our sample.  There we also plot the $f_\mathrm{esc}^\mathrm{Ly\alpha}$ vs. $\sigma_0^\mathrm{obs}$ pairings of our sample colour coded by $E(B-V)$.  We see that the majority of our low-$E(B-V)$ objects are commensurate with Eq.~\eqref{eq:11} for $N_\mathrm{H} \sim 10^{19}$\,cm$^{-2}$ and that high-$E(B-V)$ objects would require lower neutral columns.  However, most of these low-$f_\mathrm{esc}^\mathrm{Ly\alpha}$ high-$E(B-V)$ objects show Ly$\alpha$/H$\alpha$ ratios indicative of a clumpy dust distribution (Sect.~\ref{sec:interpl-betw-kinem}).

We caution not to over overinterpret the comparison to the here presented analytic approximation, as Eq.~\eqref{eq:11} requires a highly idealised set-up.  Our values on $\sigma_0^\mathrm{obs}$ trace the frequency distribution of the intrinsic Ly$\alpha$ radiation field, which is thus far from monochromatic, and the static plane-parallel slab oversimplifies the density and velocity distribution of the scattering medium encountered in reality.  Moreover, \cite{Munirov2023} recently highlighted the limited applicability of the micro-turbulence framework for Ly$\alpha$ radiative transfer.  These authors show how the macroscopic nature of turbulence may significantly alter $\tau_\mathrm{Ly\alpha}$ and thus also the number of scatterings and the spectral distribution Ly$\alpha$ photons emerging from a turbulent medium.  This is because turbulence leads to zones of coherent velocities whose size- and velocity distributions are statistically described.  Hence the simple convolution of the thermal- and turbulent velocity distributions used in the micro-turbulence framework may not be justified in reality.  \cite{Munirov2023} show especially that macroscopic turbulence can decrease the number of scatterings by orders of magnitude before Ly$\alpha$ photons escape the medium.  While this underpins the importance of turbulence in regulating $f_\mathrm{esc}^\mathrm{Ly\alpha}$, quantitative inferences will require observational estimates of the characteristic scale lengths of the turbulent flows.

\section{Conclusions}
\label{sec:conc}

Here we presented an analysis of the ionised gas kinematics obtained from all 42 galaxies that comprise the combined LARS+eLARS samples.  The whole sample was observed in medium spectral resolution with the PMAS integral field unit at the Calar-Alto 3.5m telescope.  We constructed line of sight velocity fields and velocity dispersion maps for each galaxy (Sect.~\ref{sec:line-sight-velocity}) and we here presented the newly obtained kinematic maps for the 28 eLARS galaxies (Figure~\ref{fig:1}).  We combined this sample with the 14 galaxies that were analysed in \citetalias{Herenz2016}.  To date, this sample is the largest kinematical sample of galaxies that are also observed in Ly$\alpha$ emission and, thanks to the comprehensive HST imaging analyses of the sample presented in \citetalias{Melinder2023}, it allowed now for more robust statistical analyses regarding possible trends that were found in \citetalias{Herenz2016}.  The reduced datacubes and the analysed velocity fields are released to the community with this publication.  Our aim here was to check whether the Ly$\alpha$ observables $\mathrm{EW}_\mathrm{Ly\alpha}$, $f_\mathrm{esc}^\mathrm{Ly\alpha}$, and $L_\mathrm{Ly\alpha}$ are affected by galaxy kinematics.
  
Our sample revealed two null results.  First, the Ly$\alpha$ observables are not affected by the morpho-kinematical nature of the velocity fields (Sect.~\ref{sec:comp-lyalpha-observ}).  Second, the inclination of a system in our sample does also not affect the Ly$\alpha$ observables (Sect.~\ref{sec:incl-depend-lyalpha}).  The first null-result is new, and requires verification with larger samples at high redshift.  The second result is in agreement with the null-result seen by photometric inclination estimates in high-$z$ studies.  Still, given the discrepancies between photometric- and kinematic inclinations, verification with kinematical samples at high-$z$ appear desirable.

We proceeded to quantify the global kinematical state of the galaxies by measurements of their intrinsic velocity dispersion, $\sigma_0^\mathrm{obs}$ (corrected for beam-smearing effects; Appendix~\ref{sec:meas-intr-disp}), and their projected maximum velocity amplitude, $v_\mathrm{shear}$ (dubbed shearing velocity, Sect.~\ref{sec:comp-glob-kinem}).  The gas kinematics of the sample are characterised by highly turbulent motions, with $30$\,km\,s$^{-1} \lesssim \sigma_0^\mathrm{obs} \lesssim 80$\,km\,s$^{-1}$.  Moreover, the majority of the shearing velocities in the sample are significantly below the values for disk galaxies.  More than half of the sample (24 galaxies) show dispersion dominated kinematics.

Our main result is that clear trends between the kinematical measurements and the Ly$\alpha$ observables emerge (Sect.~\ref{sec:corr-betw-glob}).  In particular, we find that $\mathrm{EW}_\mathrm{Ly\alpha}$ and $L_\mathrm{Ly\alpha}$ correlate with $\sigma_0^\mathrm{obs}$, whereas $f_\mathrm{esc}^\mathrm{Ly\alpha}$ anti-correlates with $v_\mathrm{shear}$ and $v_\mathrm{shear}/\sigma_0^\mathrm{obs}$.  Moreover, adopting the canonical criterion of $\mathrm{EW}_\mathrm{Ly\alpha} \geq 20$\,\AA{} to define a LAE, 75\,\% of the LAEs in our sample are dispersion dominated systems, i.e. they are characterised by $v_\mathrm{shear}/\sigma_0^\mathrm{obs} < 1.86$.  This result is in agreement with the recent study of high-$z$ galaxies by \cite{Foran2023}.
We also find that $\sigma_0^\mathrm{obs}$ strongly correlates with $L_\mathrm{Ly\alpha}$.  However, this correlation is driven by the normalisation of $L_\mathrm{Ly\alpha}$ with $L_\mathrm{H\alpha}$ in our sample.  We also find a mild, but statistically not significant, trend between $\sigma_0^\mathrm{obs}$ and $L_\mathrm{Ly\alpha}$.  The here uncovered trends are commensurate with our initial hypothesis, that the kinematics play an important role in regulating the Ly$\alpha$ observability of a galaxy.

We then tried to quantify the relative importance of the kinematical parameters with respect to other known observational and physical parameters of the galaxy sample.  Our statistical analyses in Sect.~\ref{sec:mp} and Sect.~\ref{sec:cp} reveals that $\sigma_0^\mathrm{obs}$ emerges as a dominant variable in regulating $\mathrm{EW}_\mathrm{Ly\alpha}$.  We also find that $\sigma_0^\mathrm{obs}$ is of significant importance regarding $f_\mathrm{esc}^\mathrm{Ly\alpha}$ and $L_\mathrm{Ly\alpha}$.  However, no clear trends regarding the importance of $v_\mathrm{shear}$ or $v_\mathrm{shear} / \sigma_0^\mathrm{obs}$ emerged.

We remark that the here used metrics for discussing the parameter importance are relatively simplistic.  Moreover, the here used step-wise approach and also the marginal approach are actually criticised and disfavoured in the modern statistical literature \citep{Groemping2015,Smith2018}.  Future analyses of the problem would benefit from more refined importance metrics, e.g. those reviewed by \cite{Groemping2015}.  Adopting these methods was beyond the scope of the present analysis.  Nevertheless, we suggest that especially the analysis of simple one-to-one correlations might lead to wrong conclusions or over-simplifications regarding the astrophysics of Ly$\alpha$ radiative transport.

While the empirical results are clear, as of yet we do not know what is the dominating mechanism for injecting turbulence and thereby regulating $\sigma_0^\mathrm{obs}$.  This complicates the interpretation why $\sigma_0^\mathrm{obs}$ is found to be of dominant importance in regulating $\mathrm{EW}_\mathrm{Ly\alpha}$.  Motivated by results from stellar-population analysis we discussed a simple scenario in which the age of the starburst simultaneously regulates $\sigma_0^\mathrm{obs}$, $f_\mathrm{esc}^\mathrm{Ly\alpha}$ and $EW_\mathrm{Ly\alpha}$.  However, this scenario does not agree with our data.

That $\sigma_0^\mathrm{obs}$ is of dominating importance for regulating $\mathrm{EW}_\mathrm{Ly\alpha}$ while not being of similar importance for regulating $f_\mathrm{esc}^\mathrm{Ly\alpha}$ deserved a closer look, especially because of the highly significant linear relationship between $\mathrm{EW}_\mathrm{Ly\alpha}$ and $f_\mathrm{esc}^\mathrm{Ly\alpha}$.  In Sect.~\ref{sec:astr-interpr} we discussed that galaxies that do not follow this linear relationship are often better described by a clumpy dust-screen.  But those galaxies were also found to exhibit high $\sigma_0^\mathrm{obs}$.  This suggests a connection between ionised gas kinematics and small-scale dust distribution.

Lastly, we discussed how already in a highly idealised medium a complicated interplay between $N_\mathrm{H}$, $E(B-V)$, and $\sigma_0$ is expected to influence $f_\mathrm{esc}^\mathrm{Ly\alpha}$.  We also highlighted, that the conventional treatment of turbulence as a microscopic effect may underestimate the actual quantitative influence of $\sigma_0$ on $f_\mathrm{esc}^\mathrm{Ly\alpha}$.  Given the here uncovered results it appears worthwhile that future analysis investigate the here uncovered trends, especially the role of turbulence in facilitating Ly$\alpha$ escape, from a modelling and radiative transfer perspective.  Moreover, future spatially resolved kinematic studies of high-$z$ galaxies with and without Ly$\alpha$ in emission appear worthwhile to understand whether our results are indeed applicable to the early universe.


\begin{acknowledgements}
  Based on observations collected at the Centro Astron\'{o}mico Hispano-Alem\'{a}n (CAHA) at Calar Alto, operated jointly by Junta de Andaluc\'{i}a and Consejo Superior de Investigaciones Científicas (IAA-CSIC).  The authors acknowledge financial support by the Centre National D'Etudes
  Spatiales (CNES). \vspace{0.2em} \\
  We acknowledge the use of the following software: ds9 \citep{Joye2003}, QFitsView \citep{Ott2012}, Astropy \citep{AstropyCollaboration2022}, Numpy \citep{Harris2020}, SciPy \citep{Virtanen2020}, matplotlib \cite{Hunter2007}, seaborn \citep{Waskom2021}, cmasher \citep{vanderWelden2020}, GalPak$^\mathrm{3D}$ \citep{Bouche2015}, dust\_extinction
  \citep{gordon_2024_11235336}, and p3d \citep{Sandin2010}.  \vspace{0.2em} \\
  MH is fellow of the Knut \& Alice Wallenberg Foundation. AS acknowledges the support by Jörg Main and the hospitality at Institute for Theoretical Physics 1 (Stuttgart University) to carry out this research. AS's contribution to this project was partly made possible by funding from the Carl-Zeiss-Stiftung.  We thank especially Daniel Kunth for organising meetings at Institut Astrophysique de Paris and for continuous support with the project.  ECH and AS acknowledge travel funds provided by the IAP for attending these meetings.  We thank N. Bouch\'{e} for support regarding the use of GalPak$^\mathrm{3D}$.  We thank Rikke Saust as well as the motivated staff at Calar-Alto observatory for help with the preparation and execution of the visitor mode observations.  ECH acknowledges the generous hospitality of Leiden University for carrying out parts of this research.  ECH acknowledges funding of an internship for AS within the DAAD RISE scheme. ECH dedicates this work in loving memory to a cat named Pixel.
\end{acknowledgements}


\bibliographystyle{aa}
\bibliography{elars_global_kin_resub.bib}


\appendix

\section{PMAS Data Reduction}
\label{sec:pmas-data-reduction}

The raw observational data was reduced with \texttt{p3d}\footnote{Version 2.6.4 obtained from \url{https://p3d.sourceforge.io/.}}  \citep{Sandin2010,Sandin2012} with post-processing by custom developed Python routines.  We outline the steps of the data reduction in the following, noting that a more comprehensive description can be found in \citetalias{Herenz2016}.

First, we create a master bias from the individual bias frames for each night.  Next, we create trace masks and dispersion masks for all continuum- and arc-lamp exposures.  The trace masks are then used to extract 1D spectra for each spectral trace of the target exposures.  For this extraction of the spectral traces from the detector we use the optimal weighting scheme from \cite{Horne1986}.  For the spectral re-binning in the extraction we opt for the ``Drizzle'' algorithm by \cite{Fruchter2002}.  The dispersion masks that control the relevant mapping from detector pixel to wavelength space during the extraction are found by fitting 4\textsuperscript{th} order polynomials to the HgNe lines from the arc-lamp exposures.  Cosmic-rays hits on the detector are rejected with the L.A.Cosmic algorithm of \citet[][]{vanDokkum2001} using the same parameters as in \citetalias{Herenz2016}.  In the final step the extracted data are flat-fielded with the sky-flats that were extracted in the same manner as the science data.

Different to \citetalias{Herenz2016} here we do not flux calibrate the eLARS PMAS data, since an absolute flux calibration is not needed for the kinematic analysis.  For the same reason we also do not subtract telluric emission from the extracted spectra, except for galaxies where the redshifted H$\alpha$ line is in close proximity or even overlapping with a sky line (eLARS 2, and eLARS 22 to 28).

The data for targets where open-shutter time was accumulated over multiple exposures is stacked via simple addition.  The resulting data products from above reduction chain are a row-stacked representation of the 256 spaxels for each science target exposure in analogue-digital-units.  \texttt{p3d} also propagates errors through each step of the reduction chain and stores them in a separate row-stacked spectra array.  These arrays were then reformatted to 3D cuboids using the lookup table that comes with \texttt{p3d}.  The unvignetted spectral range covers $\sim$ 6100\,\AA{} to 7400\,\AA{}, and the spectral sampling is 0.46\,\AA{} per wavelength bin.

We registered the world coordinate system \citep[WCS;][]{Greisen2002} of all eLARS PMAS cuboids against the HST H$\alpha$ images from \cite{Melinder2023}.  We did this visually\footnote{The website \url{https://www.anisotropela.dk/work/lars/hst2pmas} documents this effort.}, taking inspiration from the alignment procedure outlined in \cite{Ferruit2010}.  For the visual alignment we relied on two images.  First, we created surface brightness contours that outline important morphological features visible in H$\alpha$.  These contours were overlaid over a H$\alpha$ narrow-band image that was extracted from the PMAS datacube at a given field of view position.  Secondly, we created a simulated PMAS H$\alpha$ flux map from the HST H$\alpha$ image at a given field of view position.  The field of view position was then adjusted manually until the contours and the simulated observations matched satisfactorily with the observations. The final field of view positions with respect to the HST imaging data are shown in Appendix~A.1 of \cite{Schaible2023}.  We estimate that the relative offset between HST imaging WCS and the so determined PMAS cuboid WCS is better than 0.25\arcsec{}.

For the analysis in the present article we only use this WCS information to resample the datacuboids of the three galaxies that required two pointings onto a common grid (eLARS 3, 5, and 26).  This is achieved with the \texttt{reproject} python package \citep{Robitaille2020}.  However, the presence of a WCS in the PMAS datacuboids of LARS and eLARS, that is relatively accurate with respect to the HST imaging data, may be exploited in future analysis that aim at investigating local relations between kinematics and Ly$\alpha$ observables.

\section{Kinematic map creation}
\label{sec:deta-descr-kinem}

In order to create the kinematic maps shown in Figure~\ref{fig:1} we first subtract a running median in spectral direction (width = 151 wavelength bins) from each spaxel of the datacube.  This removes telluric- and stellar continuum and only the emission line profiles from the galaxies remain.  Next, we model the H$\alpha$ profile in each spaxel with a simple Gaussian profile
\begin{equation}
  \label{eq:12}
  f_i(\lambda \, | \, A_i, \lambda_{0,i}, \sigma_i) = A_i \exp ( - [(\lambda -
\lambda_{0,i})/\sigma_i]^2) \;\text{.}
\end{equation}
The best-fitting $A_i$, $\lambda_{0,i}$, and $\sigma_i$ for each spaxel $i$ are found with the Levenberg-Marquardt algorithm.  For the fitting we use the framework provided by \texttt{astropy.modelling} \citep{AstropyCollaboration2022}.  The initial parameter guesses for the algorithm are derived from the data itself -- for $\lambda_{0,i}$ and $\sigma_i$ via the first- and second moment and for $A_i$ via the peak value of the observed line profile.  For each spaxel the best-fit $\lambda_{0,i}$ and $\sigma_i$ are then converted into $v_{\mathrm{los},i}$ and $\sigma_{\mathrm{obs},i}$ in km\,s$^{-1}$ via
\begin{equation}
  \label{eq:13}
 v_{\mathrm{los},i} = c \, (\lambda_{0,i} - \overline{\lambda}) / \lambda_{0,i} 
\end{equation}
and
\begin{equation}
  \label{eq:14}
 \sigma_{\mathrm{obs},i} = c \,\sqrt{(\sigma_i^2 -
  \sigma_{\mathrm{LSF},i}^2)/\lambda_{0,i}^2} \;\text{,} 
\end{equation}
where $\overline{\lambda}$ is the average of all reliable $\lambda_{0,i}$ fits and $\sigma_{\mathrm{LSF},i} = \mathrm{FWHM}_{\mathrm{LSF},i} / (2 \sqrt{2 \ln 2})$ relates to resolving power $R_i = \lambda_{0,i} / \mathrm{FWHM}_{\mathrm{LSF},i} $ at $\lambda_{0,i}$ for each spaxel $i$.  We remark that cosmic ray hits severely hampered the line fits for some galaxies that were observed only with a single exposure.  We mask the affected spaxels manually.  This results in a few gaps in the velocity field (see, e.g., eLARS 4 and eLARS 11 in Figure~\ref{fig:1} for the worst cases).

To measure $R_i$ we first extract the HgNe arc-lamp exposures that flanked each observation.  Then, we fit Gaussians to the Ne lines of the lamp that are blue- and red of the galaxies H$\alpha$ lines (6717\,\AA{} and 6929\,\AA{}) for each spaxel.  Lastly, we interpolate linearly between these so determined widths to the wavelength $\lambda_{0,i}$ of the galaxies H$\alpha$ emission.  In Table~\ref{tab:log} we list for each target galaxy the average $R$ over all spaxels that have reliable H$\alpha$ profiles.  As already demonstrated in \citetalias{Herenz2016}, $R$ varies typically by a factor 1.5 over the FoV, and the pattern depends on where the telescope was pointing \citepalias[cf. Figure~2 in][]{Herenz2016}.

To quantify whether a fit is reliable or not we estimate the fit parameter uncertainties via a Monte-Carlo simulation.  Here each spaxel is perturbed according to the corresponding error vector from the data reduction and then fitted again.  This process is repeated $10^3$ times and then the standard deviation of the resulting distributions for each parameter $A$, $v_\mathrm{los}$, and $\sigma_0$ define the associated uncertainties $\Delta A$, $\Delta v_\mathrm{los}$, and $\Delta \sigma_0$, respectively.  We define the signal-to-noise ratio for each fit as $S/N = A \sigma / (\Delta A \Delta \sigma)$.  By visually inspecting the quality of the fits we find that spaxels with $S/N \geq 6$ provide us with reliable fits over the whole sample.  By evaluating the statistics of all those reliable fits (3473) via linear regression in $\left (\log_{10}(S/N)\,,\,\log_{10}(\Delta \sigma_v) \right )$-space we obtain the scaling relations $\Delta \sigma_0 = 45.8 \;\mathrm{km\,s}^{-1} \times (S/N)^{-0.93}$ and $\Delta v_\mathrm{los} = 44.6 \;\mathrm{km\,s}^{-1} \times (S/N)^{-0.91}$ for the statistical uncertainties of the line-of-sight kinematical parameters.  These scaling laws can be used in Figure~1 to visually estimate the uncertainties in the $v_\mathrm{los}$ and $\sigma_0$ maps by using their corresponding $S/N$ map.

Moreover, as shown by \cite{Landman1982}, the statistical uncertainties $\Delta \sigma_0$ and $\Delta v_\mathrm{los}$ from least-square fitting of Gaussians to noisy intrinsic Gaussian profiles are expected to be related to $S/N$ and $\sigma_0$ via
\begin{equation}
  \label{eq:15}
 \Delta v_\mathrm{los} \propto \Delta \sigma_0 \propto \sqrt{\sigma_0}
(S/N)^{-1} 
\end{equation}
\citep[see also][]{Lenz1992}.  We test whether these scaling laws hold for our data by linear regression to the relations
\begin{equation}
  \label{eq:16}
  \Delta v_\mathrm{los} = A_{v_\mathrm{los}}  \left( \frac{1}{2}
  \log_{10}(\sigma_0 [\mathrm{km\,s}^{-1}]) - \log_{10}(S/N) \right ) +
B_{v_\mathrm{los}}
\end{equation}
and
\begin{equation}
  \label{eq:17}
  \Delta \sigma_0 = A_{ \sigma_0}  \left( \frac{1}{2} \log_{10}(\sigma_0
    [\mathrm{km\,s}^{-1}]) - \log_{10}(S/N) \right ) + B_{ \sigma_0}   
\end{equation}
to the results from all 3473 reliable fits from all eLARS galaxies.  The linear regressions result in
\begin{equation}
  \label{eq:18}
 (A_{\sigma_0},B_{\sigma_0}) = (1.02,1.00) 
\end{equation}
and
\begin{equation}
  \label{eq:19}
  (A_{v_\mathrm{los}},B_{v_\mathrm{los}}) = (0.99,0.99) \;\text{.}
\end{equation}
The relations of Eq.~\eqref{eq:16} and Eq.~\eqref{eq:17} with the determined coefficients in Eq.~\eqref{eq:18} and Eq.~\eqref{eq:19} are in near perfect agreement with the expected relations of Eq.~\eqref{eq:15}.

Lastly, the adequateness of the Gaussian profiles is also visually checked for each of the 3473 fits.  As anticipated from above steps, for most of the galaxies, the information content in the observed H$\alpha$ profiles per PMAS spaxel is optimally summarised by the three Gaussian parameters.  Exceptions are the galaxies eLARS 7 and eLARS 24 where numerous spaxels in both galaxies show clear double component H$\alpha$ profiles.  These complicated profiles lead to inflated $\sigma_0$ values in the fits and, moreover, this kinematical complexity is lost in our $v_\mathrm{los}$ and $\sigma_0$ maps.  Accordingly, we also exclude LARS 9 and LARS 13 from the \citetalias{Herenz2016} sample, as both of these galaxies also exhibit double component profiles.  We excluded all galaxies with double component profiles from statistical analyses that involve $\sigma_0^\mathrm{obs}$ and $v_\mathrm{shear} / \sigma_0^\mathrm{obs}$.

\section{Measurement of the intrinsic dispersion}
\label{sec:meas-intr-disp}

The analysis presented in this work requires measurements of the intrinsic velocity dispersion of the H$\alpha$ emitting gas.  This quantity, $\sigma_0$, is defined as a galaxy's average isotropic velocity dispersion.  Simple empirical and model-independent estimates of $\sigma_0$ from observed velocity dispersion maps, $\sigma_{x,y}$, are the mean,
\begin{equation}
  \label{eq:20}
  \sigma_\mathrm{m,u} = \frac{1}{N} \sum_{x,y} \sigma_{x,y} \;\text{,}
\end{equation}
or the weighted mean,
\begin{equation}
  \label{eq:21}
  \sigma_\mathrm{m,w} = \frac{1}{\sum_{x,y} w_{x,y}} \sum_{x,y} w_{x,y} \cdot \sigma_{x,y} \;\text{,}
\end{equation}
where $N$ is the number of spaxels that can be used.  As weights one can adopt the fluxes or the $S/N$'s (since $S/N \sim $ flux) of the emission line in each spaxels.  The flux weighted mean was used in \citetalias{Herenz2016} for the LARS galaxies.

The finite size of a spaxel and the convolution of the galaxy's observable H$\alpha$ emission with the atmospheric point spread function lead to a broadening of the measured $\sigma_i$, when the spaxel $i$ is at the position of a steep velocity gradient.  In analogy to radio astronomy, where a broadening of line widths occurs due to the finite size of the beam, this broadening is sometimes named ``beam smearing effect'' \citep[][]{Davies2011,Varidel2016}.  The simple measures $\sigma_\mathrm{m,u}$ and $\sigma_\mathrm{m,w}$ are thus biased high for galaxies that show strong gradients.  Especially in disk galaxies the intrinsic rotation is steepest near the bright centre of the galaxy, hence $\sigma_\mathrm{m,u}$ appears not to be an ideal measure for the combined LARS and eLARS samples due to its significant fraction of disk- (RD) and disk-like (PR) galaxies (73\,\%; Sect.~\ref{sec:meth-class}).

Kinematical modelling, that accounts for the spatial and instrumental broadening of the observed emission line signal in the datacube, appears to be the most reliable method to obtain a measure of $\sigma_0$ \citep{Davies2011,Bouche2015,Varidel2019}.  However, for CK and PR galaxies such modelling can not produce reliable results, as their kinematics are not captured by simple parametric models\footnote{This problem is illustrated by the application of a rotating disk model to the complex velocity field of the star-burst SBS\,0335-052E by \cite{Isobe2023}.  The recovered kinematical parameters and, especially, the kinematical position angle differ significantly compared to a modified disk model that \cite{Moiseev2010} used for this galaxy.  Their model accounts for a kinematical disturbance caused by an expanding super-bubble.  Such modified models, however, need to be defined on an object by object basis (see especially \citealt{Sardaneta2020} for an idea of a model for eLARS01), which is not practical for a statistical analysis.  }.  Thus, the method can only be applied to less than half of our sample, as it contains 57\% CKs and PRs (Sect.~\ref{sec:meth-class}).  Our approach is therefore to rely on an empirical method that performs best when compared against the results from kinematical modelling of the RD subsample.  For the modelling of the RDs we use the GalPak$^\mathrm{3D}$ software \citep{Bouche2015} and we assert that the recovered dispersion of the model, $\sigma_\mathrm{GP}$, represents the ground truth.

The details on our GalPak$^\mathrm{3D}$ modelling are given in Appendix~\ref{sec:galpak3d-modelling}.  We then present several empirical methods in Appendix~\ref{sec:empirical-methods} before we asses their performance in Appendix~\ref{sec:which-empir-meth}.

\subsection{GalPak$^\mathrm{3D}$ Modelling}
\label{sec:galpak3d-modelling}

\begin{table}
  \caption{Kinematical parameters from GalPak$^\mathrm{3D}$ modelling}
  \label{tab:gppar}
   \centering
  \begin{tabular}{rcccc}
    \hline \hline \noalign{\smallskip} 
    ID & $\sigma_\mathrm{GP,0}$ & $V_\mathrm{max}$ & $R_e$ & $R_t$ \\
    {} & [km\,s$^{-1}$]  & [km\,s$^{-1}$]  & [\arcsec{}] & [\arcsec{}] \\
    \hline \noalign{\smallskip}
    L08 & $43.3 \pm 0.2$ & $210.3 \pm 0.4$ & $3.89 \pm 0.01$ & $2.23 \pm 0.01$ \\
    L11 & $53.6 \pm 0.6$ & $124.2 \pm 0.8$ & $5.42 \pm 0.03$ & $1.38 \pm 0.04$ \\
    E04 & $37.4 \pm 0.3$ & $162.8 \pm 1.5$ & $3.09 \pm 0.01$ & $2.05 \pm 0.03$ \\
    E06 & $25.5 \pm 0.2$ & $99.6 \pm 0.4$  & $3.10 \pm 0.01$ & $0.89 \pm 0.01$ \\
    E08 & $13.9 \pm 0.6$ & $167.0 \pm 2.2$ & $3.77 \pm 0.02$ & $3.00 \pm 0.01$ \\ 
    E10 & $30.4 \pm 0.3$ & $147.4 \pm 0.4$ & $5.49 \pm 0.02$ & $2.99 \pm 0.01$ \\
    E11 & $47.3 \pm 0.2$ & $169.2 \pm 2.2$ & $2.95 \pm 0.01$ & $2.99 \pm 0.01$ \\
    E12 & $30.4 \pm 0.3$ & $147.4 \pm 0.4$ & $5.97 \pm 0.02$ & $2.21 \pm 0.01$  \\
    E14 & $39.0 \pm 0.1$ & $204.5 \pm 2.0$ & $1.34 \pm 0.01$ & $2.99 \pm 0.03$ \\
    E16 & $29.3 \pm 0.4$ & $61.9 \pm 0.7$  & $3.89 \pm 0.04$ & $3.00 \pm 0.01$ \\
    E17 & $20.8 \pm 0.8$ & $110.8 \pm 2.1$ & $3.37 \pm 0.08$ & $2.37 \pm 0.10$ \\
    E18 & $28.5 \pm 0.4$ & $89.0 \pm 0.6$  & $4.55 \pm 0.04$ & $2.89 \pm 0.02$ \\
    E19 & $25.6 \pm 0.2$ & $67.1 \pm 1.1$  & $1.04 \pm 0.01$ & $2.99 \pm 0.01$ \\
    E21 & $25.0 \pm 0.7$ & $50.7 \pm 1.1$  & $4.05 \pm 0.08$ & $2.98 \pm 0.03$ \\
    E27 & $26.7 \pm 0.6$ & $47.3 \pm 0.8$  & $5.99 \pm 0.02$ & $0.03 \pm 0.01$ \\
    \hline \hline
  \end{tabular}
\end{table}

\begin{table}
  \caption{Summarising statistics of the residual maps $v_\mathrm{obs} - v_\mathrm{GP}$ and $\sigma_v - \sigma_\mathrm{GP}$ shown in Figure~\ref{fig:gpcomp}.}
  \label{tab:gpcomp}
  \centering
  \begin{tabular}{crrrr}
    \hline \hline \noalign{\smallskip}
    ID & $\Delta_{v,\mathrm{GP}}$ & $\Delta_{\sigma,\mathrm{GP}}$ & $\mathrm{std}_{v,\mathrm{GP}}$ & $\mathrm{std}_{\sigma,\mathrm{GP}}$ \\
       & [$\mathrm{km\,s^{-1}}$] & [$\mathrm{km\,s^{-1}}$] & [$\mathrm{km\,s^{-1}}$] & [$\mathrm{km\,s^{-1}}$] \\ \hline \noalign{\smallskip}
    L08 & 20.5 & -5.4 & 15.3 & 14.9 \\
    L11 & 30.0 & 2.1 & 21.0 & 16.8 \\
    E04 & 13.2 & 0.6 & 7.1 & 8.8 \\
    E06 & 8.8 & -3.5 & 7.5 & 10.4 \\
    E08 & 7.9 & -1.1 & 7.9 & 10.7 \\
    E10 & 22.3 & -5.0 & 23.0 & 28.5 \\
    E11 & 10.8 & -5.0 & 8.9 & 12.8 \\
    E12 & 11.9 & -2.8 & 15.5 & 12.0 \\
    E14 & 13.2 & -11.6 & 9.4 & 5.7 \\
    E16 & 13.7 & -7.5 & 9.8 & 8.4 \\
    E17 & 8.4 & -2.5 & 5.9 & 7.2 \\
    E18 & 7.4 & -8.3 & 6.1 & 6.6 \\
    E19 & 17.2 & 2.3 & 7.6 & 7.7 \\
    E21 & 10.7 & -0.9 & 7.8 & 9.9 \\
    E27 & 22.6 & -2.6 & 10.1 & 8.3 \\
    \hline \hline
  \end{tabular}
  \tablefoot{$\Delta_{v,\mathrm{GP}}$ is the mean of the
    absolute deviation over all spaxels,
    $\Delta_{v,\mathrm{GP}} = \langle | v_{\mathrm{obs},i} - v_{\mathrm{GP},i} |
    \rangle$, and $\mathrm{std}_{v,\mathrm{GP}}$ is the associated standard
    deviation.  Moreover,
    $\Delta_{\sigma,\mathrm{GP}} = \langle \sigma_{\mathrm{obs},i} -
    \sigma_{\mathrm{GP},i} \rangle$, and $\mathrm{std}_{\sigma,\mathrm{GP}}$ is
    the associated standard deviation.}
\end{table}

\begin{figure}
  \centering
  eLARS 4 
  \includegraphics[width=0.5\textwidth]{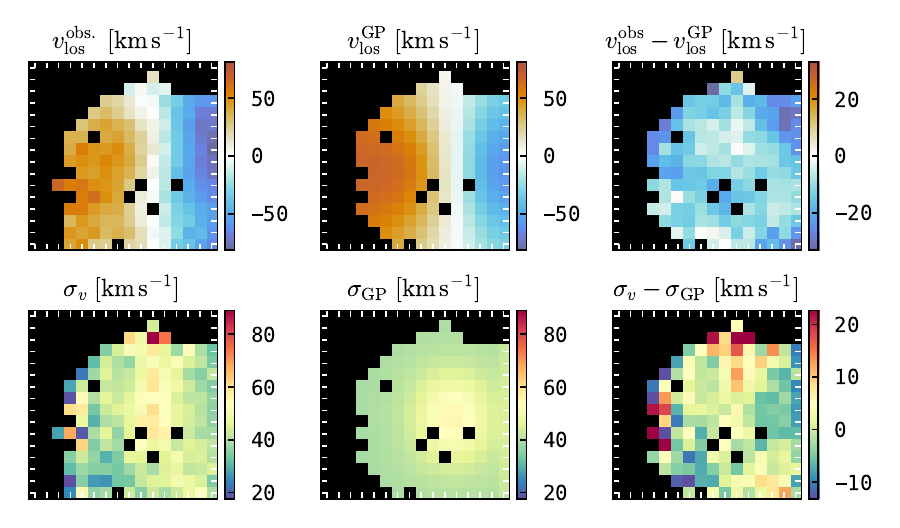}
  eLARS 6
  \includegraphics[width=0.5\textwidth]{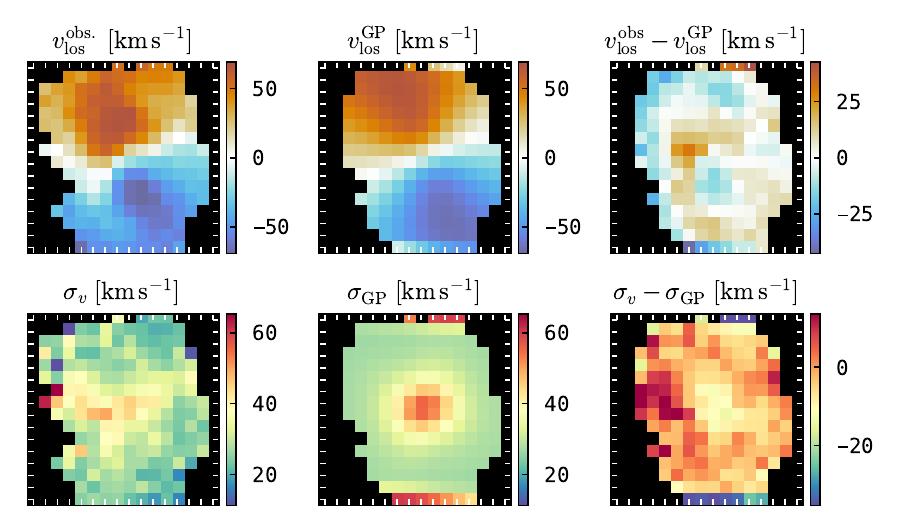}
  eLARS 8
  \includegraphics[width=0.5\textwidth]{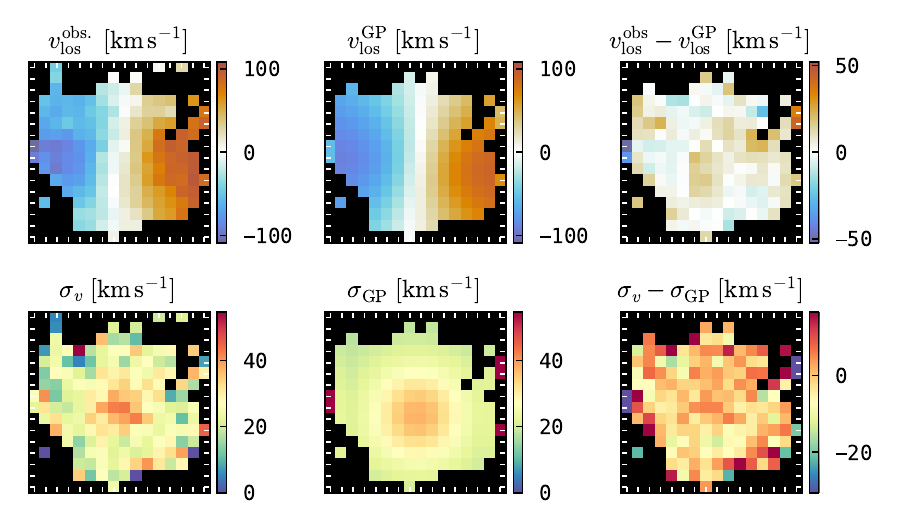}
  eLARS 10
  \includegraphics[width=0.5\textwidth]{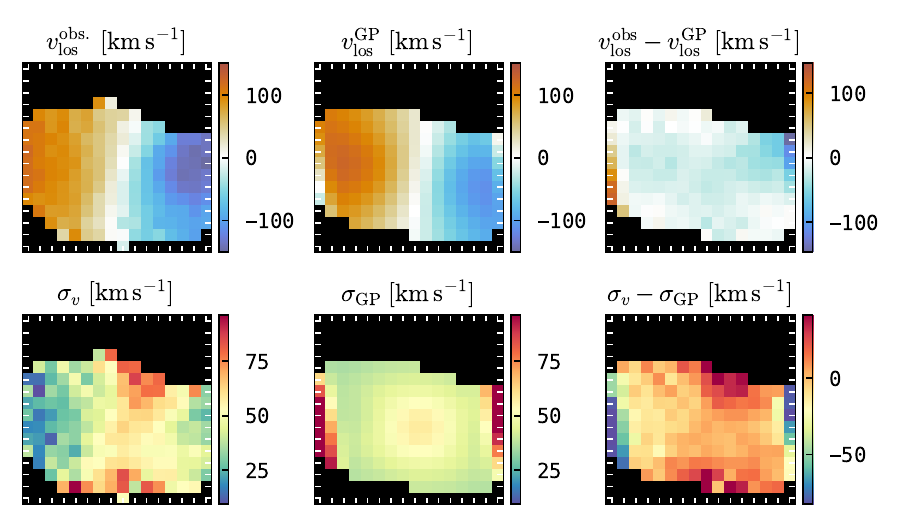}
  \vspace{-1em}
  \caption{Comparison between observed (left column) and modelled (middle
    column) line-of-sight velocity field (top row for each galaxy) and velocity
    dispersion maps (bottom row for each galaxy). The difference between
    observations and model are shown in the right column. }
  \label{fig:gpcomp}
\end{figure}

\begin{figure}
  \centering
  eLARS 11
  \includegraphics[width=0.5\textwidth]{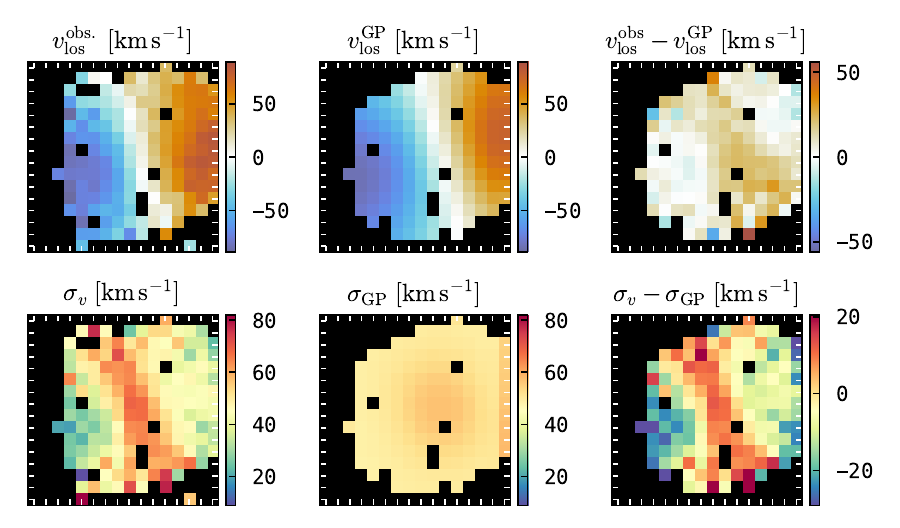}
  eLARS 12
  \includegraphics[width=0.5\textwidth]{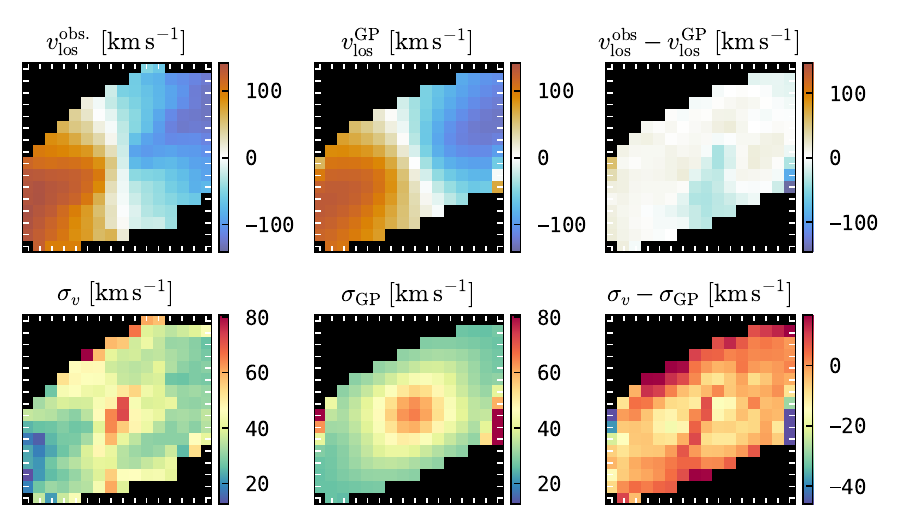}
  eLARS 14
  \includegraphics[width=0.5\textwidth]{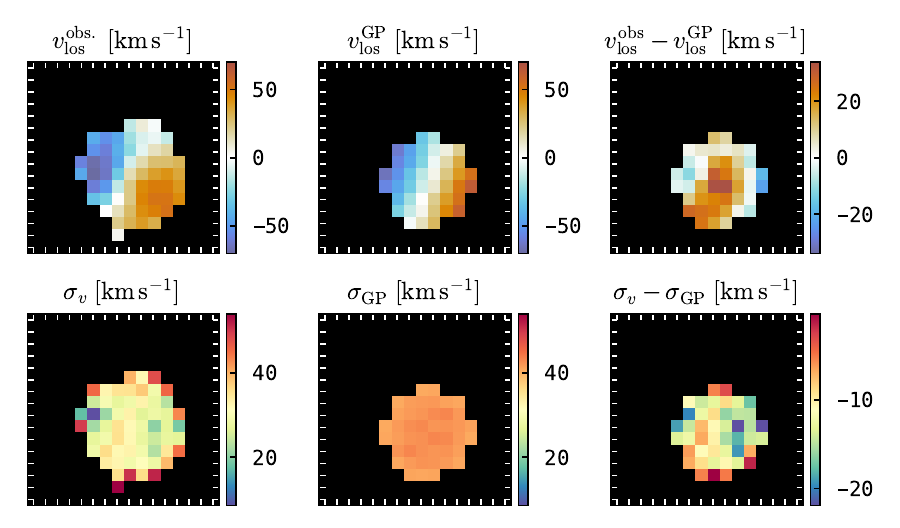}
  eLARS 16
  \includegraphics[width=0.5\textwidth]{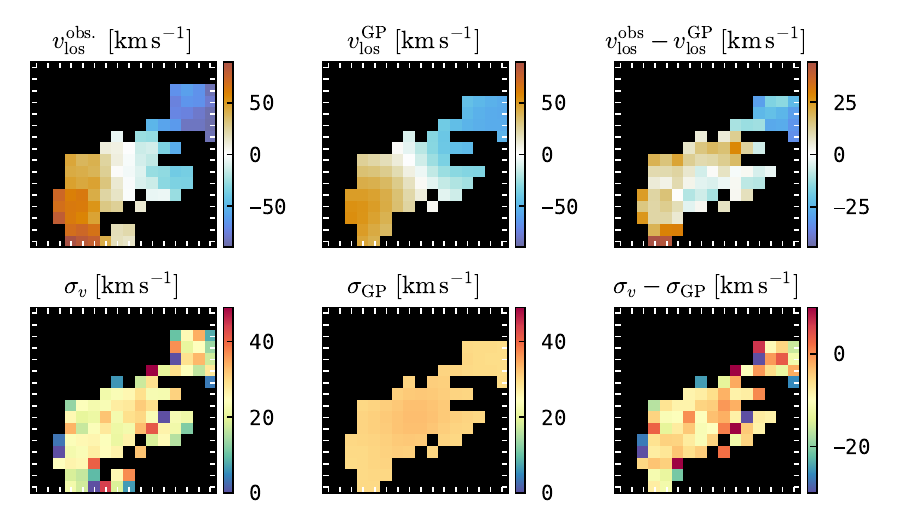}
  \setcounter{figure}{\value{figure}-1}
  \vspace{-1.6em}
  \caption{continued.}
\end{figure}

\begin{figure}
  \centering
  eLARS 17
  \includegraphics[width=0.5\textwidth]{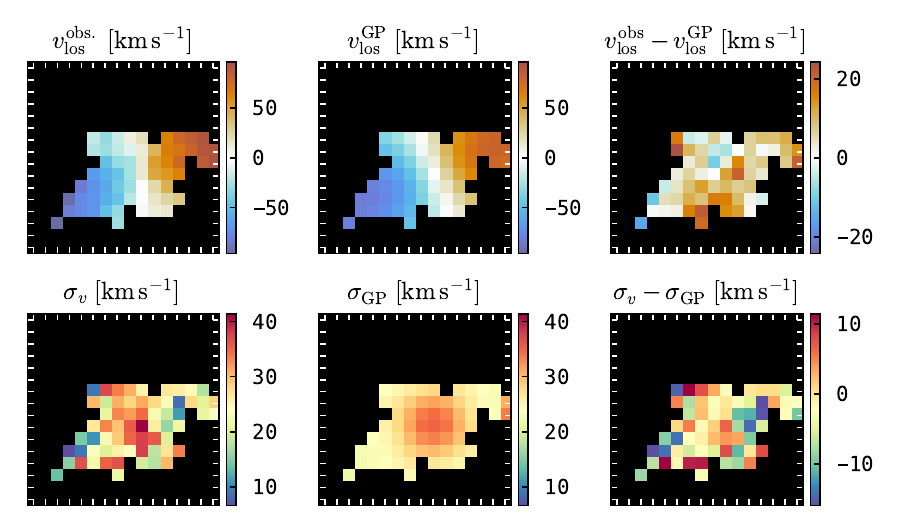}
  eLARS 18
  \includegraphics[width=0.5\textwidth]{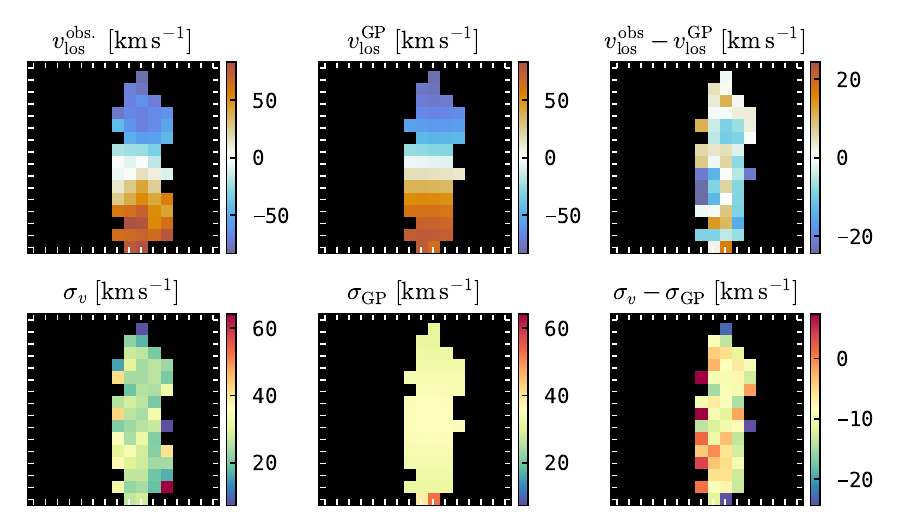}
  eLARS 19
  \includegraphics[width=0.5\textwidth]{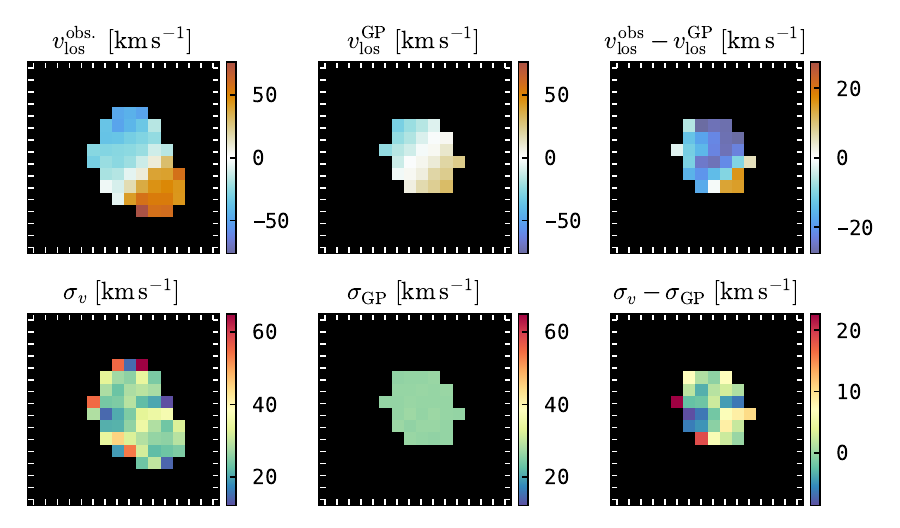}
  eLARS 21
  \includegraphics[width=0.5\textwidth]{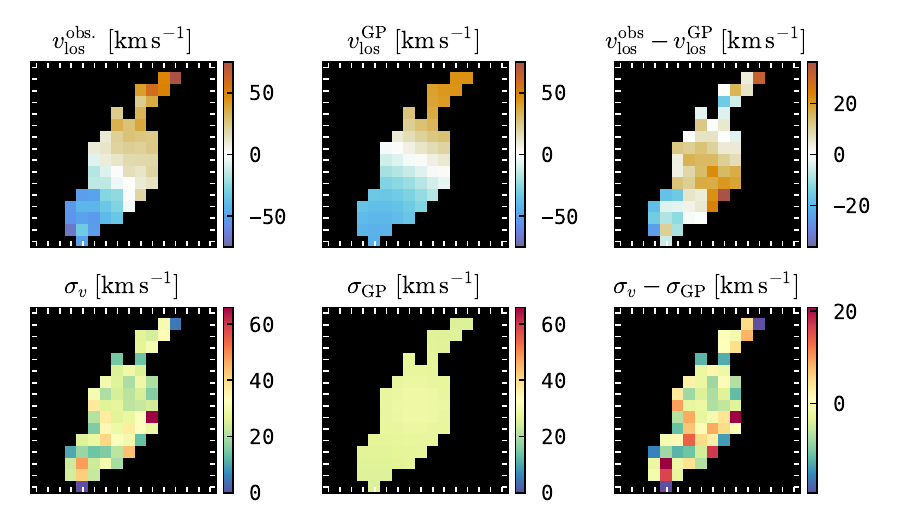}
  \setcounter{figure}{\value{figure}-1}
  \vspace{-1.6em}
  \caption{continued.}
\end{figure}

\begin{figure}
  \centering
  eLARS 27
  \includegraphics[width=0.5\textwidth]{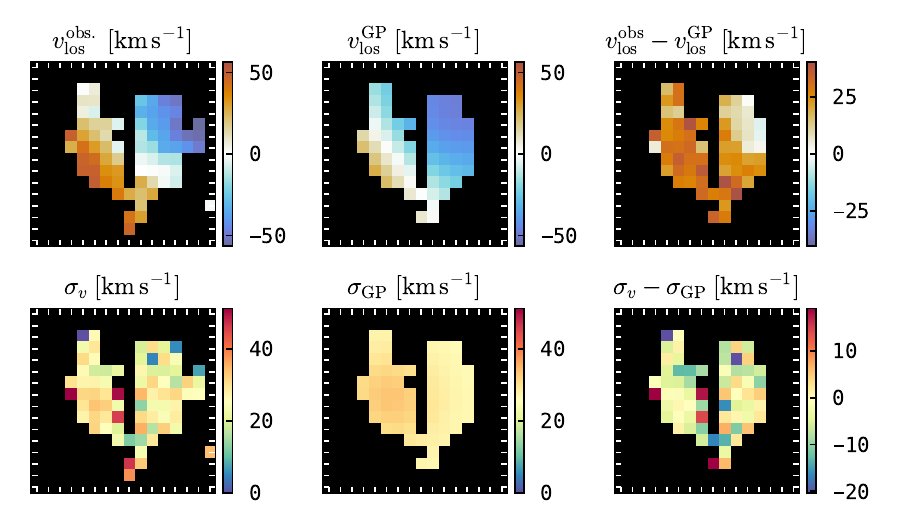}
  LARS 8
  \includegraphics[width=0.5\textwidth]{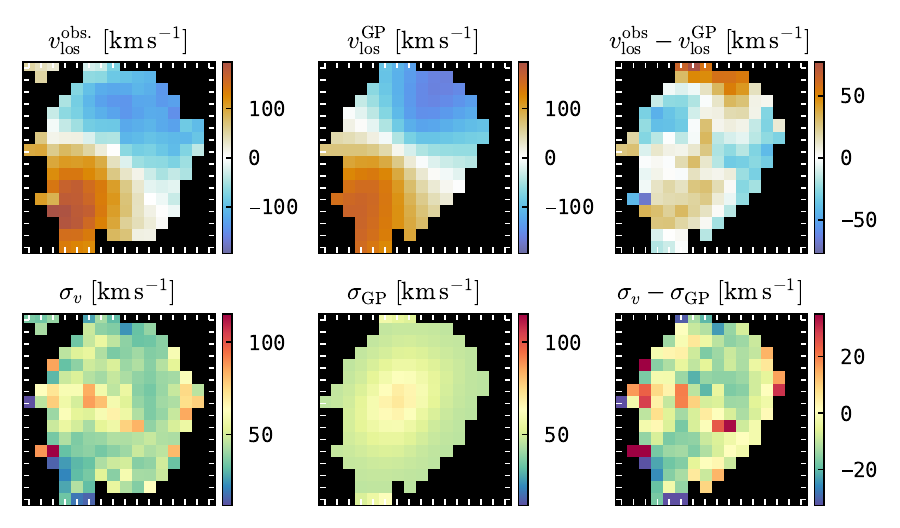}
  LARS 11
  \includegraphics[width=0.5\textwidth]{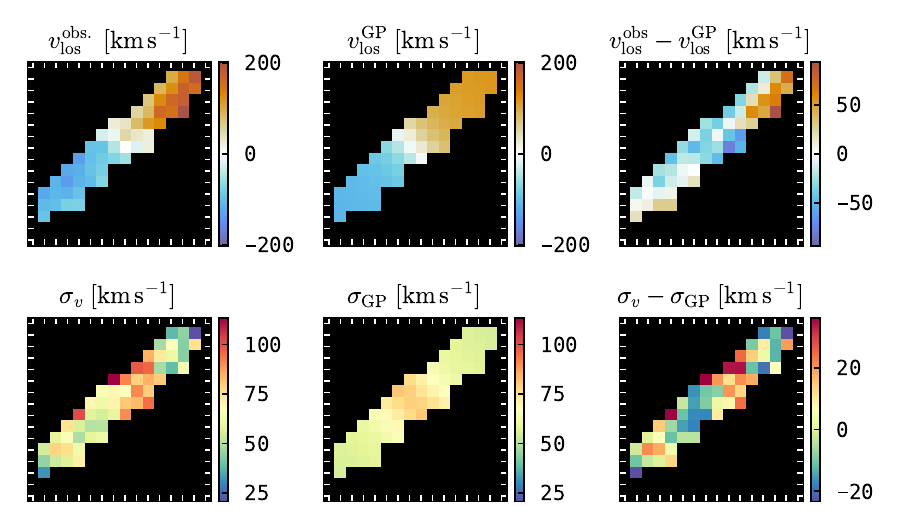}  
  \setcounter{figure}{\value{figure}-1}
  \vspace{-1.6em}
  \caption{continued.}
\end{figure}

GalPak$^\mathrm{3D}$ \citep{Bouche2015} uses a Markov Chain Monte Carlo technique to find the parameter uncertainties of a disk model given the line emission in the IFS datacube.  The defining feature of GalPak$^\mathrm{3D}$ is that the comparison between model and data is directly evaluated on the datacube.  Broadening by the instrument's line spread-function (LSF) and the effect of seeing are accounted for in the modelling by convolution with a 1D and 2D Gaussian, respectively.  The width parameters of these Gaussians have to be provided as input parameters for the model.  As mentioned in Appendix~\ref{sec:deta-descr-kinem}, the width of the LSF varies slightly over the PMAS field-of-view.  We here use the mean LSF width as input parameter (according to the $R$ values provided in Table~\ref{tab:log}).  For the 2D Gaussian that mimics the seeing, we use the average DIMM measure for each observation, also provided in Table~\ref{tab:log}.  For simplicity, we also chose to model only galaxies that fit within one PMAS 16\arcsec{}$\times$16\arcsec{} pointing, i.e. we excluded eLARS 05 from the modelling.  We furthermore exclude eLARS 24 from the modelling due to its double component H$\alpha$ profiles (Sect.~\ref{sec:line-sight-velocity}).

The parametric model assumes the radial light distribution of the disk to follow an exponential profile, $I(r) = I_e \exp \left ( -b_1 [(r/R_e) - 1 ] \right )$, with free parameters $R_e$ and $I_e$, i.e. the half-light radius and the surface-brightness at $R_e$ (this is ensured via the constant $b_1 = 1.678$; \citealt{Graham2005}).  The vertical light distribution of the disk is assumed as exponential, $I(z) \propto \exp( -z / h_z)$, where $h_z$ is fixed to $h_z = 0.15 R_e$.  As rotation curve we chose the empirical tanh-profile, $v(r) = V_\mathrm{max} \tanh (r/R_t)$, with the maximum velocity, $V_\mathrm{max}$, and the turnover radius, $R_t$, being the free parameters.  The velocity distribution of the disk is modelled in GalPak$^\mathrm{3D}$ as three terms that are added in quadrature, $\sigma_\mathrm{GP} = \sqrt{ \sigma_\mathrm{GP,g}^2 + \sigma_\mathrm{GP,m}^2 + \sigma_\mathrm{GP,0}^2}$, where the first term is the local isotropic velocity dispersion driven by self-gravity, $\sigma_\mathrm{GP,g}(r) / h_z = v(r) / r $, the second term, $\sigma_\mathrm{GP,m}$, is a mixing term that arises due to projection effects, and the third term is the for our analysis required isotropic and homogeneous dispersion term $\sigma_\mathrm{GP,0}$.  The model contains three additional free parameters, namely the 3D coordinates of the rotating disks centre.

We list the final results of the disk parameters obtained with the modelling in Table~\ref{tab:gppar} and in Figure~\ref{fig:gpcomp} we compare the observed line of sight velocity and velocity dispersion maps against the maps produced by GalPak$^\mathrm{3D}$.  Moreover, we summarise the level of congruity between the line of sight velocity and dispersion maps in Table~\ref{tab:gpcomp}.  There we list the the mean of the absolute deviations over all spaxels, $\Delta_{v,\mathrm{GP}} = \langle | v_{\mathrm{obs},i} - v_{\mathrm{GP},i} | \rangle$, as well as the standard deviation, $\mathrm{std}_{v,\mathrm{GP}} = \mathrm{std}(| v_{\mathrm{obs},i} - v_{\mathrm{GP},i} |)$, to quantify the match between observed and modelled velocity field.  To quantify the match between the observed dispersion maps and the dispersion maps recovered from the model, we provide the mean, $\Delta_{\sigma,\mathrm{GP}} = \langle \sigma_{\mathrm{obs},i} - \sigma_{\mathrm{GP},i} \rangle$, and the standard deviation, $\mathrm{std}_{\sigma,\mathrm{GP}} = \mathrm{std} ( \sigma_{\mathrm{obs},i} - \sigma_{\mathrm{GP},i} )$, over the residual maps in Table~\ref{tab:gpcomp}.

As can be seen from this comparison, for most of the galaxies the simple parametric model results in line of sight velocity maps that capture qualitatively the large scale kinematics of the H$\alpha$ emitting gas.  However, for none of our galaxies the residuals appear as spatially randomly distributed noise, instead they appear to indicate genuine kinematical features that are not captured by the simple parametric model.  The amplitude of these deviations from the simple disk model are, however, significantly smaller than the overall observed shearing amplitudes.  With $\Delta_{v,\mathrm{GP}} < 10$\,km\,s$^{-1}$, the best level of congruities are found in eLARS 6, eLARS 8, eLARS 17, and eLARS 18, whereas the largest incongruities are found for LARS 11 with $\Delta_{v,\mathrm{GP}} \approx 30$\,km\,s$^{-1}$.  The situation is slightly different for the velocity dispersion maps.  While here a few galaxies display also structures within the residuals that are likely kinematical features not captured by the simple model (eLARS10 and eLARS 11 are prominent examples), for other galaxies the residual maps appear to indicate indeed random noise (e.g., eLARS 8 and eLARS 17).  Nevertheless, the overall match between observed and modelled dispersions is satisfactory, with the absolute mean $|\Delta_{\sigma,\mathrm{GP}}| \lesssim 5$\,km\,s$^{-1}$ for most of the sample, albeit the model tends to result in slightly higher dispersions.  The standard deviation of the residual maps is typically 10\,km\,s$^{-1}$.

The main application of GalPak$^\mathrm{3D}$ is the modelling of spatially unresolved high-$z$ disks.  Given that the here modelled low-$z$ galaxies are spatially resolved into individual sub-structures, a perfect match between model and observations is not expected.  A better fit might be achieved by a model that incorporates the complex gas structure \citep[i.e. BLOBBY3D;][]{Varidel2019}.  Still, the overall congruence between recovered and observed line of sight velocity fields and velocity dispersion maps is quite encouraging.  This is in line with the recent application of GalPak$^\mathrm{3D}$ modelling of low-$z$ low-metallicity star-forming galaxies by \cite{Isobe2023}.  We will thus use $\sigma_{0,\mathrm{GP}}$ values as our ground truth and, given the presented statistics of the residual maps, consider an empirical method as successful in recovering the truth, if it is on average within 10\,km\,s$^{-1}$ of the dispersion values from the model.

\subsection{Empirical Methods}
\label{sec:empirical-methods}

\begin{table}
  \caption{Empirical estimates of the intrinsic velocity dispersion (in km\,s$^{-1}$) obtained with the methods described in Appendix~\ref{sec:empirical-methods}.}
  \label{tab:emp}
  \centering
  \addtolength{\tabcolsep}{-0.2em}
  \footnotesize
  \begin{tabular}{cccccccccc}
    \hline \hline \noalign{\smallskip}
      ID & $\sigma_\mathrm{m,u}^\mathrm{V16}$ & $\sigma_\mathrm{m,w}^\mathrm{V16}$ & $\sigma^\mathrm{Yu21}$ & $\sigma_\mathrm{m,w}$ & $\sigma_\mathrm{m,u}$ & $\sigma^\mathrm{grad}_\mathrm{m,w}$ & $\sigma^\mathrm{grad}_\mathrm{m,u}$ & $N$ & $N_\mathrm{mask}$\\ \noalign{\smallskip} \hline \noalign{\smallskip}
L01 & 13.0 & 44.0 & 43.5  & 46.7 & 47.2 & 46.7 & 47.2 & 106 & 0 \\
L02 & 3.1 & 30.9 & \dots & 37.9 & 38.2 & 37.9 & 38.2 & 45 & 0 \\
L03 & 9.8 & 45.5 & 70.9 & 88.9 & 101.1 & 54.0 & 77.3 & 69 & 47 \\
L04 & 14.4 & 36.7 & \dots & 43.6 & 43.1 & 43.5 & 43.0 & 134 & 1 \\
L05 & 19.2 & 38.0 & 37.9 & 45.6 & 52.0 & 45.4 & 51.2 & 155 & 6 \\
L06 & 1.4 & 15.2 & \dots & 27.2 & 26.2 & 27.2 & 26.2 & 43 & 0 \\
L07 & 10.8 & 53.9 & 60.8 & 58.3 & 60.6 & 58.3 & 60.6 & 74 & 0 \\
L08 & 13.7 & 28.3 & 35.6 & 47.1 & 46.6 & 40.2 & 41.2 & 160 & 71 \\
L09 & \dots & \dots & \dots & 56.5 & 58.8 & 55.5 & 56.1 & 232 & 76 \\
L10 & 6.2 & 28.2 & 36.4 & 36.5 & 36.5 & 36.5 & 36.5 & 76 & 0 \\
L11 & 4.4 & 30.6 & 54.1 & 65.3 & 66.3 & 58.2 & 60.9 & 55 & 28 \\
L12 & 7.9 & 54.5 & \dots & 70.0 & 70.6 & 66.3 & 64.8 & 59 & 16 \\
L13 & \dots & \dots & \dots & 65.5 & 62.0 & 65.4 & 61.0 & 100 & 43 \\
L14 & 3.2 & 50.3 & \dots & 65.2 & 59.4 & 65.5 & 59.0 & 30 & 3 \\
E01 & 29.7 & 45.1 & \dots & 57.2 & 56.2 & 56.8 & 55.6 & 213 & 4 \\
E02 & 5.5 & 27.6 & 32.7 & 32.8 & 31.2 & 32.8 & 31.2 & 74 & 0 \\
E03 & 8.8 & 20.4 & \dots & 48.3 & 45.0 & 42.1 & 40.7 & 321 & 41 \\
E04 & 13.8 & 25.4 & 39.5 & 45.1 & 43.7 & 45.1 & 43.7 & 161 & 0 \\
E05 & 9.7 & 18.4 & \dots & 46.9 & 41.8 & 39.4 & 37.7 & 321 & 21 \\
E06 & 12.0 & 20.4 & 25.4 & 30.7 & 29.6 & 29.0 & 28.7 & 179 & 12 \\
E07 & 4.9 & 16.3 & \dots & 38.2 & 40.2 & 29.1 & 30.6 & 109 & 34 \\
E08 & 8.0 & 16.5 & 26.3 & 26.6 & 25.5 & 26.4 & 25.4 & 159 & 2 \\
E09 & 3.2 & 22.7 & 30.7 & 32.0 & 32.8 & 32.0 & 32.8 & 54 & 0 \\
E10 & 8.4 & 14.4 & 31.7 & 47.5 & 47.4 & 43.4 & 43.7 & 161 & 26 \\
E11 & 11.9 & 23.7 & 33.2 & 50.4 & 47.2 & 50.2 & 46.7 & 164 & 7 \\
E12 & 12.1 & 19.6 & 26.2 & 37.7 & 37.6 & 33.5 & 35.4 & 185 & 27 \\
E13 & 2.6 & 47.1 & \dots & 63.0 & 61.7 & 63.0 & 61.7 & 27 & 0 \\
E14 & 3.5 & 21.1 & 24.9 & 29.6 & 32.0 & 28.4 & 31.9 & 62 & 9 \\
E15 & 3.6 & 20.7 & 26.0 & 35.5 & 32.1 & 34.9 & 31.1 & 74 & 6 \\
E16 & 3.3 & 14.3 & 20.9 & 24.1 & 23.9 & 24.1 & 23.9 & 87 & 0 \\
E17 & 1.7 & 8.8 & 18.5 & 27.0 & 25.2 & 26.2 & 24.7 & 65 & 3 \\
E18 & 2.7 & 17.0 & 22.5 & 26.0 & 26.4 & 26.0 & 26.4 & 58 & 0 \\
E19 & 2.6 & 19.3 & 29.9 & 26.9 & 28.9 & 26.6 & 28.2 & 50 & 3 \\
E20 & 6.7 & 21.9 & 33.9 & 34.8 & 36.9 & 34.8 & 36.9 & 101 & 0 \\
E21 & 2.8 & 15.2 & 23.6 & 26.4 & 26.6 & 26.4 & 26.6 & 61 & 0 \\
E22 & 2.8 & 28.6 & \dots & 38.2 & 40.4 & 37.7 & 39.5 & 41 & 9 \\
E23 & 5.5 & 17.9 & 17.6 & 23.1 & 23.9 & 23.1 & 23.9 & 103 & 17 \\
E24 & 6.6 & 54.6 & 70.1 & 99.5 & 90.8 & 77.3 & 78.0 & 45 & 26 \\
E25 & 6.6 & 23.6 & 29.5 & 33.8 & 34.1 & 33.8 & 34.1 & 113 & 15 \\
E26 & 2.5 & 19.9 & \dots & 35.0 & 33.2 & 32.0 & 30.9 & 124 & 10 \\
E27 & 2.2 & 12.5 & 27.2 & 27.1 & 26.6 & 27.1 & 26.6 & 75 & 0 \\
E28 & 5.9 & 20.6 & 25.1 & 26.7 & 25.8 & 26.7 & 25.8 & 102 & 13 \\
 \hline \hline
\end{tabular}
\tablefoot{Typical statistical uncertainties on these measurements are $\sim 0.5$\,km\,s$^{-1}$.  The last two columns indicate the number of spaxels with reliable $\sigma_{x,y}$ measurements, $N$, and the number of spaxels masked, $N_\mathrm{mask}$, according Eq.~\eqref{eq:22} and Eq.~\eqref{eq:23} with $v_{g,\mathrm{thresh}} = 75$\,km\,s$^{-1}$.}
\end{table}

\begin{figure*}
  \sidecaption
  \begin{minipage}{12cm}
    \includegraphics[width=12cm]{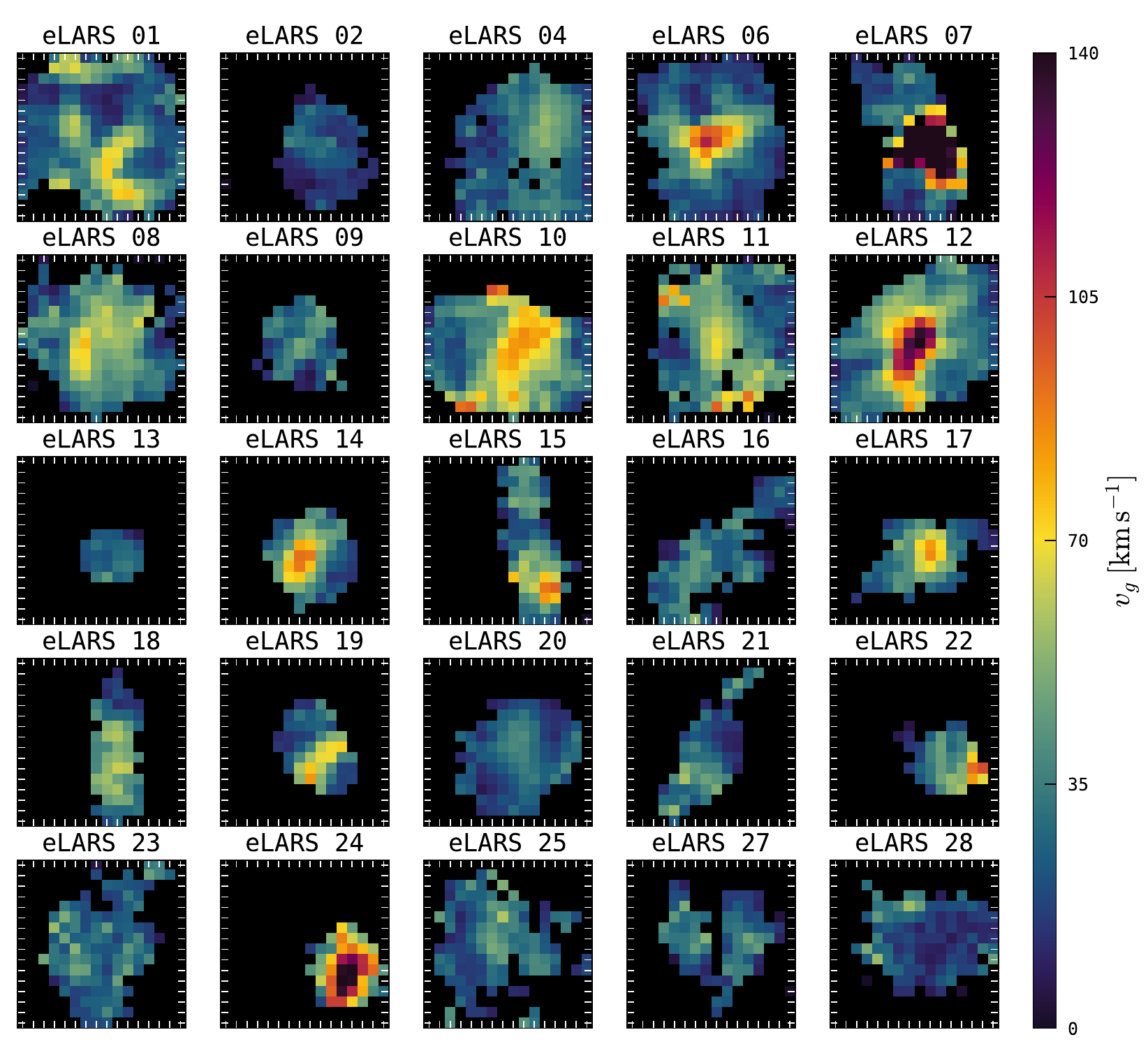}
    \includegraphics[align=c,width=3.9cm,trim=50 0 120 0,clip=true]{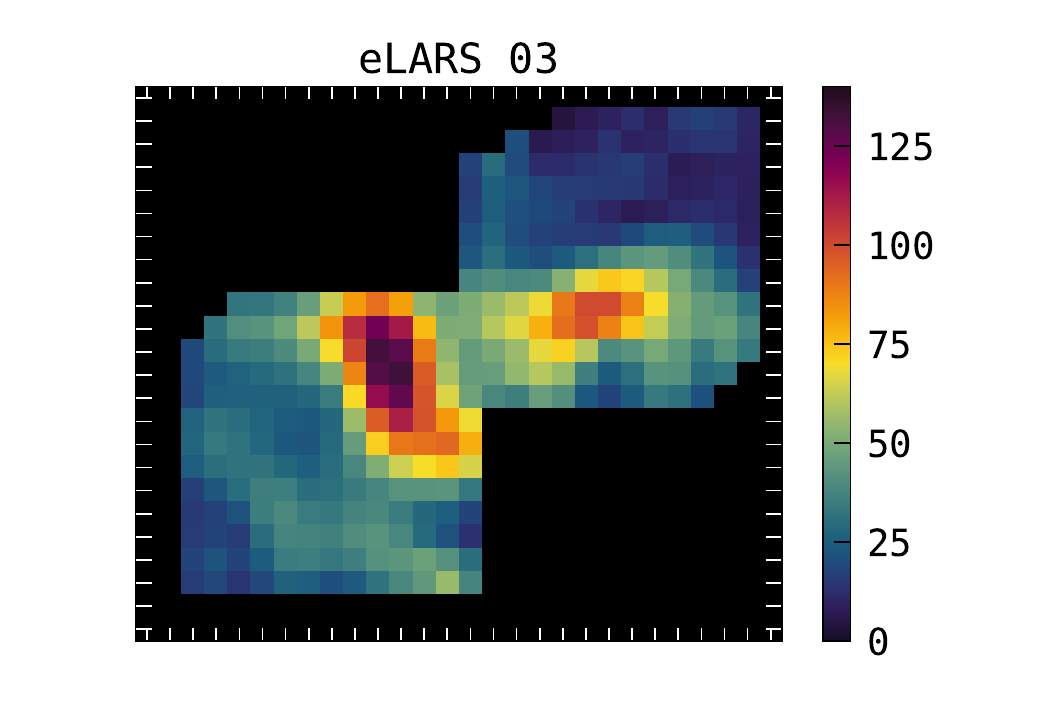}
    \includegraphics[align=c,width=3.9cm,trim=50 0 120 0,clip=true]{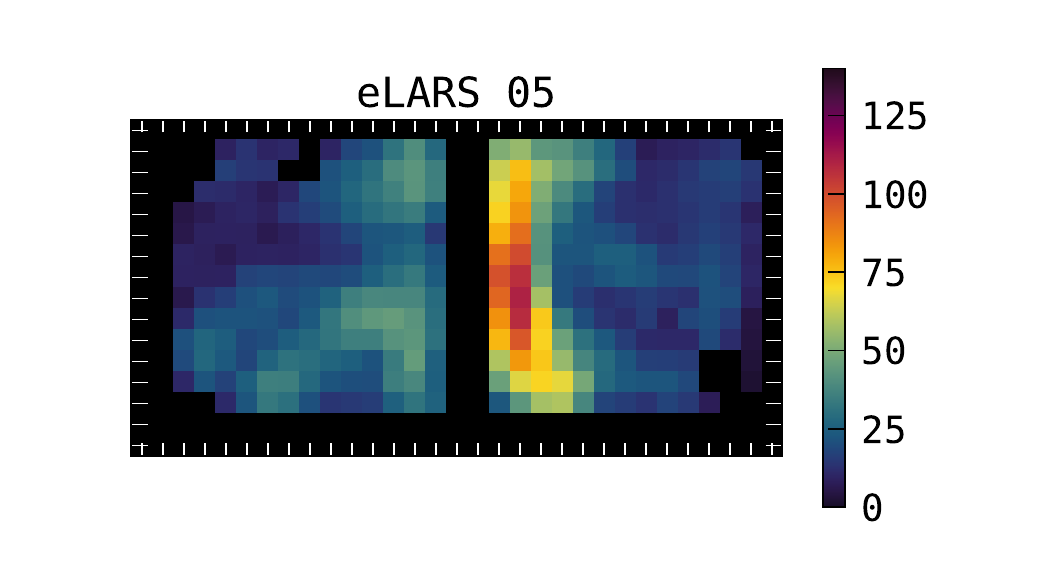}
    \includegraphics[align=c,width=3.9cm,trim=50 0 120 0,clip=true]{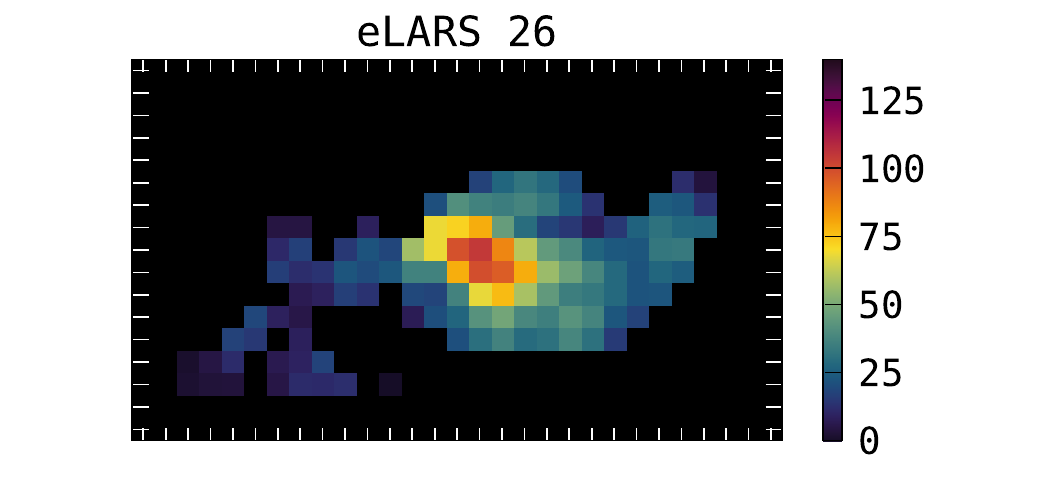}
  \end{minipage}
  \caption{Gradient maps according Eq.~\eqref{eq:22} for the eLARS galaxies.}
  \label{fig:grad}
\end{figure*}

\begin{figure*}
  \centering
  \large{RD} \\
  \includegraphics[width=0.95\textwidth]{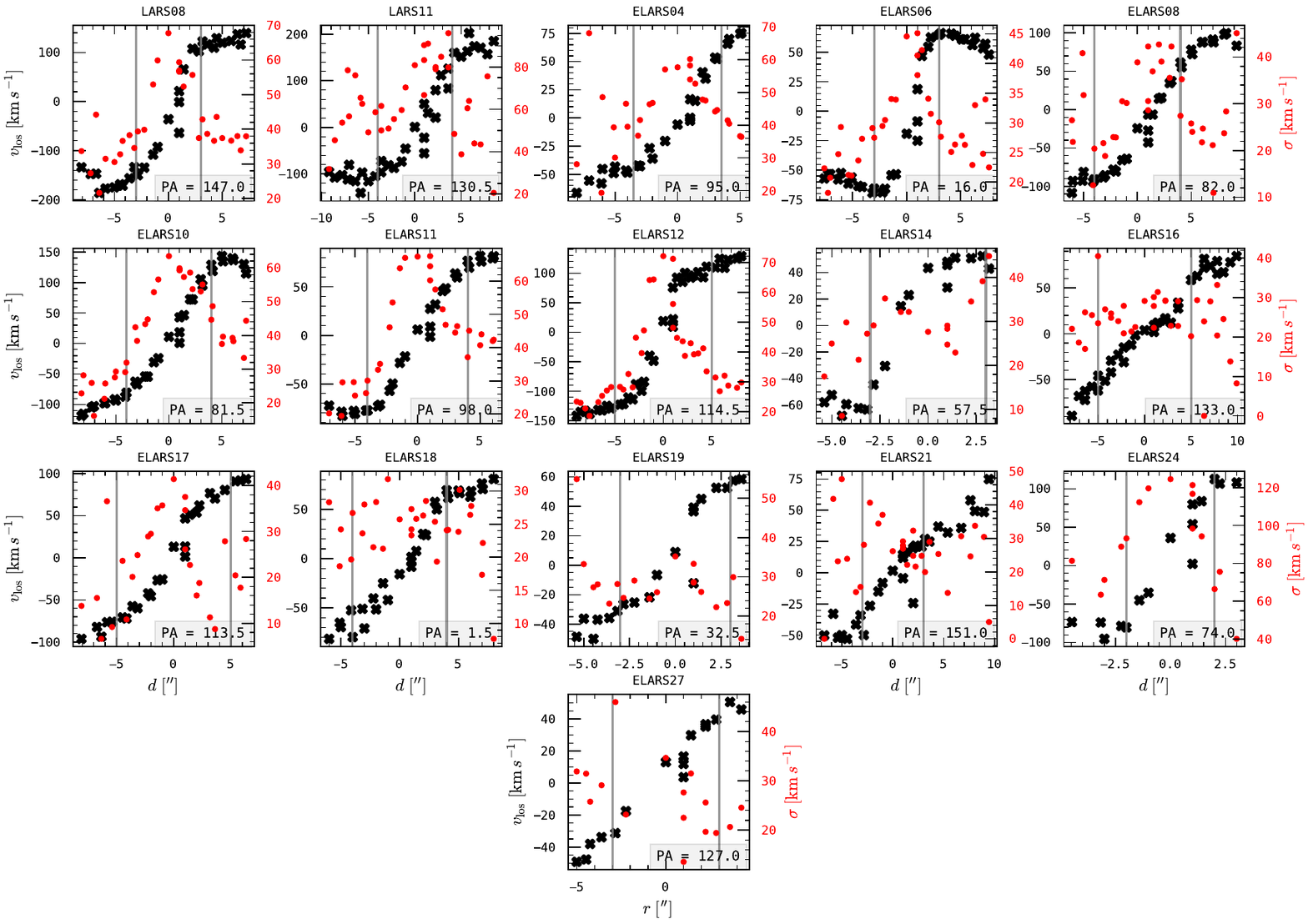} \\
  \large{PR} \\
  \includegraphics[width=0.95\textwidth]{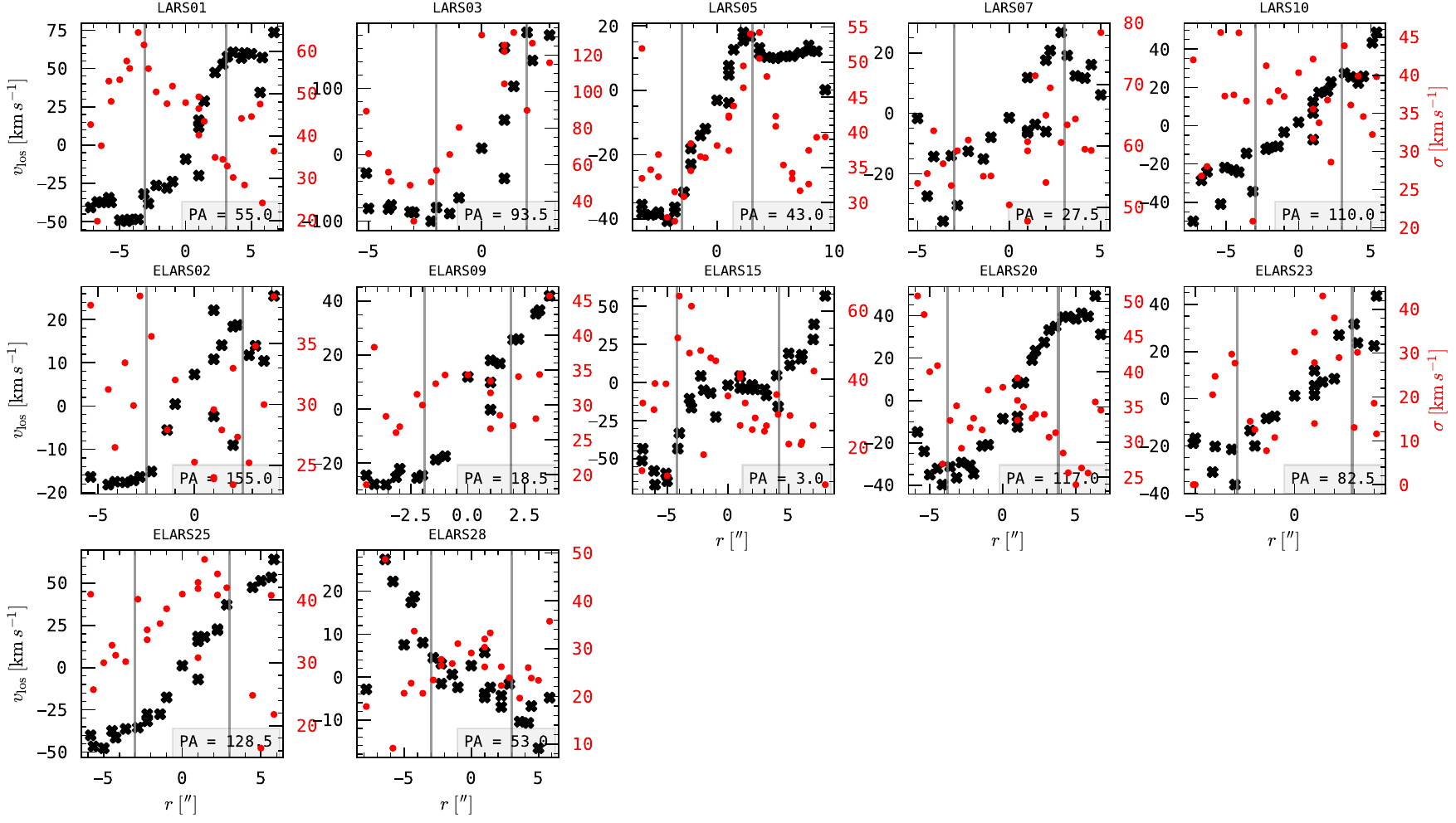}
  \caption{Radial line of sight velocity profiles (left axes, black crosses) and
    velocity dispersion profiles for (right axes, red circles) for the RD (top
    panels) and PR (bottom panels) galaxies.  The position angle is shown in the
    bottom right of each panel.  Vertical lines show the radius beyond which the
    velocity dispersion measurements were not deemed to be affected by PSF
    smearing.  }
  \label{fig:yu21}
\end{figure*}

The simplest heuristic method for estimating the intrinsic velocity dispersion are the means introduced already in Eq.~\eqref{eq:20} and Eq.~\eqref{eq:21}. We list in Table~\ref{tab:emp} $\sigma_\mathrm{m,u}$ and $\sigma_\mathrm{m,w}$ according to those equations.  For $\sigma_\mathrm{m,w}$ we use the $S/N$ as weights in the calculation.  We also provide in Table~\ref{tab:emp} the number of spaxels, $N$, that contribute to the respective means.

As mentioned, the simple- and the weighted mean can be biased high due to beam smearing.  A conceptually straight forward workaround is masking out spaxels in the calculation of $\sigma_\mathrm{m,w}$ and $\sigma_\mathrm{m,u}$ that are significantly affected by the PSF induced line broadening.  As these spaxels are in the vicinity of strong velocity gradients, we here introduce a gradient-like measure for each spaxel, $v_g$, calculated with respect to all eight neighbouring spaxels on the observed line of sight velocity map.  Our gradient-like map is thus defined by
\begin{equation}
  \label{eq:22}
  v_g(x,y) = \sqrt{ \sum_{i = (-1, 0, +1)} \sum_{j = (-1, 0, +1)} [v_{x-i,y-j} -
    v_{x,y}]^2 } \;\text{.}
\end{equation}
Empty spaxels, i.e. spaxels that did not meet the signal-to-noise criterion or were affected by cosmic-ray hits (Sect.~\ref{sec:line-sight-velocity}), are set to zero in the calculation.  We show the so obtained $v_g(x,y)$ maps for the eLARS galaxies in Figure~\ref{fig:grad}.  We then define a binary mask,
\begin{equation}
  \label{eq:23}
  m(x,y) =
  \begin{cases}
    1 &\Leftrightarrow v_g(x,y) \leq v_{g,\mathrm{thresh}} \\
    0 &\Leftrightarrow v_g(x,y) > v_{g,\mathrm{thresh}}
  \end{cases}\; \text{,}
\end{equation}
to calculate masked versions of the mean (Eq.~\ref{eq:20}),
\begin{equation}
  \label{eq:24}
  \sigma^\mathrm{grad}_\mathrm{m,u} = \frac{1}{\sum_{x,y} m(x,y)} \sum_{x,y} m(x,y) \cdot \sigma_{x,y} \;\text{,}
\end{equation}
and the weighted mean (Eq.~\ref{eq:21}),
\begin{equation}
  \label{eq:25}
  \sigma^\mathrm{grad}_\mathrm{m,w} = \frac{1}{\sum_{x,y} m(x,y) \cdot w_{x,y}} \sum_{x,y} m(x,y) \cdot w_{x,y} \cdot \sigma_{x,y} \;\text{,}
\end{equation}
respectively.  Based on visual inspection of Figure~\ref{fig:grad}, and by iteratively assessing the performance of $\sigma^\mathrm{grad}_\mathrm{m,u}$ and $\sigma^\mathrm{grad}_\mathrm{m,w}$ against $\sigma_\mathrm{GP,0}$ from the GalPak$^\mathrm{3D}$ modelling (Sect.~\ref{sec:which-empir-meth}), we settled on a masking threshold of $v_\mathrm{g,\mathrm{thresh}} = 70$\,km\,s$^{-1}$.

We list in Table~\ref{tab:emp} the $\sigma^\mathrm{grad}_\mathrm{m,u}$ and $\sigma^\mathrm{grad}_\mathrm{m,w}$ computed with Eq.~\eqref{eq:24} and Eq.~\eqref{eq:25}, respectively, alongside the number of masked spaxels, $N_\mathrm{masked}$, for which $v_g(x,y) > 70$\,km\,s$^{-1}$.  For galaxies without large shearing amplitudes within a small solid angle we have $v_g(x,y) < 70$\,km\,s$^{-1}$ for all spaxels, hence $N_\mathrm{masked} = 0$ and therefore $\sigma^\mathrm{grad}_\mathrm{m,u} \equiv \sigma_\mathrm{m,u}$ and $\sigma^\mathrm{grad}_\mathrm{w,u} \equiv \sigma^\mathrm{grad}_\mathrm{m,u}$.  For galaxies where $N_\mathrm{mask} > 0$, we always have $\sigma^\mathrm{grad}_\mathrm{m,w} < \sigma_\mathrm{m,w}$ and $\sigma^\mathrm{grad}_\mathrm{m,u} < \sigma_\mathrm{m,u}$.  In these cases the corrections of both means from the masking are mostly small ($\lesssim 5$\,km\,s$^{-1}$) with the exceptions being LARS 3 and eLARS 24, where the correction amounts to 24\,km\,s$^{-1}$ (35\,km\,s$^{-1}$) and 12\,km\,s$^{-1}$ (22\,km\,s$^{-1}$) for the weighted (unweighted) mean.  The latter galaxy shows strongly double peaked H$\alpha$ emission, however the mask actually covers the spaxels, resulting in a strong downward correction due to the exclusion of the artificially broadened spaxels.  LARS 3 (cf. Figure~5 in \citetalias{Herenz2016}), on the other hand, is a perturbed rotator with a large shearing amplitude and here numerous spaxels that appear to be severely affected by beam smearing get masked.

\cite{Varidel2016} made an alternative suggestion to correct empirically for the beam smearing effect.  Their method starts by calculating a gradient-like measure similar to Eq.~\eqref{eq:22}, but only uses the 4-connected topology, i.e.,
\begin{equation}
  \label{eq:26}
  v_g^\mathrm{V16}(x,y) =
  \sqrt{\left(v_{x+1,y} - v_{x-1,y} \right )^2 + \left (v_{x,y+1} - v_{x,y-1}
    \right)^2} \; \text{.}
\end{equation}
This measure is then fitted to a linear 2D model that expresses $\sigma_{x,y}$ as combination of a constant term, a linear term in flux, and a linear term in $v_g^\mathrm{V16}(x,y)$:
\begin{equation}
  \label{eq:27}
  \sigma_{x,y} = m_\mathrm{H\alpha} \cdot \log_{10} (F_\mathrm{H\alpha})_{x,y} +
  m_{v_g} \cdot v_g^\mathrm{V16}(x,y) + c  \; \mathrm{.}
\end{equation}
The rationale beyond Eq.~\eqref{eq:27} is, according to \cite{Varidel2016}, that beam smearing will be dependent on the received flux and the velocity gradient.  A beam-smearing corrected map of velocity dispersion is then obtained by fitting Eq.~\eqref{eq:27} to each spaxel, before creating a beam-smearing corrected map from the fits by setting $m_{v_g} \equiv 0$: $\sigma_{x,y}^\mathrm{V16} = m_\mathrm{H\alpha} \cdot \log_{10} (F_\mathrm{H\alpha})_{x,y} + c$.  Then, the mean, $\sigma_\mathrm{m,u}^\mathrm{V16}$, and the weighted mean, $\sigma_\mathrm{m,w}^\mathrm{V16}$, are calculated via Eq.~\eqref{eq:20} and Eq.~\eqref{eq:21}, where $\sigma_{x,y}^\mathrm{V16}$ is used instead of the observed $\sigma_{x,y}$.  We list the so obtained values for $\sigma_\mathrm{m,u}^\mathrm{V16}$ and $\sigma_\mathrm{m,w}^\mathrm{V16}$ in Table~\ref{tab:emp}.  Both the $\sigma_\mathrm{m,u}^\mathrm{V16}$ and $\sigma^\mathrm{V16}_\mathrm{m,w}$ values are significantly smaller than the other empirical measures introduced above.

As final empirical method we here test the scheme put forward by \cite{Yu2021} \citep[see also][]{Wisnioski2015}.  The ansatz of this method is to avoid spaxels affected by beam smearing via a visual inspection of the radial line of sight velocity profile along the kinematical major axis.  Following \cite{Yu2021}, we extract the kinematical major axis from our line of sight velocity fields with the tool \texttt{fit\_kinematic\_pa}\footnote{\url{https://pypi.org/project/pafit/}} \citep{Krajnovic2006}.  The line of sight velocity profiles and velocity dispersion profiles are then created by considering spaxels $\pm 1.5$\arcsec{} above and below the kinematical major axis.  We show the so obtained line of sight velocity profiles and velocity dispersion profiles for the RDs and PDs in our sample in Figure~\ref{fig:yu21}.  \cite{Yu2021} suggest to visually inspect the profiles to define a radius where beam smearing is not deemed to be relevant, i.e. ideally where the rotation curve becomes flat.  We follow this method, but note that for some galaxies we do not sample the flat part of the rotation curve.  However, for those galaxies the velocity gradient also appears to be shallower, and thus the spaxels are not as strongly affected by PSF smearing.  Averaging the measured dispersions (red points in Figure~\ref{fig:yu21}) in the outer part of the velocity dispersion profiles then provides us with empirical velocity dispersion that we denominate $\sigma^\mathrm{Yu21}$.  We tabulate $\sigma^\mathrm{Yu21}$ alongside the other empirical velocity dispersion measures in Table~\ref{tab:emp}.  For most of the analysed galaxies $\sigma^\mathrm{Yu21}$ is larger than the weighted and unweighted dispersions according to the \citet{Varidel2016} method, but smaller than the other empirical dispersion measures.

\subsection{Which empirical method performs best?}
\label{sec:which-empir-meth}

\begin{figure*}
  \centering
 \includegraphics[width=0.9\textwidth]{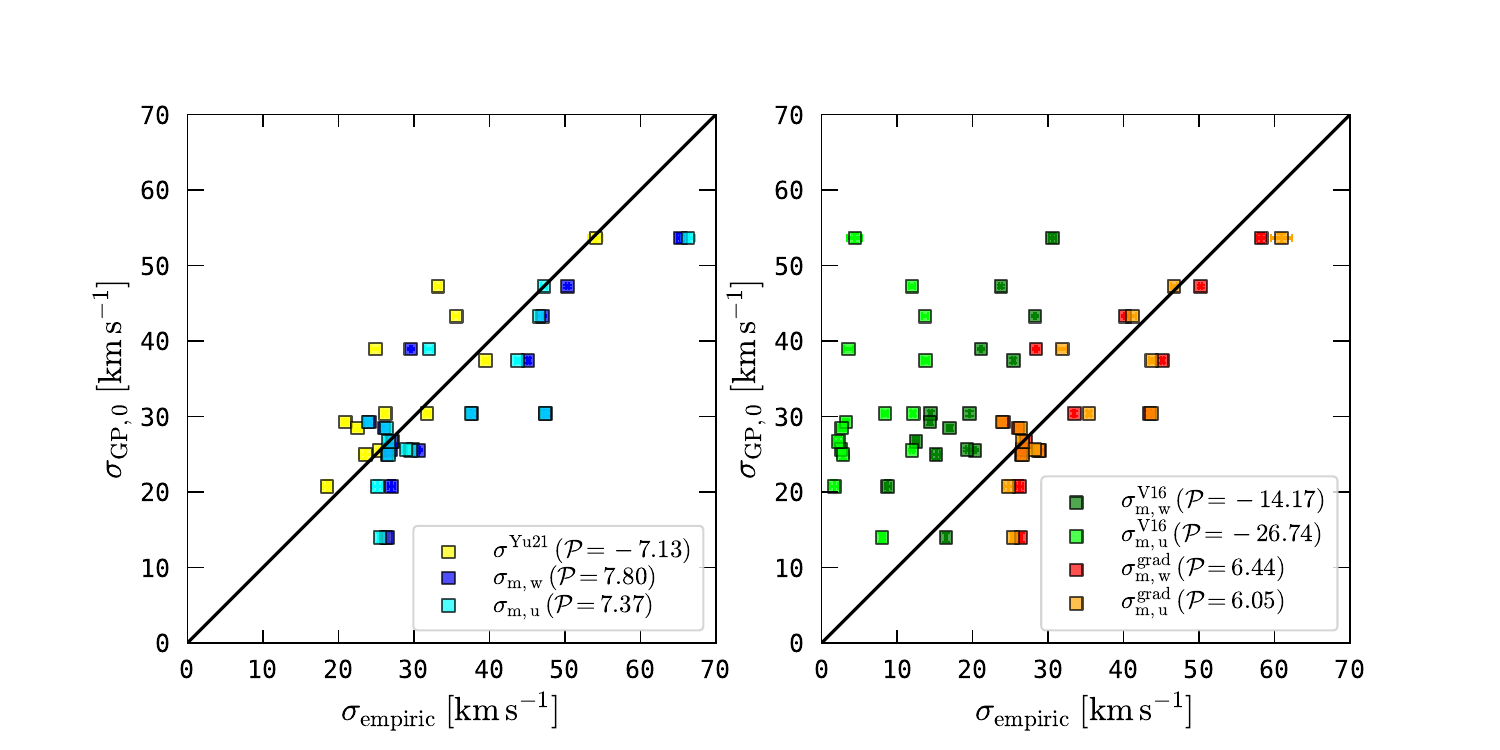}
  \vspace{-1em}
  \caption{Comparison between velocity dispersions calculated with the empirical
    methods and the velocity dispersion from the GalPak$^\mathrm{3D}$ modelling.
    For each method the $\mathcal{P}$ measure from Eq.~\eqref{eq:a1} is given
    in the legend.}
  \label{fig:cmpgp}
\end{figure*}

\begin{table*}
  \caption{Difference (in km\,s$^{-1}$) between velocity dispersions calculated
    with the empirical methods and the intrinsic velocity dispersion
    $\sigma_{\mathrm{GP},0}$ of the GalPak$^\mathrm{3D}$ models.  The last row
    shows $\mathcal{P}$ according to Eq.~\eqref{eq:a1}.}
  \label{tab:sigdiff}
  \centering
   \addtolength{\tabcolsep}{-0.2em}
  \begin{tabular}{lrrrrrrr}
    \hline \hline \noalign{\smallskip}
ID &    $\sigma_\mathrm{m,w}^\mathrm{V16} - \sigma_{\mathrm{GP},0}$ & $\sigma_\mathrm{m,u}^\mathrm{V16} - \sigma_{\mathrm{GP},0}$  & $\sigma^\mathrm{Yu21} - \sigma_{\mathrm{GP},0}$ & $\sigma_\mathrm{m,w} - \sigma_{\mathrm{GP},0}$ & $\sigma_\mathrm{m,u} - \sigma_{\mathrm{GP},0}$ & $\sigma_\mathrm{m,w}^\mathrm{grad} - \sigma_{\mathrm{GP},0}$ & $\sigma_\mathrm{m,u}^\mathrm{grad} - \sigma_{\mathrm{GP},0}$ \\ \noalign{\smallskip} \hline \noalign{\smallskip}
L08 & -15.0 & -29.6 & -7.7 & 3.8 & 3.3 & -3.1 & -2.1 \\
L11 & -23.1 & -49.2 & 0.5 & 11.7 & 12.7 & 4.6 & 7.3 \\
E04 & -12.0 & -23.7 & 2.1 & 7.7 & 6.3 & 7.7 & 6.3 \\
E06 & -5.2 & -13.6 & -0.1 & 5.1 & 4.1 & 3.4 & 3.2 \\
E08 & 2.5 & -5.9 & 12.4 & 12.6 & 11.6 & 12.4 & 11.4 \\
E10 & -16.0 & -22.0 & 1.3 & 17.1 & 16.9 & 13.0 & 13.3 \\
E11 & -23.5 & -35.3 & -14.0 & 3.1 & -0.0 & 3.0 & -0.5 \\
E12 & -10.8 & -18.3 & -4.2 & 7.3 & 7.2 & 3.1 & 5.0 \\
E14 & -17.8 & -35.4 & -14.1 & -9.4 & -7.0 & -10.6 & -7.1 \\
E16 & -14.9 & -26.0 & -8.4 & -5.2 & -5.3 & -5.2 & -5.3 \\
E17 & -12.0 & -19.1 & -2.2 & 6.3 & 4.4 & 5.5 & 4.0 \\
E18 & -11.5 & -25.8 & -6.0 & -2.4 & -2.1 & -2.4 & -2.1 \\
E19 & -6.3 & -23.1 & 4.3 & 1.3 & 3.3 & 0.9 & 2.6 \\
E21 & -9.8 & -22.1 & -1.4 & 1.5 & 1.7 & 1.5 & 1.7 \\
E27 & -14.2 & -24.5 & 0.5 & 0.4 & -0.1 & 0.4 & -0.1 \\ \hline \noalign{\smallskip}
$\mathcal{P}$ & -14.2 & -26.7 & -7.1 & 7.8 & 7.3 & 6.4 & 6.1 \\   
    \hline \hline
\end{tabular}
\end{table*}

We asses the performance of the empirical methods by comparing them to the velocity dispersion from the GalPak$^\mathrm{3D}$ model.
We use the metric introduced by
\cite{Davies2011} for this purpose:
\begin{equation}
  \label{eq:a1}
  \mathcal{P} = \mathcal{S} \left [ \frac{1}{n} \sum_{k=1}^{n}
    \left ( \sigma_\mathrm{emp}^k - \sigma_\mathrm{GP,0}^k \right )^2  \right
  ]^{1/2} \;\text{,}
\end{equation}
where $\mathcal{S} = \langle \sigma_\mathrm{emp}^k - \sigma_\mathrm{GP}^k \rangle / | \langle \sigma_\mathrm{emp}^k - \sigma_\mathrm{GP}^k \rangle |$ is either $+1$ or $-1$ depending on whether the empirical velocity dispersions are, on average, an underestimate or overestimate of the true velocity dispersion, respectively.  The sum in Eq.~\eqref{eq:a1} runs over all galaxies.  For each galaxy we list in Table~\ref{tab:sigdiff} the individual differences, $\sigma_\mathrm{emp} - \sigma_\mathrm{GP,0}$ for every empirical method discussed in the previous section.  A graphical comparison is given in Figure~\ref{fig:cmpgp}.  The values for $\mathcal{P}$ are given both in the legend of Figure~\ref{fig:cmpgp} and in the last row of Table~\ref{tab:sigdiff}.

As discussed above (Appendix~\ref{sec:galpak3d-modelling}), we consider methods as useful if they result in $|\mathcal{P}| < 10$\,km\,s$^{-1}$.  Our comparison shows that the method proposed by \citet{Varidel2016} severely underestimates the velocity dispersion.  The ad-hoc parameterisation of the beam-smearing effect proposed by those authors appears thus to over correct the actual bias.  On the other hand, the method adopted by \cite{Yu2021} performs excellent for most galaxies, with $|\sigma^\mathrm{Yu21} - \sigma_{\mathrm{GP},0}| < 5$\,km\,s$^{-1}$ for 9 out of 15 galaxies, and $\mathcal{P} = -7.1$\,km\,s$^{-1}$.  This method thus fulfils our acceptance criterion of 10\,km\,s$^{-1}$, but unfortunately it is not applicable to the CK galaxies of our sample.

Fortunately, we find an acceptable positive bias for the weighted and unweighted mean, and the bias even slightly decreases by 1\,km\,s$^{-1}$ when our masking scheme (Eqs.~\ref{eq:22}--\ref{eq:25}, with $v_{g,\mathrm{thresh}} = 70$\,km\,s$^{-1}$) is applied (see last four columns in Table~\ref{tab:sigdiff}).  Based on this analysis, we thus adopt $\sigma_\mathrm{m,u}^\mathrm{grad}$ as our observational estimate, $\sigma_0^\mathrm{obs}$, of the intrinsic velocity dispersion for our analysis.

\section{Robustness of the correlations}
\label{sec:robustn-corr}

\begin{figure}
  \centering
  \includegraphics[width=0.249\textwidth]{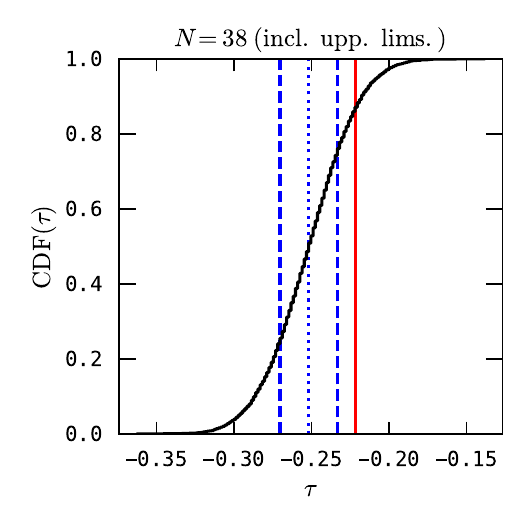}\includegraphics[width=0.249\textwidth]{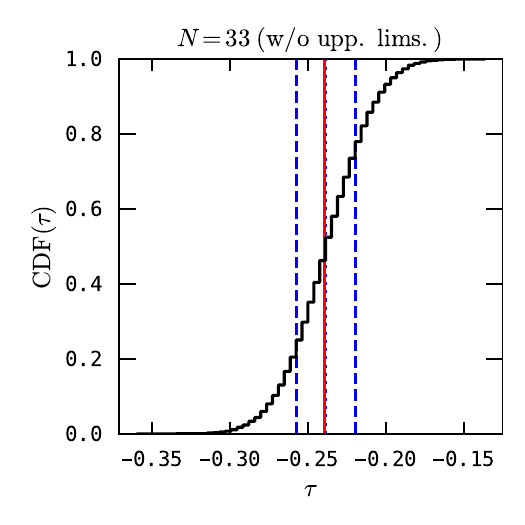}
  \caption{Examples of the adopted methodology to assess the robustness of the
    correlations presented in Appendix~\ref{sec:robustn-corr}.  Both panels show
    the cumulative distribution of Kendall's $\tau$ obtained from
    $10^4$ Monte-Carlo simulations, where for each realisation the involved
    variables were perturbed according to their error-bars.  The green dotted
    line shows the median of the distribution, which is adopted to calculate
    $p_0$, whereas the blue dashed lines show the upper- and lower quartile.
    The solid red line shows the critical $\tau$ value according to
    Eq.~\eqref{eq:30}.  Both panels show the cumulative distribution of $\tau$
    for the relation between $v_\mathrm{shear}/\sigma_0^\mathrm{obs}$ and
    $f_\mathrm{esc}^\mathrm{Ly\alpha}$, but the distribution in the left-hand
    panel is obtained for the sample of $N=38$ galaxies analysed in
    Sect.~\ref{sec:corr-betw-glob} (exclusion of galaxies with double component
    H$\alpha$ profiles), whereas the distribution in the right-hand
    panel is obtained for the sub-sample of $N=33$ galaxies that is analysed in
    Sect.~\ref{sec:mp} (galaxies without upper limits in their Ly$\alpha$
    observables).}
  \label{fig:rob}
\end{figure}

Throughout we used the Kendall's $\tau$ \citep[e.g.,][]{Puka2011} to measure the strength of a possible monotonic relation between two sets of observables.  In order to treat upper limits in these calculation we follow the formalism of \cite{Isobe1986}.  To derive error-bars for the correlation coefficients and to asses their robustness we follow \cite{Curran2014} and perform a Monte-Carlo simulation\footnote{Both the computation of the generalised Kendall's $\tau$ and the Monte-Carlo simulation were implemented in python and are available at \url{https://github.com/Knusper/kendall}.}.

In our simulation we generate $10^4$ new sets of observations of our sample with
each datum being perturbed according to its uncertainty.  We then analyse the
resulting distributions of $\tau$-values from this simulation.  The quoted
$\tau$ values always refer to the median of the distribution.  Their error-bars
are derived from the interquartile-range, $q_{75} - q_{25}$, via the relation
$\sigma_\tau = (q_{75} - q_{25}) / (2 \sqrt{2} \mathrm{erf}^{-1} (1/2))$, where
$\mathrm{erf}^{-1}(x)$ denotes the inverse of the error function
$\mathrm{erf}(x)$.  The $p_0$ value is calculated from the median of the
distribution in $\tau$'s from the Monte-Carlo experiment.  We use the
approximation that the distribution under the null-hypothesis of no
correlation (i.e., random pairings of the $N$ observables) in
$|\tau|\, /\sqrt{\mathrm{Var}(\tau)}$ is normal.  In this case the expected
variance of $\tau$ is given by $\mathrm{Var}(\tau) = (4N+10) / (9N(N-1))$.  This
allows for a closed form expression of $p_0$ for the two-tailed test:
\begin{equation}
  \label{eq:29}
  p_0 = 1 - \mathrm{erf}\left( \frac{|\tau|}{\sqrt{2\, \mathrm{Var}(\tau)}}\right)  \;\text{.}
\end{equation}
We deem a correlation as robust if the upper- ($q_{75}$) or lower quartile
($q_{25}$) does not overlap with the threshold in $\tau$ that corresponds to a
p-value.  Again assuming a normal distribution of
$|\tau|/\sqrt{\mathrm{Var}(\tau)}$ under the null-hypothesis, this allows for a
closed expression of the threshold:
\begin{equation}
  \label{eq:30}
  |\tau| > \sqrt{2 \mathrm{Var}(\tau)}\, \mathrm{erfc}^{-1}(p_0) \;\text{,}
\end{equation}
with $\mathrm{erfc}^{-1}(p_0)$ denoting the inverse of the complementary error function of $p_0$.  In our study we encounter the following sample sizes: $N=\{42,\, 38,\, 33,\, 15\}$.  For these $N$, and adopting $p_0=0.05$, Eq.~\eqref{eq:30} then results in the thresholds\footnote{Evaluating the exact distribution of $|\tau|$ under the null hypothesis, which can not be expressed as a simple formula, results in marginally different thresholds, $|\tau| > \{0.213,\,0.223,\, 0.242,\, 0.390\}$, but without any implication for our results.} $|\tau| > \{0.210,\,0.221,\, 0.239,\, 0.337\}$.  These different $N$ or $\tau$ values correspond to analyses involving $v_\mathrm{shear}$ as one parameter, analyses involving $\sigma_0^\mathrm{obs}$ or $v_\mathrm{shear}/\sigma_0^\mathrm{obs}$ (i.e., excluding galaxies with double component profiles) as one parameter, analyses excluding all upper limits in the Ly$\alpha$ observable (i.e., all analyses presented in Sect.~\ref{sec:how-important-are}), and correlation analyses that rely on measured kinematic inclinations (Sect.~\ref{sec:incl-depend-lyalpha}), respectively.

To exemplify this procedure we show in Figure~\ref{fig:rob} two cumulative
distributions of $\tau$'s from the Monte Carlo simulation -- one for which the
correlation is found to be robust (left panel) and one for which the correlation
is found not to be robust (right panel).

\section{Extinction Corrected Starburst Ages}
\label{sec:extinct-corr-starb}

\begin{figure}
  \centering
  \includegraphics[width=0.4\textwidth]{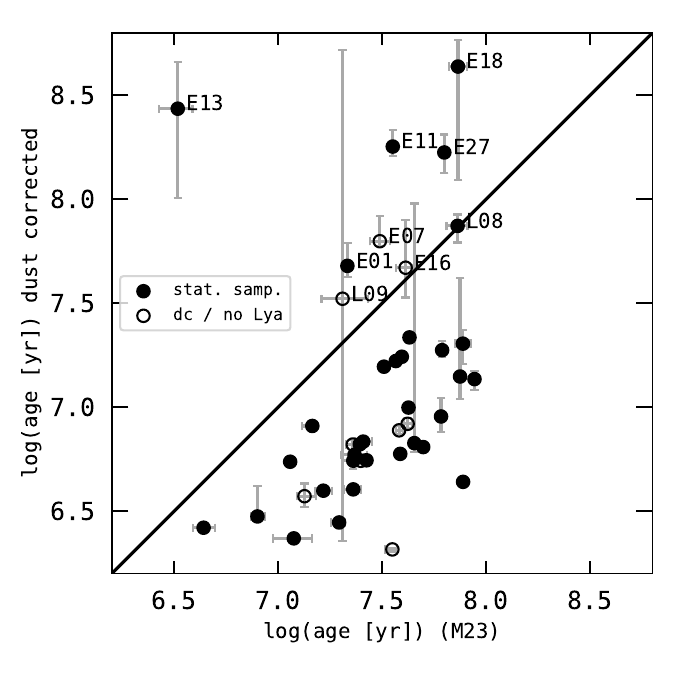}
  \caption{Comparison of extinction corrected UV luminosity weighted  starburst
    ages to UV luminosity weighted ages from \citetalias{Melinder2023}. Filled points indicate the sample used for all statistical analyses, i.e. excluding galaxies without double component H$\alpha$ profiles and galaxies without upper limits in their Ly$\alpha$ observables; the latter are shown as open symbols.  The one-to-one relation is indicated as a solid black line and we label the nine galaxies where the dust-corrected age estimate is larger than the new age estimate.}
  \label{fig:age_comp}
\end{figure}

\begin{table}
  \caption{Extinction corrected UV luminosity weighted ages.}
  \label{tab:newage}
  \centering
  \begin{tabular}{cccc} \hline \hline
    ID & $\log(\mathrm{Age\,[yr]})$ & $\Delta_+$ & $\Delta_-$ \\ \hline
    L01 & 6.475 & 0.145 & 0.018 \\
    L02 & 6.446 & 0.021 & 0.023 \\
    L03 & 6.641 & 0.011 & 0.008 \\
    L04 & 6.738 & 0.012 & 0.013 \\
    L05 & 6.421 & 0.004 & 0.004 \\
    L06 & 6.572 & 0.061 & 0.051 \\
    L07 & 6.822 & 0.015 & 0.016 \\
    L08 & 7.872 & 0.055 & 0.080 \\
    L09 & 7.522 & 1.196 & 1.166 \\
    L10 & 6.889 & 0.021 & 0.017 \\
    L11 & 7.195 & 0.016 & 0.018 \\
    L12 & 6.910 & 0.011 & 0.013 \\
    L13 & 6.316 & 0.009 & 0.010 \\
    L14 & 6.369 & 0.006 & 0.005 \\
    E01 & 7.680 & 0.111 & 0.055 \\
    E02 & 7.222 & 0.012 & 0.009 \\
    E03 & 6.745 & 0.008 & 0.009 \\
    E04 & 7.243 & 0.006 & 0.005 \\
 E05 & 6.808 & 0.014 & 0.021 \\
 E06 & 7.336 & 0.027 & 0.025 \\
 E07 & 7.798 & 0.120 & 0.021 \\
 E08 & 6.827 & 1.152 & 0.044 \\
 E09 & 6.998 & 0.013 & 0.015 \\
 E10 & 7.274 & 0.042 & 0.032 \\
 E11 & 8.254 & 0.081 & 0.046 \\
 E12 & 6.920 & 0.028 & 0.020 \\
 E13 & 8.436 & 0.225 & 0.427 \\
 E14 & 6.821 & 0.023 & 0.025 \\
 E15 & 7.148 & 0.474 & 0.106 \\
 E16 & 7.671 & 0.229 & 0.143 \\
 E17 & 6.955 & 0.089 & 0.074 \\
 E18 & 8.638 & 0.127 & 0.547 \\
 E19 & 6.771 & 0.052 & 0.048 \\
 E20 & 6.743 & 0.043 & 0.039 \\
 E21 & 7.306 & 0.068 & 0.099 \\
 E22 & 6.599 & 0.012 & 0.011 \\
 E23 & 6.776 & 0.025 & 0.022 \\
 E24 & 6.742 & 0.023 & 0.016 \\
 E25 & 6.605 & 0.020 & 0.017 \\
 E26 & 7.135 & 0.041 & 0.050 \\
 E27 & 8.225 & 0.086 & 0.100 \\
 E28 & 6.835 & 0.023 & 0.025 \\ \hline \hline
  \end{tabular}
\end{table}

In \citetalias{Melinder2023} we provided a global starburst age estimate that was weighted by the FUV luminosities.  However, this age estimate is biased towards regions that show no or little extinction.  Since star clusters form in dusty regions, this age estimate may thus biased towards older populations that have cleared their natal dust reservoirs.  When analysing the age dependence of $\sigma_0^\mathrm{obs}$, $\mathrm{EW}_\mathrm{Ly\alpha}$, and $f_\mathrm{esc}^\mathrm{Ly\alpha}$ in Sect.~\ref{sec:sigma_vs_age} we instead opted for using an age estimate that is weighted by the dust corrected UV continuum luminosity.  In Table~\ref{tab:newage}  we provide these new age estimates, and in Figure \ref{fig:age_comp} we compare the new extinction corrected age estimates to the values provided in \citetalias{Melinder2023}.  As in \citetalias{Melinder2023} we compute the error-bars on the age estimates by a Monte-Carlo simulation, where the photometric measurements in each Voronoi cell are perturbed by the corresponding error estimates.  As anticipated, for the majority of galaxies we obtain younger ages when weighing the age estimates by the dust-corrected UV luminosities.

\end{document}